\documentclass{revtex4}
\usepackage{graphicx}
\begin{document}
\title{On the modulation instability development in optical fiber systems}

\author{D.S. Agafontsev}

\affiliation{\small \textit{P.N. Lebedev Physical Institute, 53 Leninsky ave., 119991 Moscow, Russia}}

\begin{abstract}
Extensive numerical simulations were performed to investigate all stages of modulation instability development from the initial pulse of pico-second duration in photonic crystal fiber: quasi-solitons and dispersive waves formation, their interaction stage and the further propagation. Comparison between 4 different NLS-like systems was made: the classical NLS equation, NLS system plus higher dispersion terms, NLS plus higher dispersion and self-steepening and also fully generalized NLS equation with Raman scattering taken into account. For the latter case a mechanism of energy transfer from smaller quasi-solitons to the bigger ones is proposed to explain the dramatical increase of rogue waves appearance frequency in comparison to the systems when the Raman scattering is not taken into account.
\end{abstract}

\maketitle

Although the considerable progress was made toward the understanding of the physical nature of rogue waves in open ocean \cite{Dyachenko1, Dyachenko2, Kharif_review, Dysthe_review}, the theory for them is far from been complete especially in the sense of prediction of their appearance frequency. Therefore, the recent observations of similar structures in optical fibers \cite{Solli} called optical rogue waves instantly attracted much scientific interest not only because such extreme events had not been observed before in other physical systems except oceanic waves, but also because optical fibers grant almost ideal experimental conditions for their further investigation. By their origin, optical rogue waves are quasi-solitons raised during the modulation instability (MI) development which propagate with almost constant speed and shape after they exit region of interactions with other quasi-solitons. The term "extreme event" refers here to the process of their appearance: since MI development is very sensitive to noise, extremely large waves rise as rare outcomes from an almost identically prepared initial population of waves. In this sense optical rogue waves strongly differ from their oceanic counterparts, which are usually described as large waves suddenly appear from nowhere and disappear without a trace (\cite{Kharif_review, Dysthe_review, Akhmediev}). Nevertheless, recent numerical experiments revealed existence of more or less stable hydrodynamical rogue waves (see \cite{Ruban}), that gives hope that the relationship between optical and oceanic rogue waves may be more explicit as was thought hitherto.

In this work the process of MI development from initial pulse of pico-second duration in anomalous group velocity regime in photonic crystal fiber (PCF) is considered when the different linear and nonlinear terms beyond the classical NLSE are taken into account. With the help of numerical simulations it is shown that MI development in general consists of three stages: initial quasi-solitons and dispersive waves formation, their interactions and the further noninteracting propagation. The applicability of this scenario is verified for initial pulses with peak power from 50W to more than 10kW. The main attention is paid to quasi-solitons because of their direct connection to the optical rogue waves. The start of quasi-solitons formation process is shown to be very well predicted by the classical NLSE. Concerning the further propagation lengths, it is demonstrated that while the additional higher order dispersion and self-steepening terms affect MI process only through the symmetry breaking and the generation of small dispersive waves, the Raman scattering dramatically changes it's every stage. By consideration of quasi-soliton to quasi-soliton collisions it is shown that among other effects, Raman scattering include an effective mechanism of nonlinear energy transfer from smaller quasi-solitons to the bigger ones. Comparison of rogue waves appearance frequency depending on the different linear and nonlinear terms beyond the classical NLSE taken into account is also made. It is shown that in presence of Raman scattering the probability distribution function for waves heights has long non-exponential tail, while in the other cases large waves appearance frequency exponentially decays with the wave amplitude. It is supposed that the latter circumstances can be explained by the mechanism of the energy transfer from smaller quasi-solitons to the bigger ones which is provided by the Raman scattering.

To investigate the process of MI developed in PCF from initial pico-second pulse of sech-shape type there were made 3 groups of numerical experiments. Evolution of a pulse in PCF was described by the following equation (see \cite{Agrawal, DGC}):
\begin{equation}
i\frac{\partial A}{\partial z}-\hat\beta A + \gamma\bigg(1+i\tau_{shock}\frac{\partial}{\partial t}\bigg)\bigg(A\int_{-\infty}^{+\infty}R(t-t')|A(z,t')|^{2}dt'\bigg)=0,  \label{1_OpticsEnvelopeEvolution}
\end{equation}
with the same parameters for all 3 sets of experiments:
\begin{eqnarray}
&&\hat\beta = \frac{\beta_{2}}{2}\frac{\partial^{2}}{\partial t^{2}} + \hat\beta_{HD}, \quad \hat\beta_{HD} = \sum_{n=3}^{n=7}\frac{\beta_{n}}{n!}\frac{\partial^{n}}{\partial t^{n}}, \label{1_OpticsEnvelopeEvolutionCoeff} \\ 
&&\beta_2 = -12.76\times 10^{-27}s^2/m, \quad \beta_3 = 8.119\times 10^{-41}s^3/m, \quad \beta_4 = -13.22\times 10^{-56}s^4/m, \nonumber\\
&&\beta_5 = 3.032\times 10^{-70}s^5/m, \quad \beta_6 = -4.196\times 10^{-85}s^6/m, \quad \beta_7 = 2.570\times 10^{-100}s^7/m, \nonumber\\
&&\gamma = 0.125W^{-1}m^{-1}, \quad \tau_{shock}=1/\omega_{0}=0.43fs, \quad R(t)=(1-f_{R})\delta(t)+f_{R}h_{R}(t).\nonumber
\end{eqnarray}
Here $A(z,t)$ is the pulse envelope, $\beta_2<0$ is the group velocity dispersion in anomalous regime, $\gamma$ is the nonlinear coefficient (Kerr nonlinearity), $\omega_{0}=2\pi c/\lambda_{0}$ where $\lambda_{0}=806nm$ is the initial pulse carrier wavelength; linear operator $\hat\beta_{HD}$ designates higher order dispersion, while self-steepening and Raman scattering are included through operators $i\tau_{shock}\partial/\partial x$ and $h_{R}(t)$ respectively. Specific values of $f_{R}=0.18$ and $h_{R}(t)$ were determined from the experimental fused silica Raman cross-section \cite{DGC}.

\begin{figure}[t] \centering
\includegraphics[width=130pt]{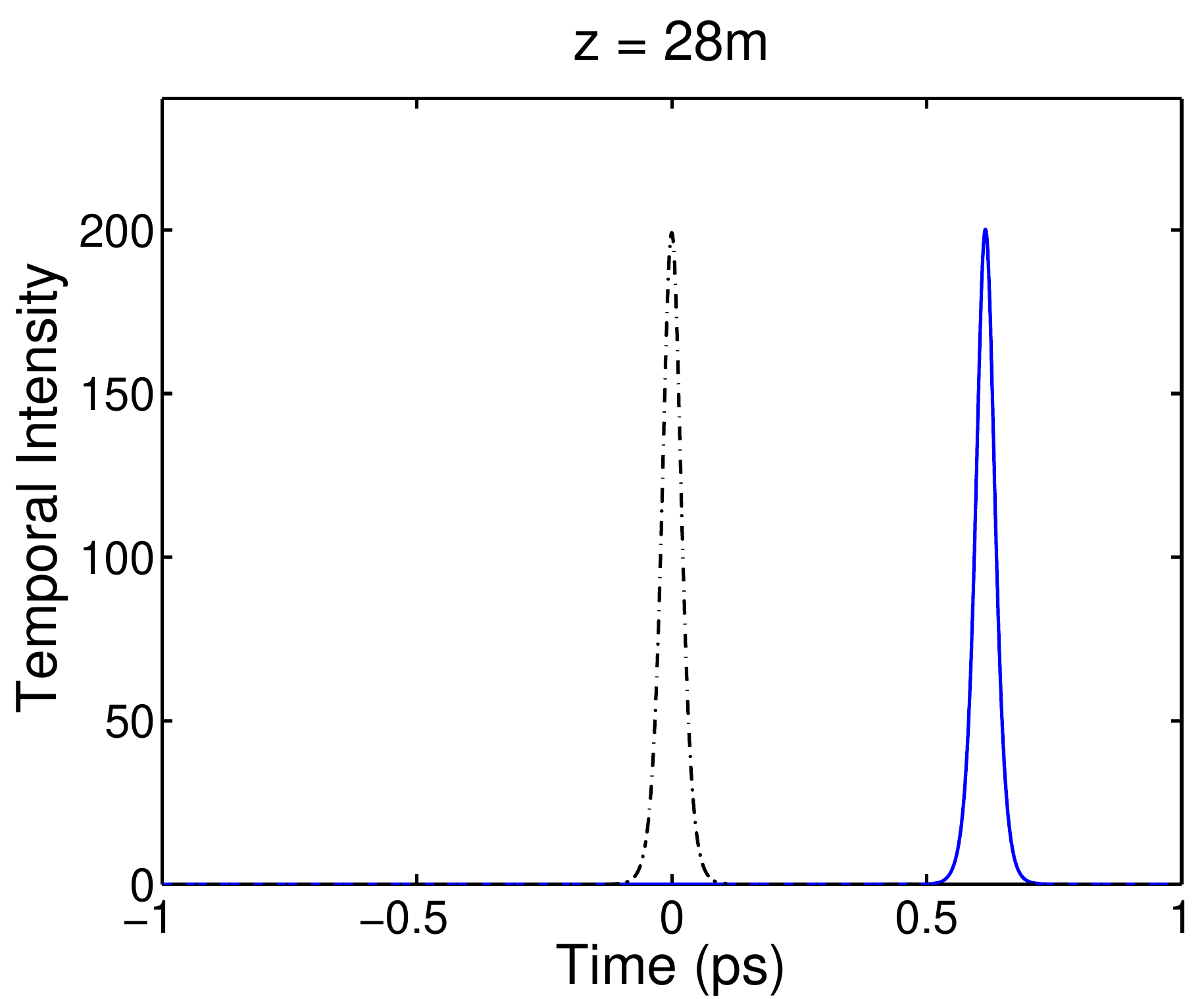}
\includegraphics[width=130pt]{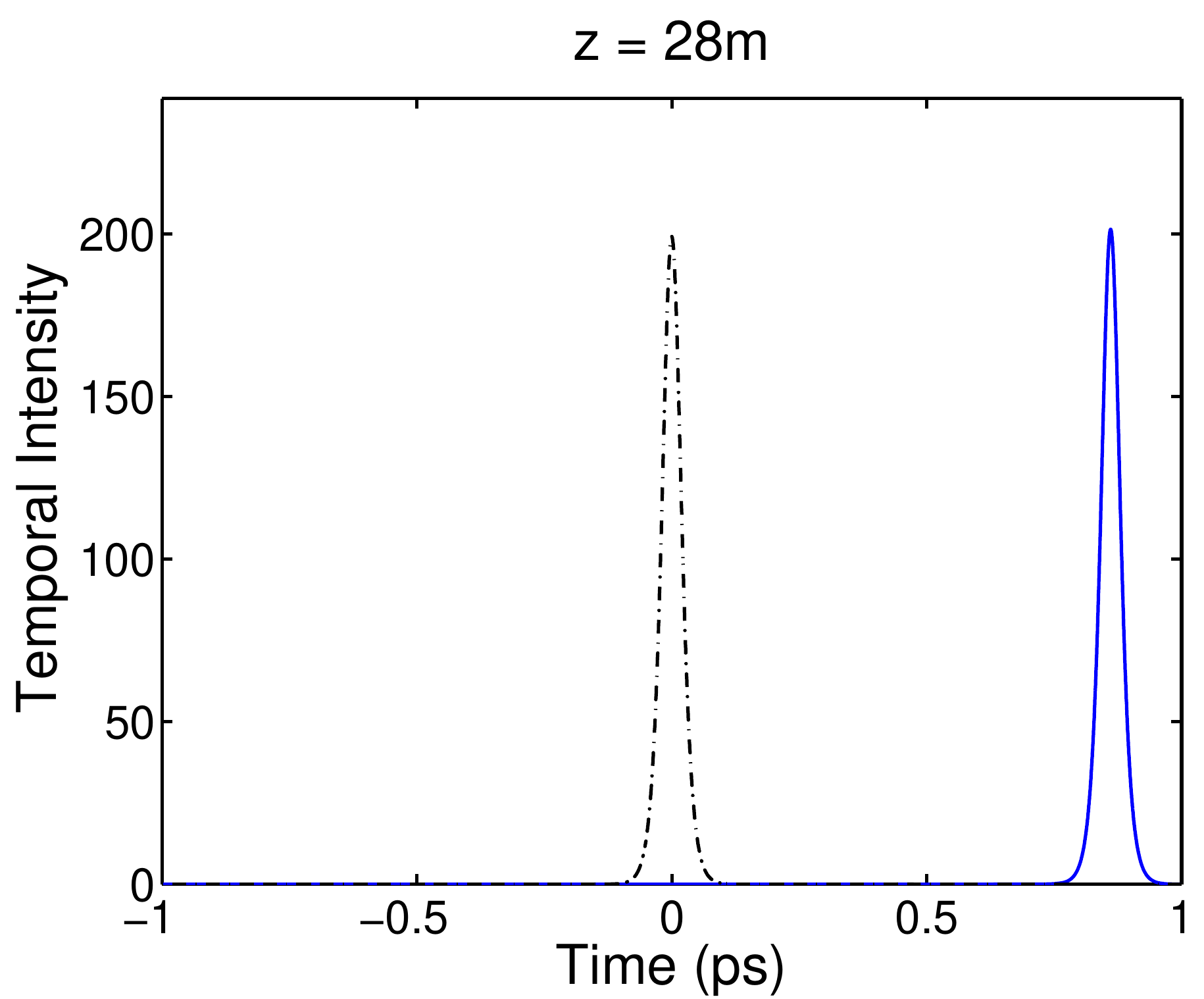}
\includegraphics[width=130pt]{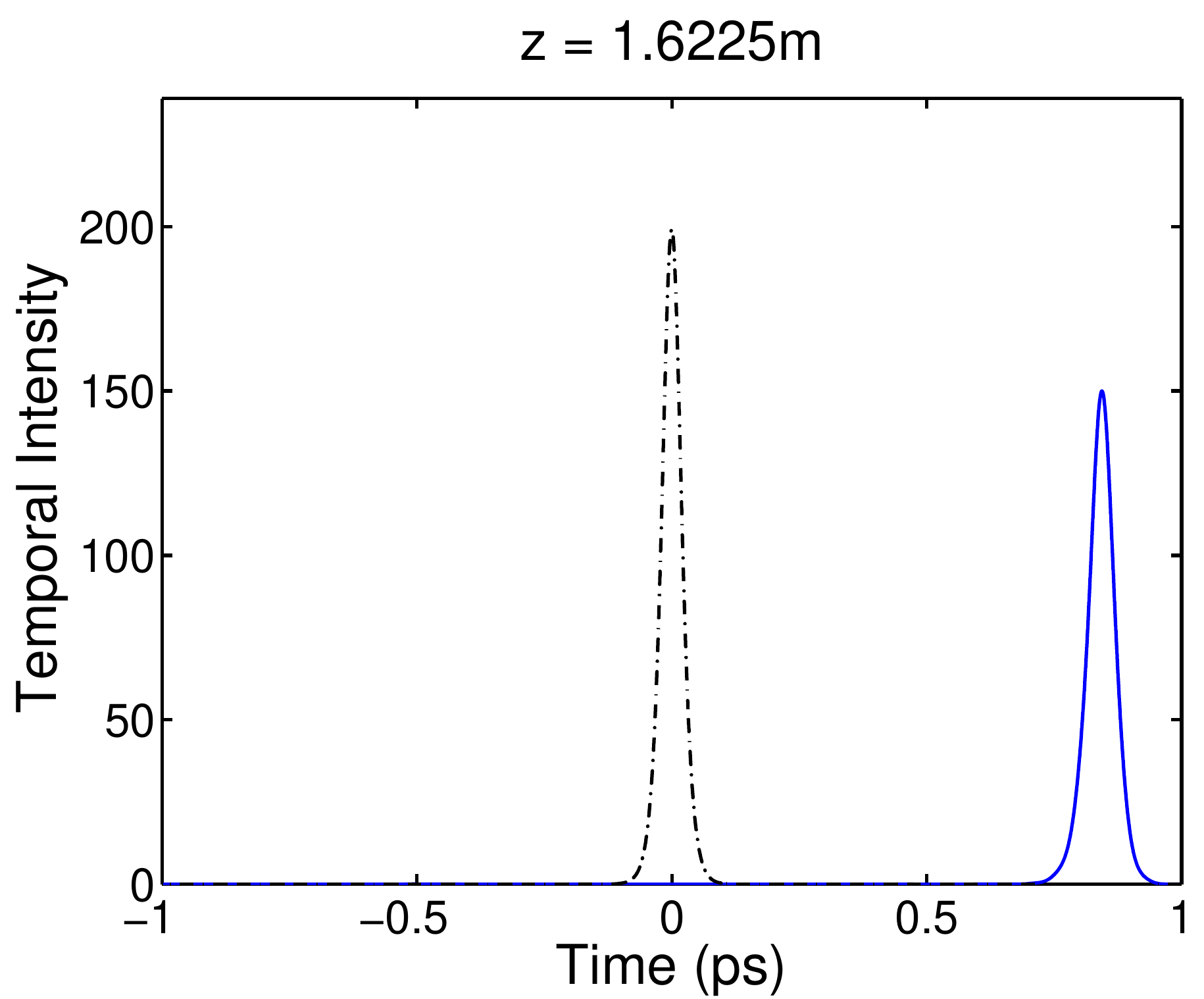}
\includegraphics[width=130pt]{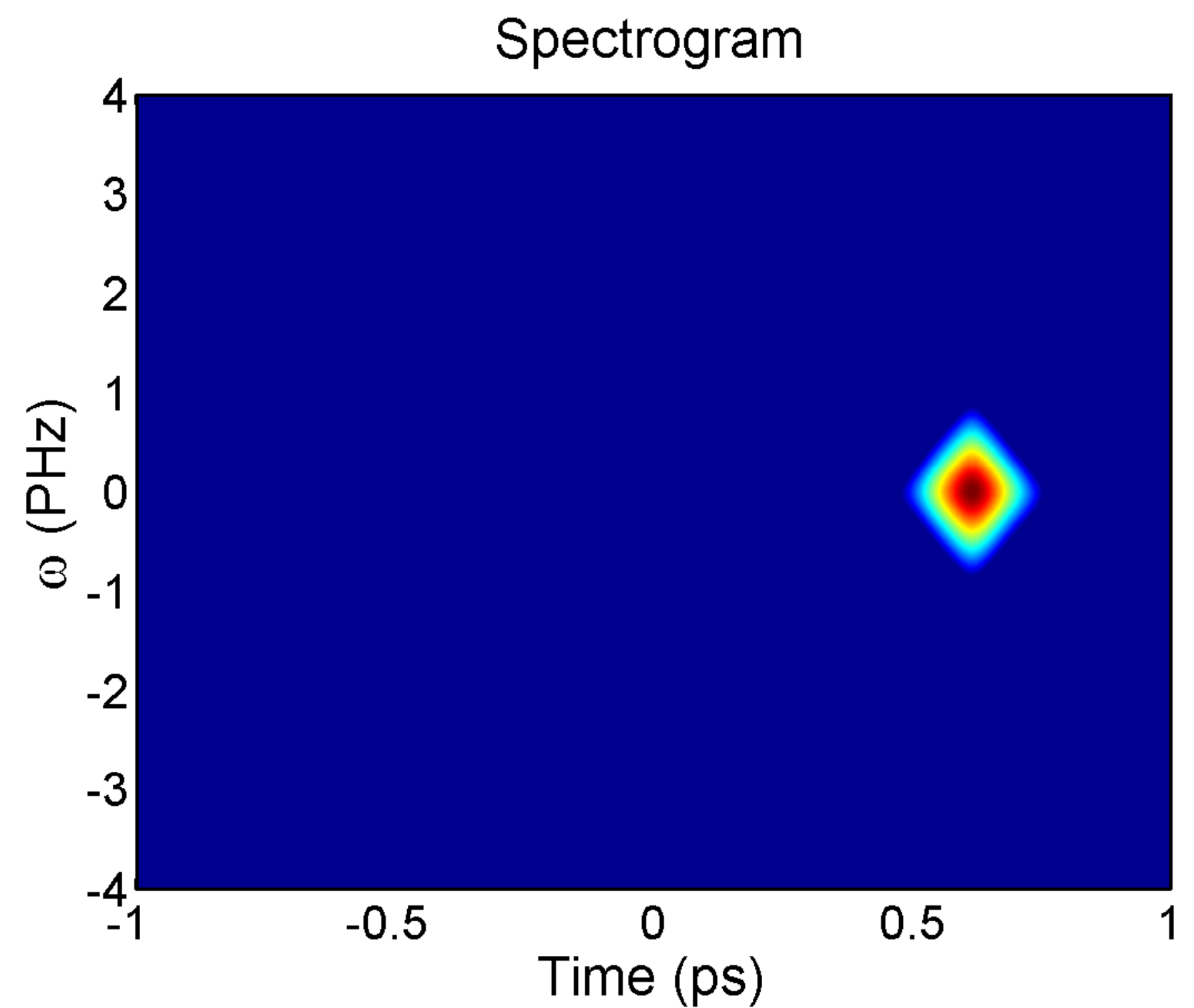}
\includegraphics[width=130pt]{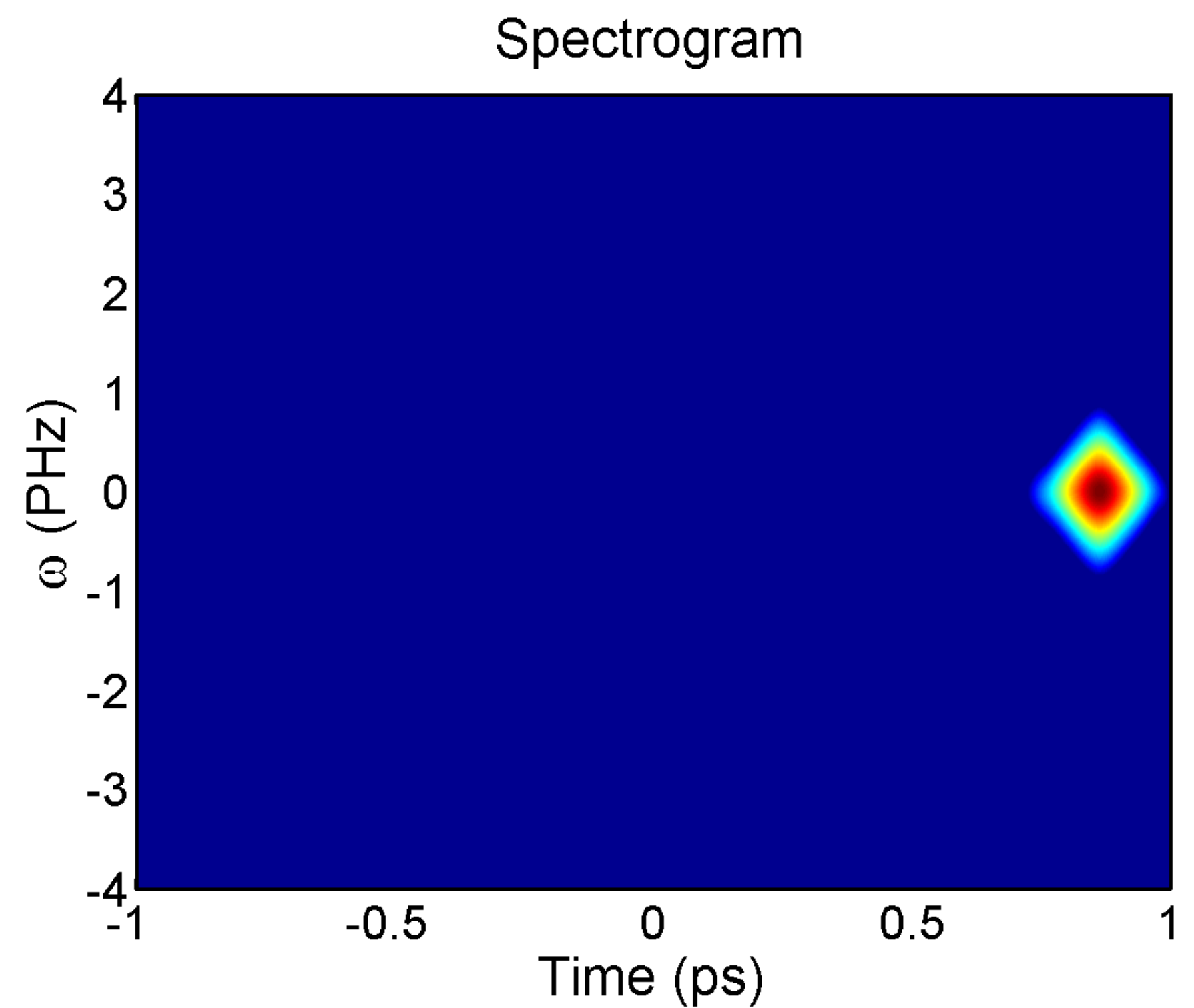}
\includegraphics[width=130pt]{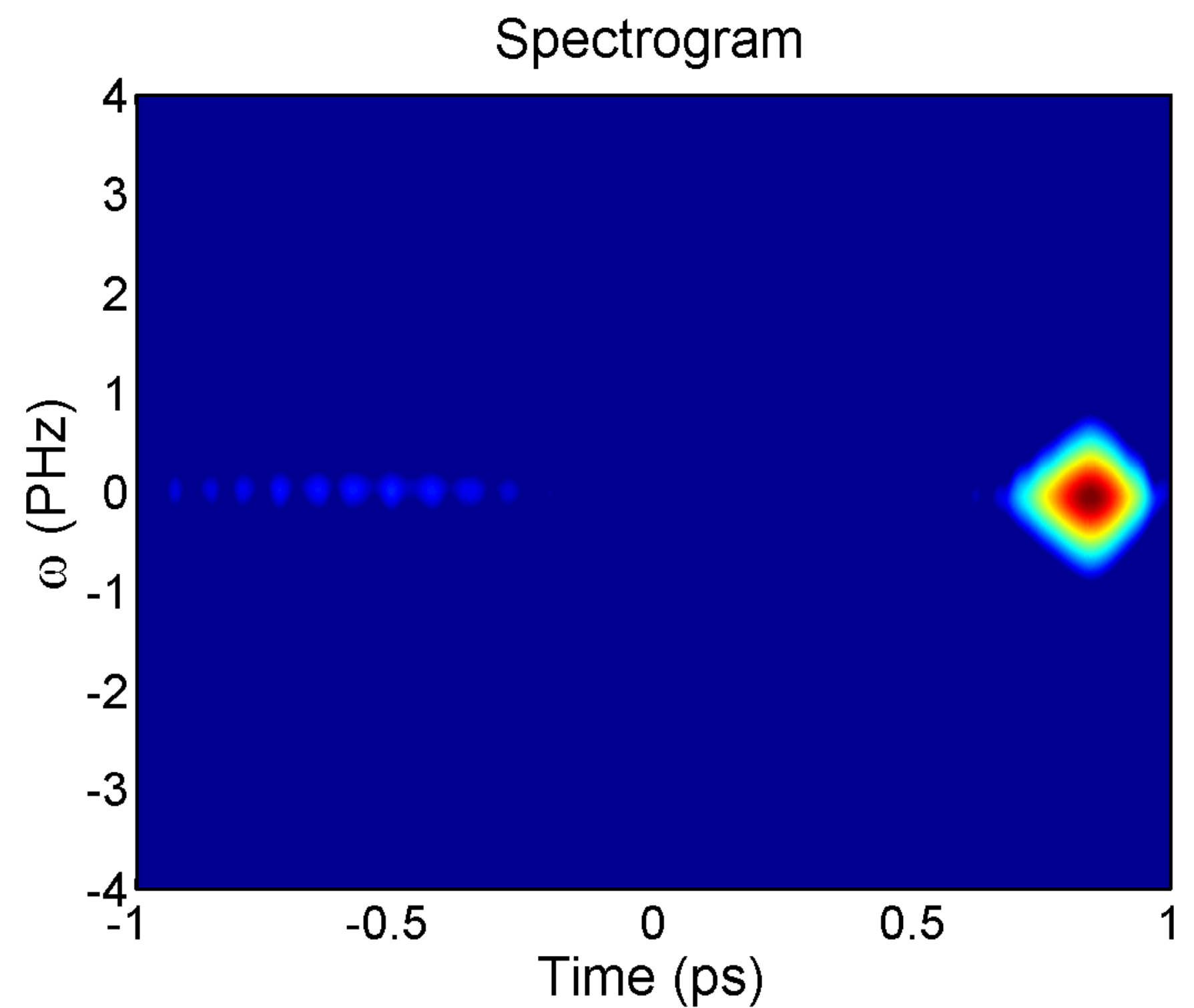}
\includegraphics[width=130pt]{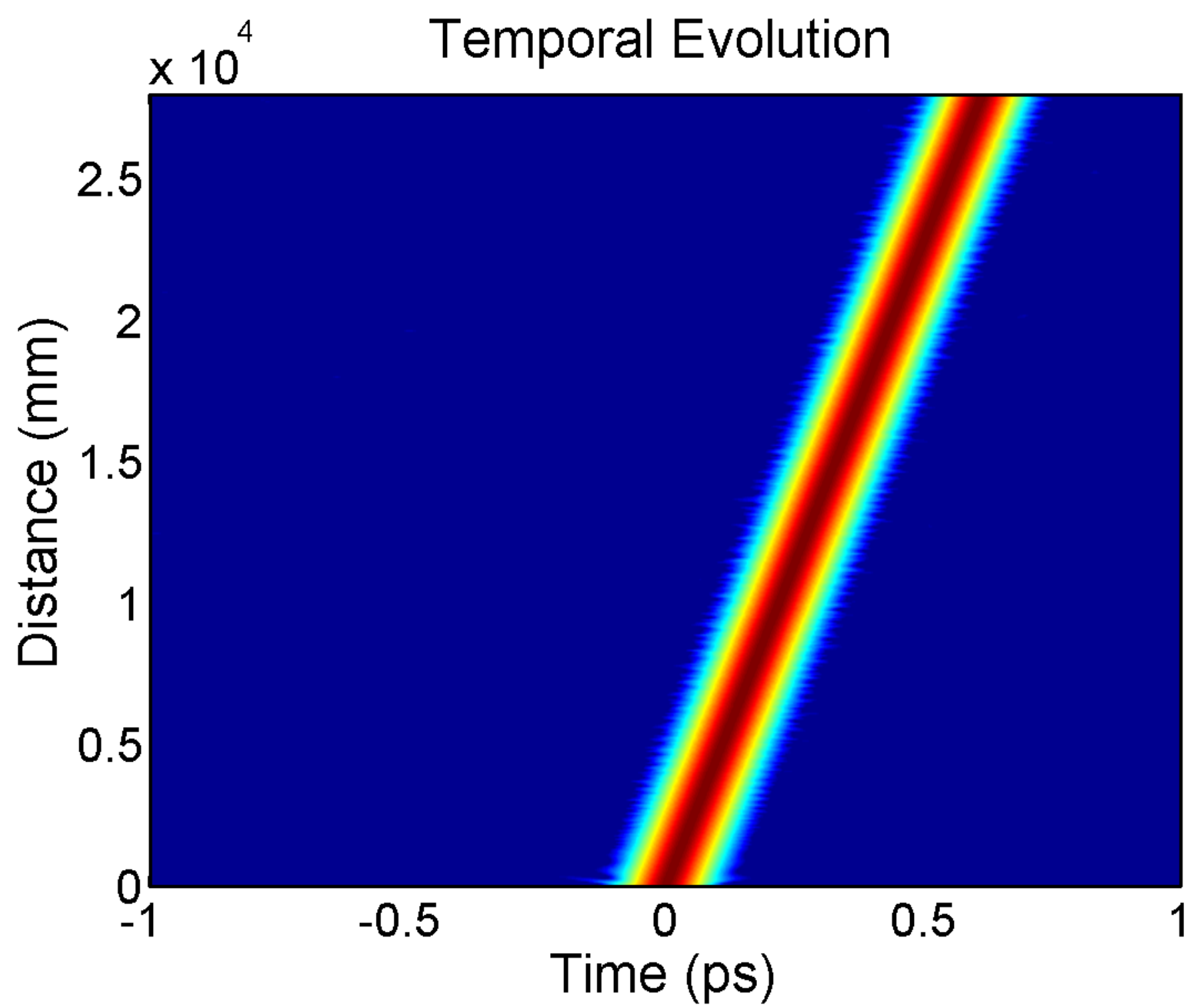}
\includegraphics[width=130pt]{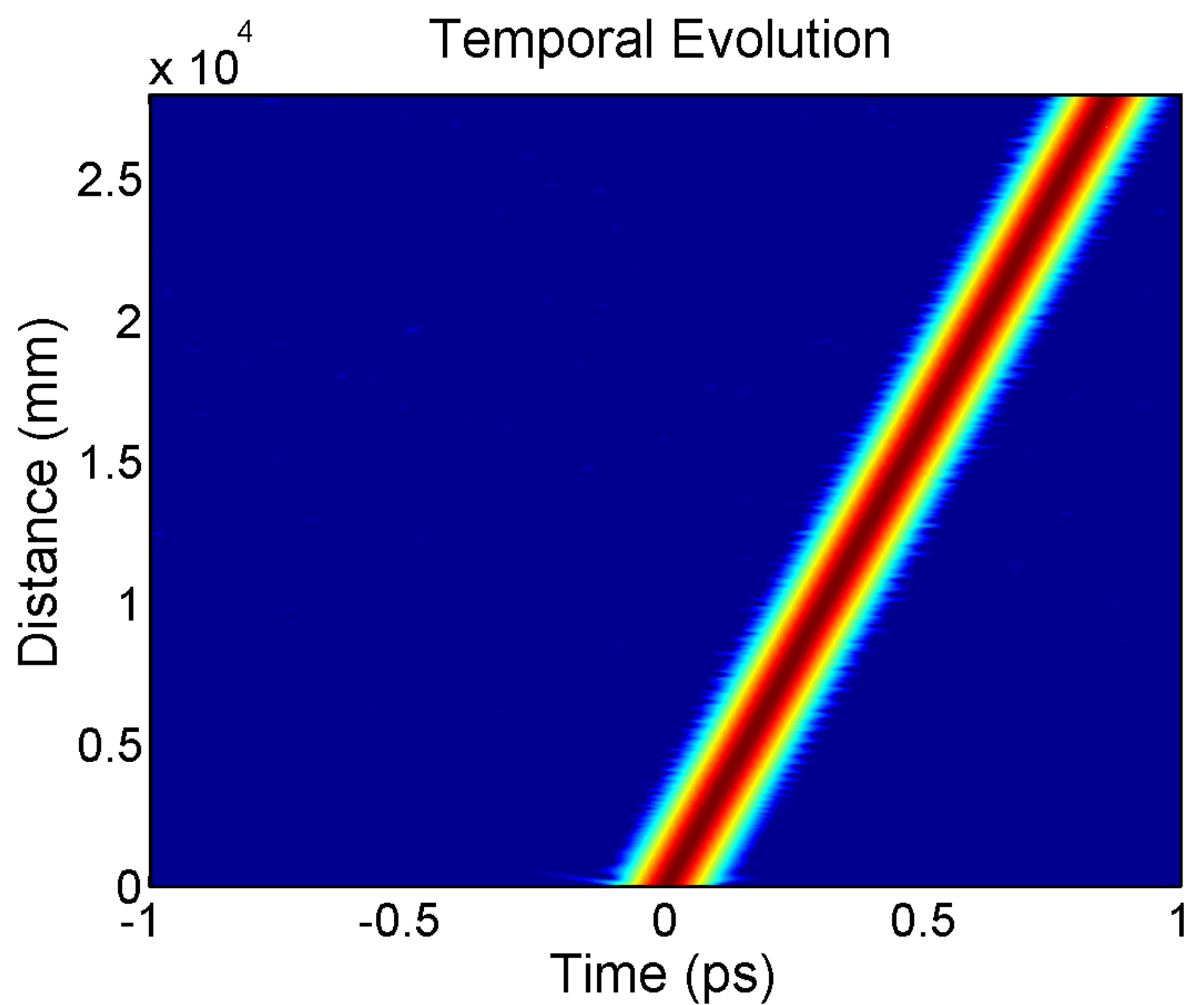}
\includegraphics[width=130pt]{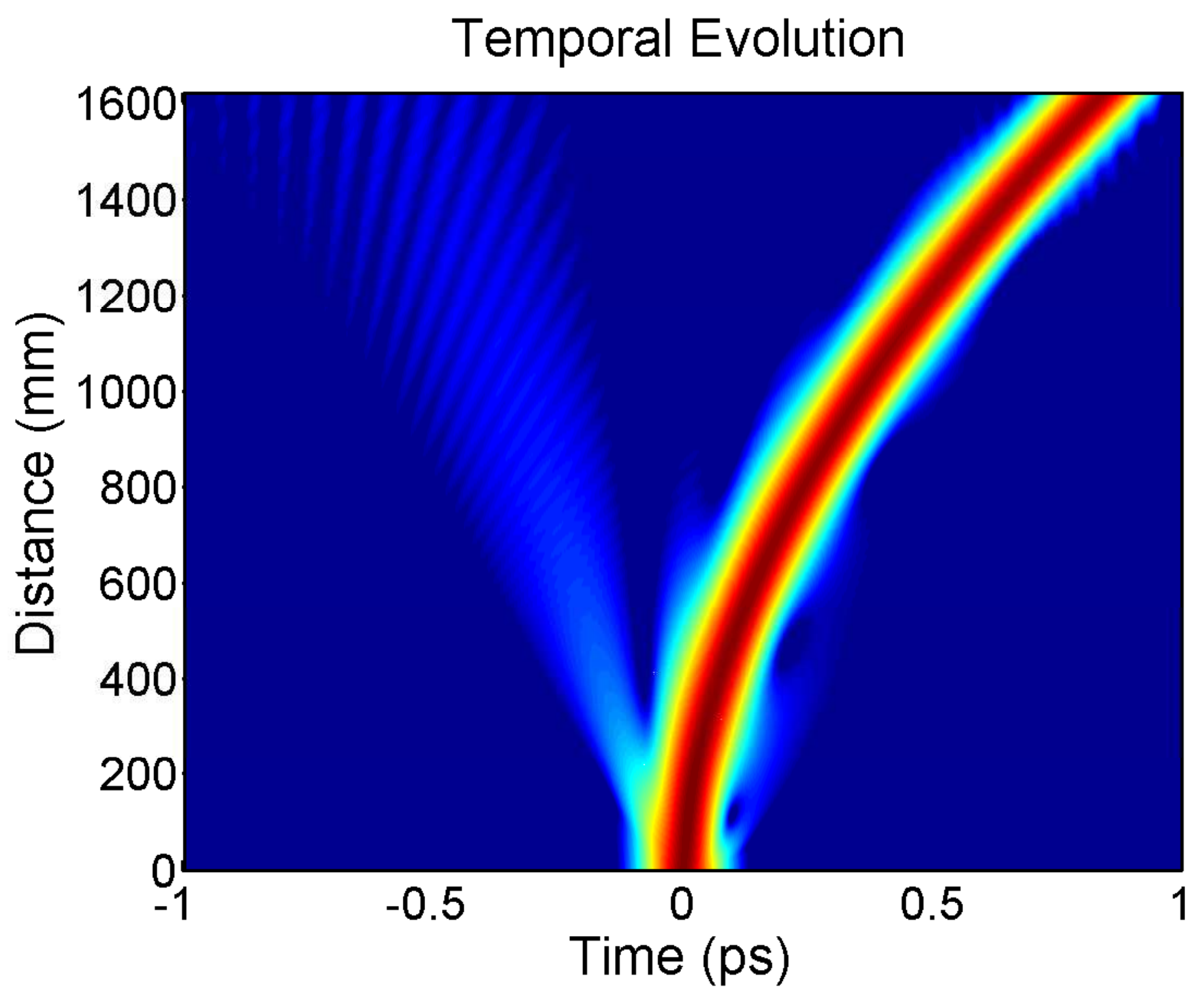}
\includegraphics[width=130pt]{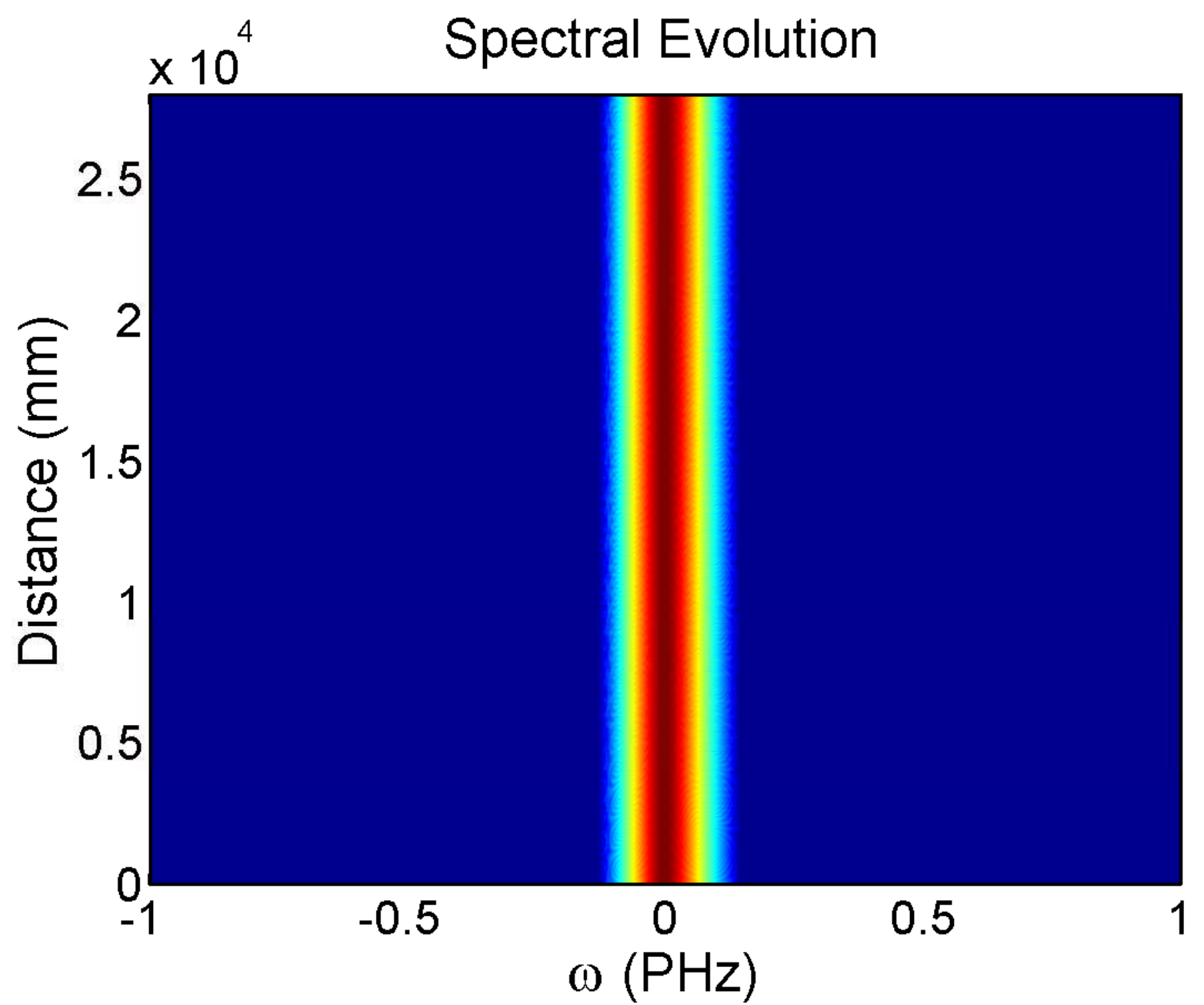}
\includegraphics[width=130pt]{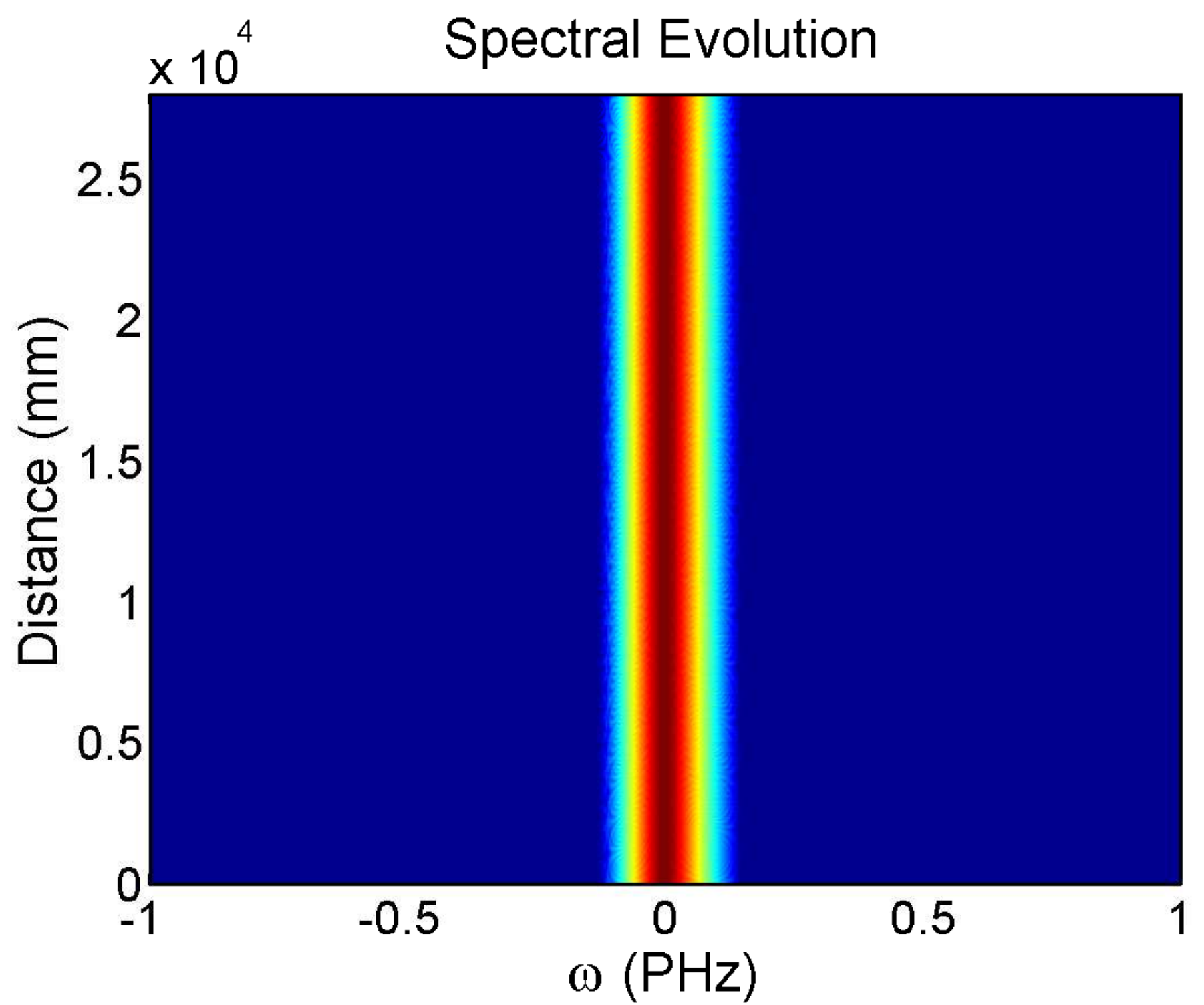}
\includegraphics[width=130pt]{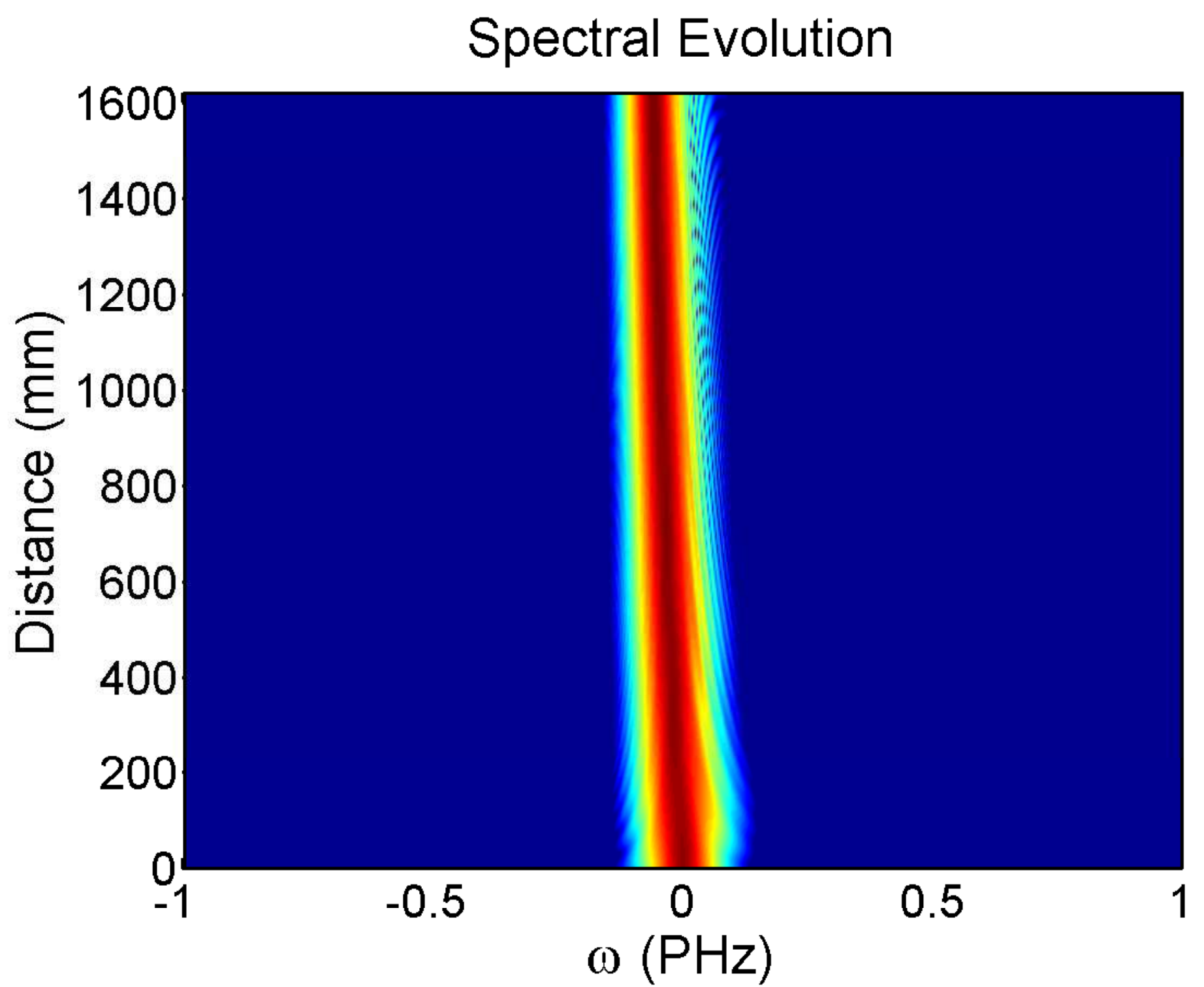}
\caption{\small {\it  (Color online) from upper to bottom row: output signal, it's spectrogram, and temporal and phase evolution during the propagation of initial pulse in the form of single classical NLS soliton inside the PCF for (from left to right) NLS+HD, NLS+HD+SS and NLS+HD+SS+RS systems respectively. If not otherwise stated, black dotted line is the initial state, while the blue solid line is the result of the motion.}}
\end{figure}

Numerical schema used for integration of Eq. (\ref{1_OpticsEnvelopeEvolution}) was split-step method in which linear and nonlinear parts of the equation were calculated separately. Linear part of Eq. (\ref{1_OpticsEnvelopeEvolution}) was solved in frequency domain. From the other hand, in order to implement Raman scattering correctly, Runge-Kutta schema of the second order was used to calculate the nonlinear part of Eq. (\ref{1_OpticsEnvelopeEvolution}) (see \cite{DGC,Cristiani}). As was shown by numerous publications (see \cite{DGC, Dudley1, Dudley3} for instance), such numerical schema give results in very good agreement with that from laboratory experiments for initial pulses in femto-seconds range. Propagation distance inside PCF was chosen to prevent significant interaction between the left and the right ends of the computational domain which occurs due to FFT usage in the numerical schema. Such choice of propagation distance corresponds to the limit of zeroth repetition rate of a real laser. The latter assumption is necessary even despite the fact that real lasers always have nonzeroth repetition rate, because MI development considered in this work is in highly incoherent regime (see \cite{DGC}), i.e. MI developed from the following laser pulse significantly differs from MI developed from the preceding one. Therefore if the interaction between the left and the right ends of the computational domain is strong, the additional coherence which has no physical meaning is brought to the system.

The first group of numerical experiments was made to analyze single NLS solitons behavior for 3 different situations: a) in presence of higher order dispersion only (NLS+HD system, $\tau_{shock}$ and $h_{R}(t)$ were equal to zero); b) in presence of self-steepening and higher order dispersion (NLS+HD+SS system, $h_{R}(t)$ was equal to zero); c) in presence of self-steepening, Raman scattering and higher order dispersion (NLS+HD+SS+RS system, full Eq. (\ref{1_OpticsEnvelopeEvolution}) with coefficients Eq. (\ref{1_OpticsEnvelopeEvolutionCoeff})). Initial pulse was taken as an exact soliton solution of the classical NLSE:
\begin{eqnarray}
i\frac{\partial A}{\partial z}-\frac{\beta_{2}}{2} A_{tt} + \gamma(1-f_{R})|A|^{2}A = 0.
\label{1_OpticsEnvelopeEvolution2}
\end{eqnarray}
As shown on Fig.1, additional terms beyond Eq. (\ref{1_OpticsEnvelopeEvolution2}) modified the group velocity of the initial pulse. While this modification was comparatively small for NLS+HD and NLS+HD+SS systems, in presence of Raman scattering the group velocity changed much more significantly both in it's absolute value and with the propagation distance. Another very important influence of Raman scattering was the Raman continuous self-frequency shift to the lower frequencies (see \cite{DGC}) which was experienced by the moving pulse as shown on the corresponding frequency evolution plot on Fig.1.

\begin{figure}[t] \centering
\includegraphics[width=130pt]{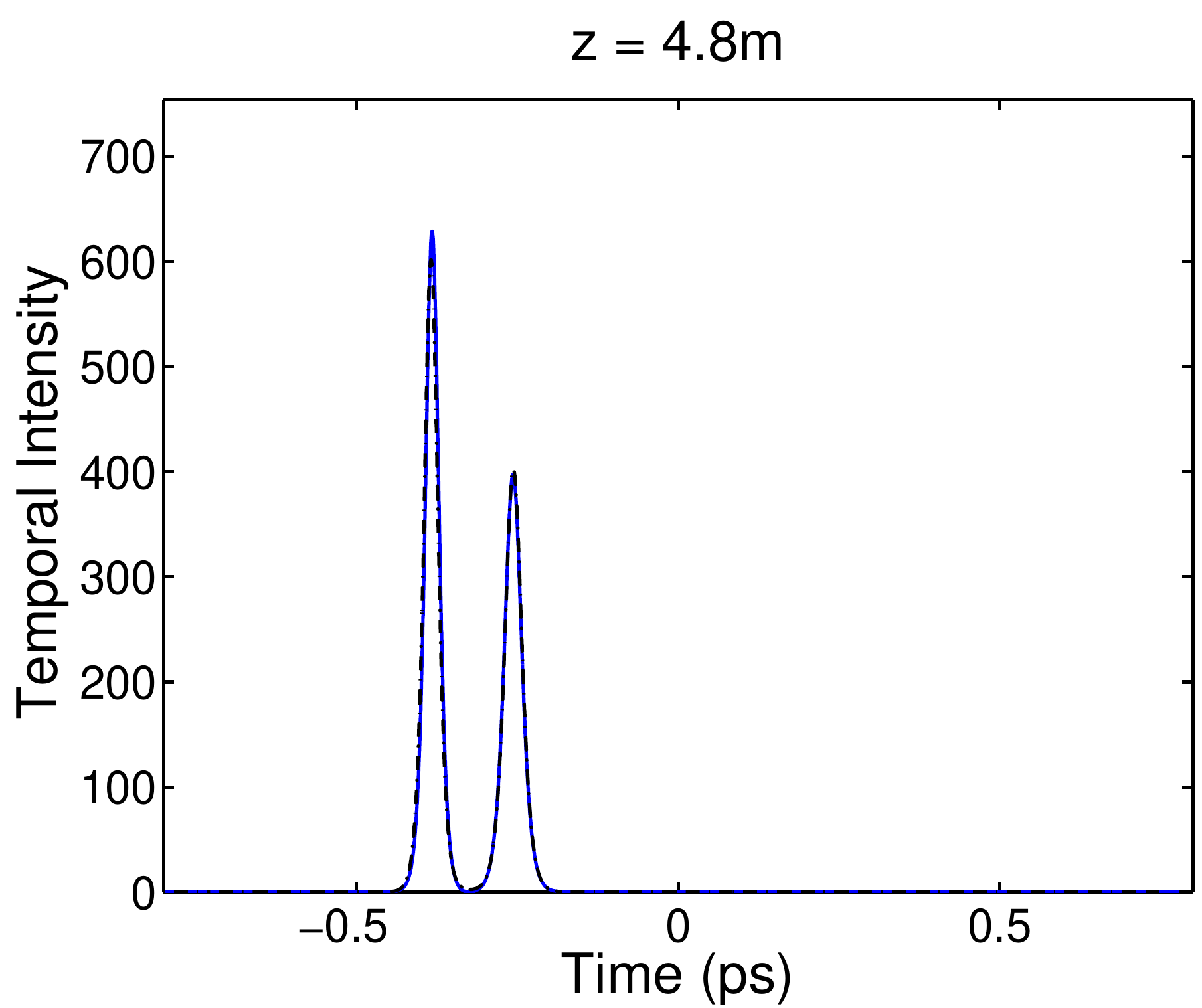}
\includegraphics[width=130pt]{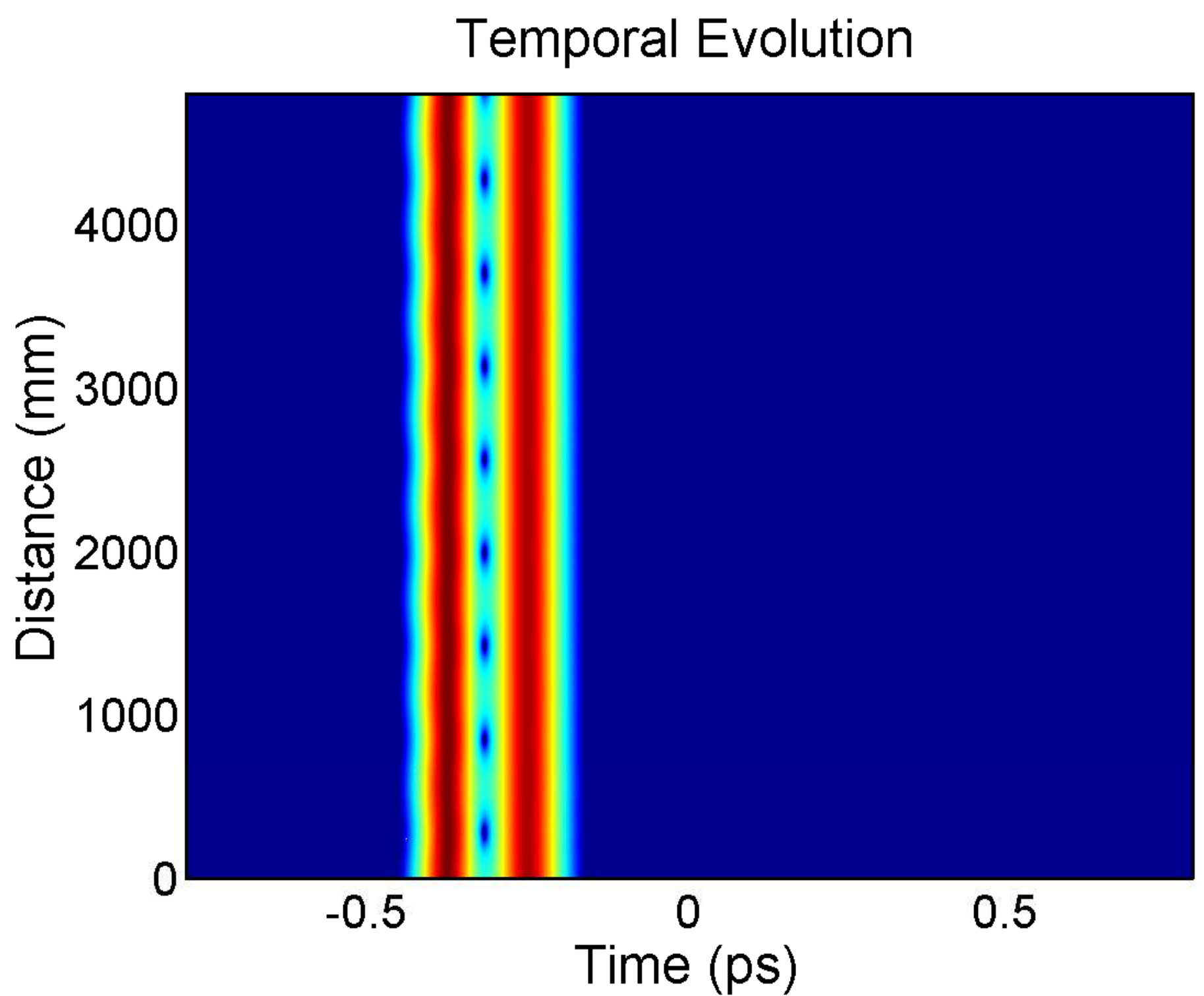}
\includegraphics[width=130pt]{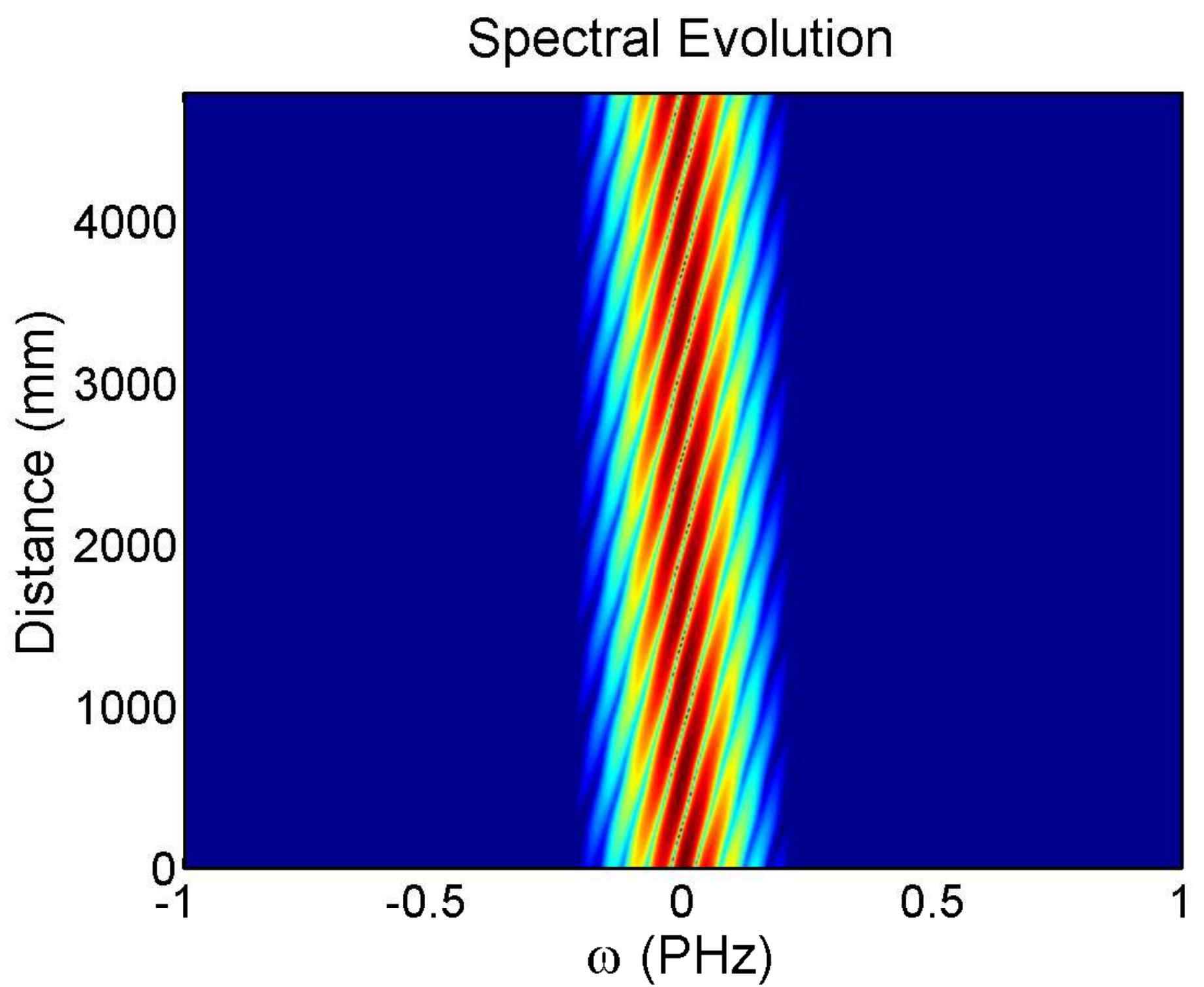}
\caption{\small {\it (Color online) from left to right: output signal and it's temporal and spectral evolution during the propagation of initial pulse in the form of superposition of two exact classical NLS solitons in PCF described by the classical NLSE.}}
\end{figure}

It turned out also that during the propagation inside the PCF solitons did not change their sech-like shape significantly in all cases a), b) and c). To analyze whether output peaks were still close to solitons of the classical NLSE or not, corresponding spectrograms were plotted. Field spectrogram allows one to completely characterize output function in both intensity and phase domains simultaneously \cite{DGC, Treacy, Dudley2}. For the given $A(t)$ field to be characterized, the spectrogram function is defined as
\begin{eqnarray*}
\Sigma(\omega,\tau) = \bigg|\int_{-\infty}^{+\infty}A(t)g(t-\tau)e^{-i\omega t}dt \bigg|^{2},
\end{eqnarray*}
where $g(t-\tau)$ is a variable-delay gate function. Spectrograms of output fields turned out to be very close to rhombus structures that means that such peaks were indeed very close to exact classical NLS solitons. In other words, if a pulse in the form of classical NLS soliton is launched inside the PCF, it evolves in such a way to stay every time very close to the family of exact soliton solutions of integrable NLS equation. 

\begin{figure}[t] \centering
\includegraphics[width=130pt]{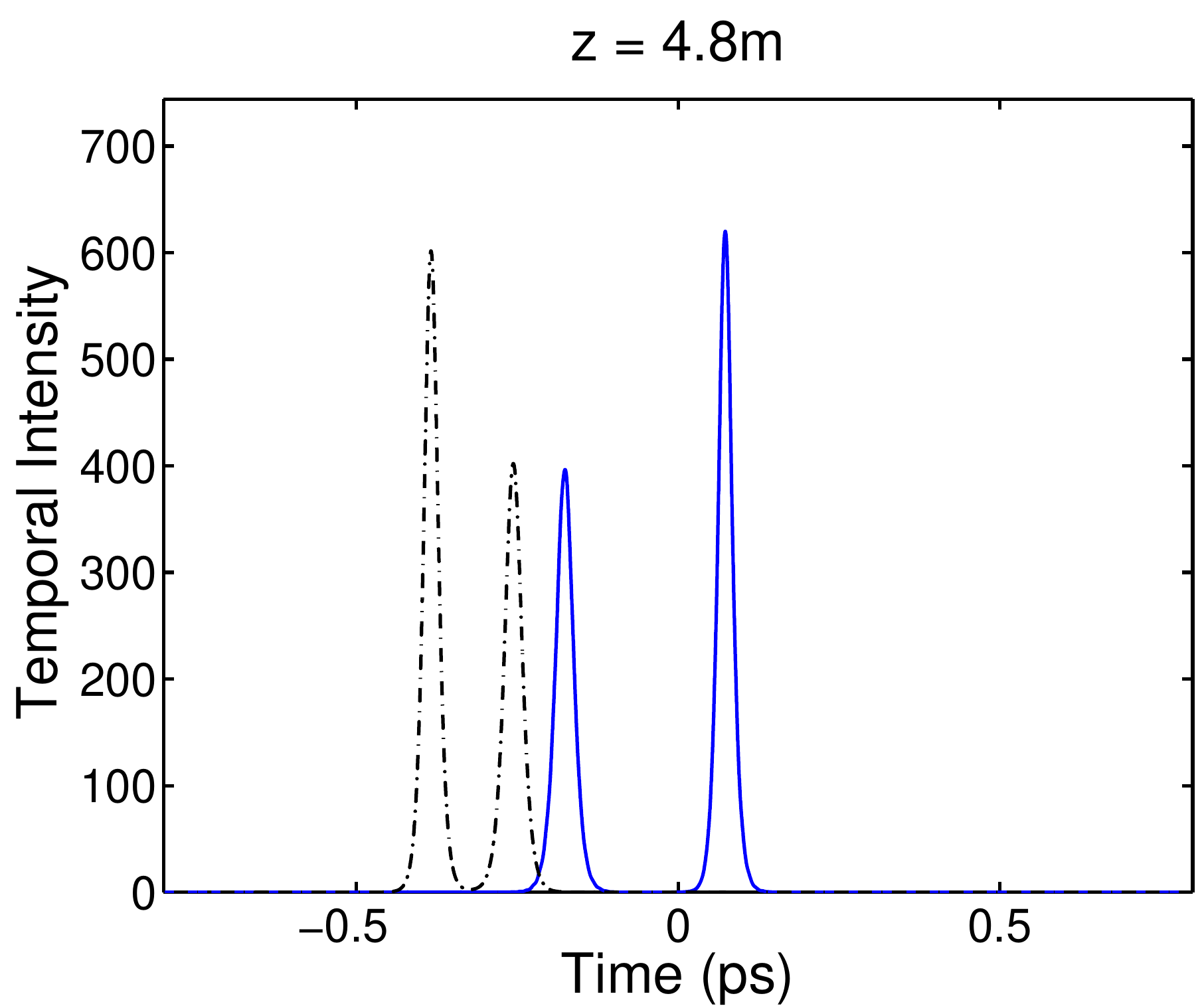}
\includegraphics[width=130pt]{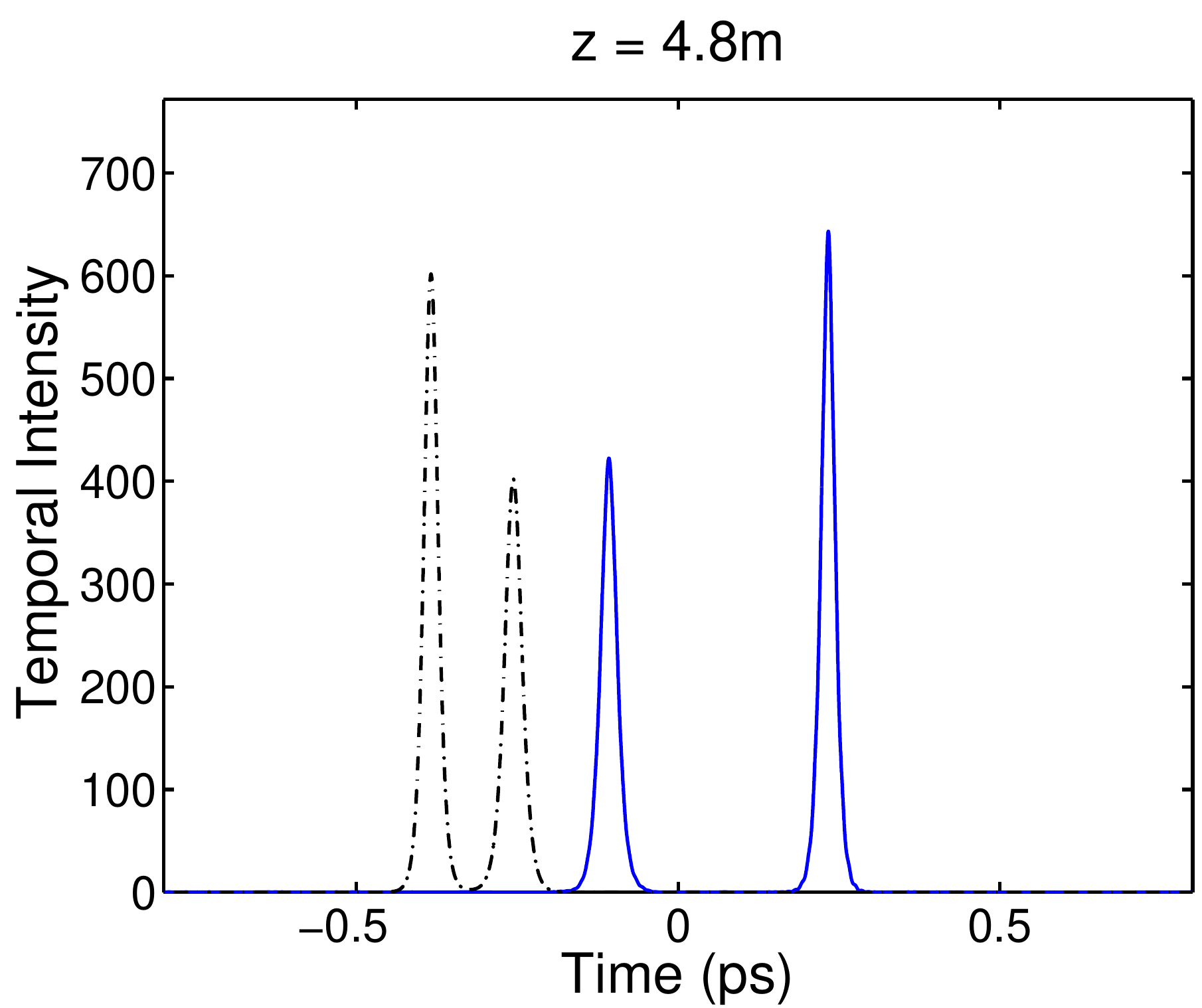}
\includegraphics[width=130pt]{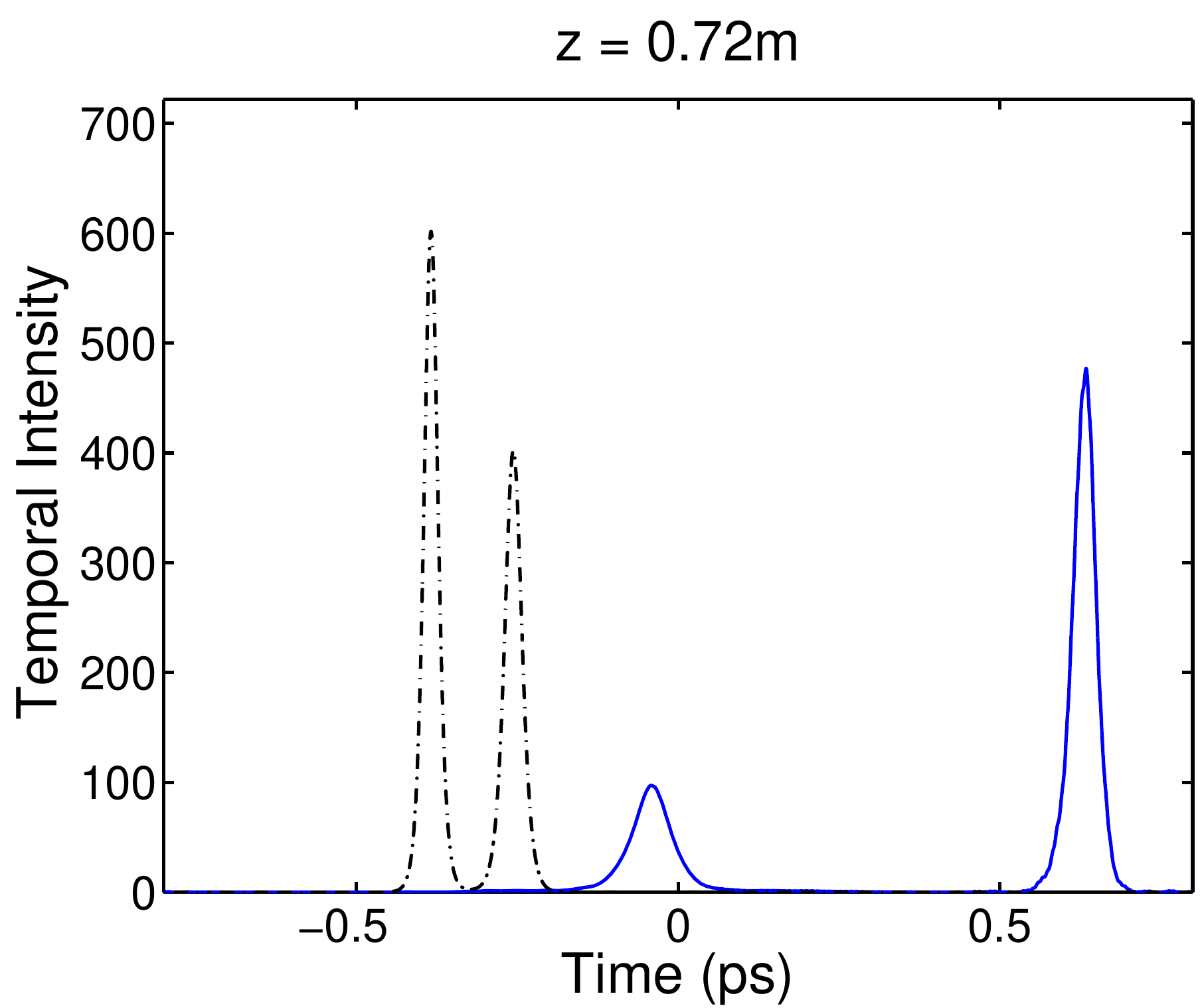}
\includegraphics[width=130pt]{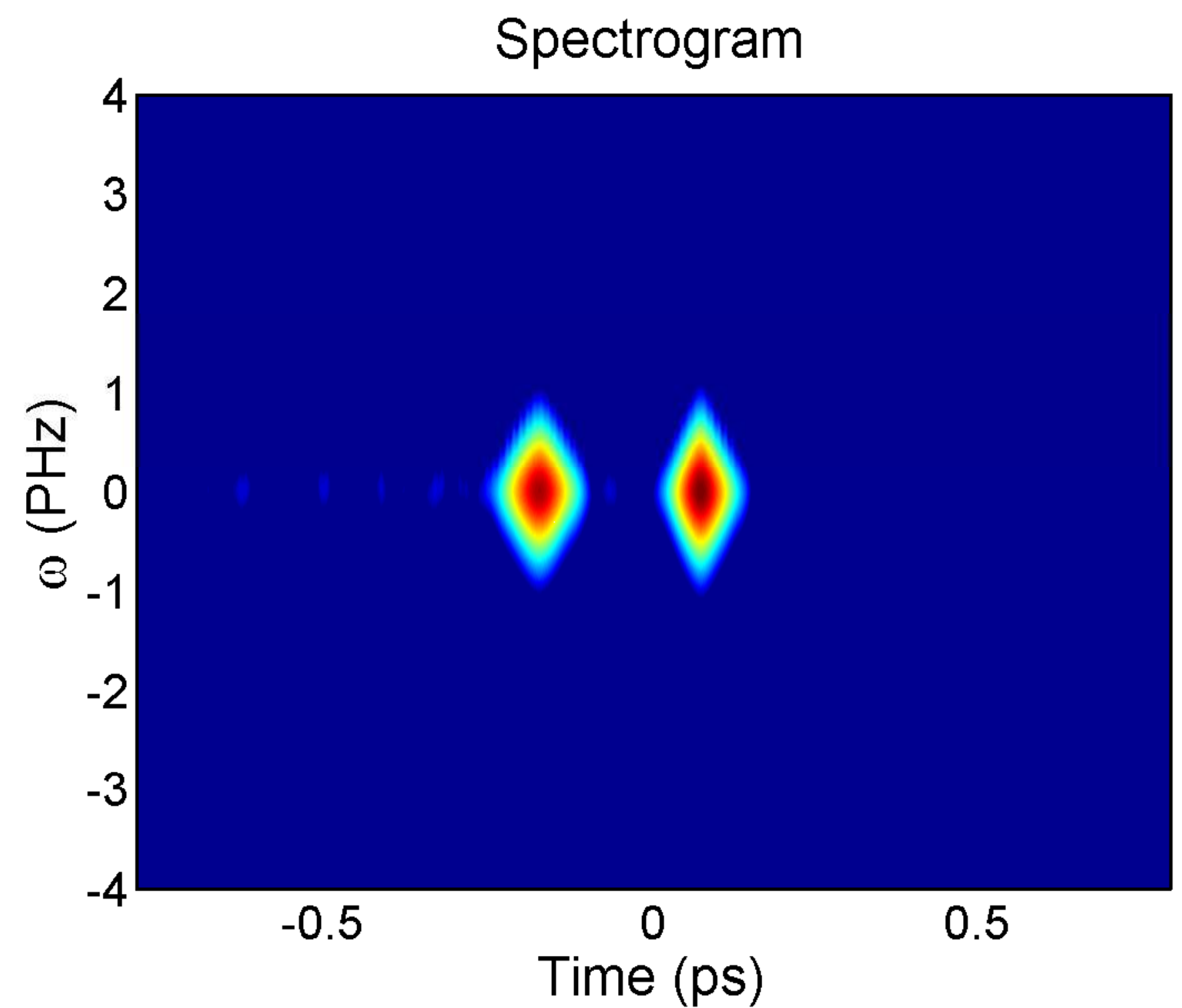}
\includegraphics[width=130pt]{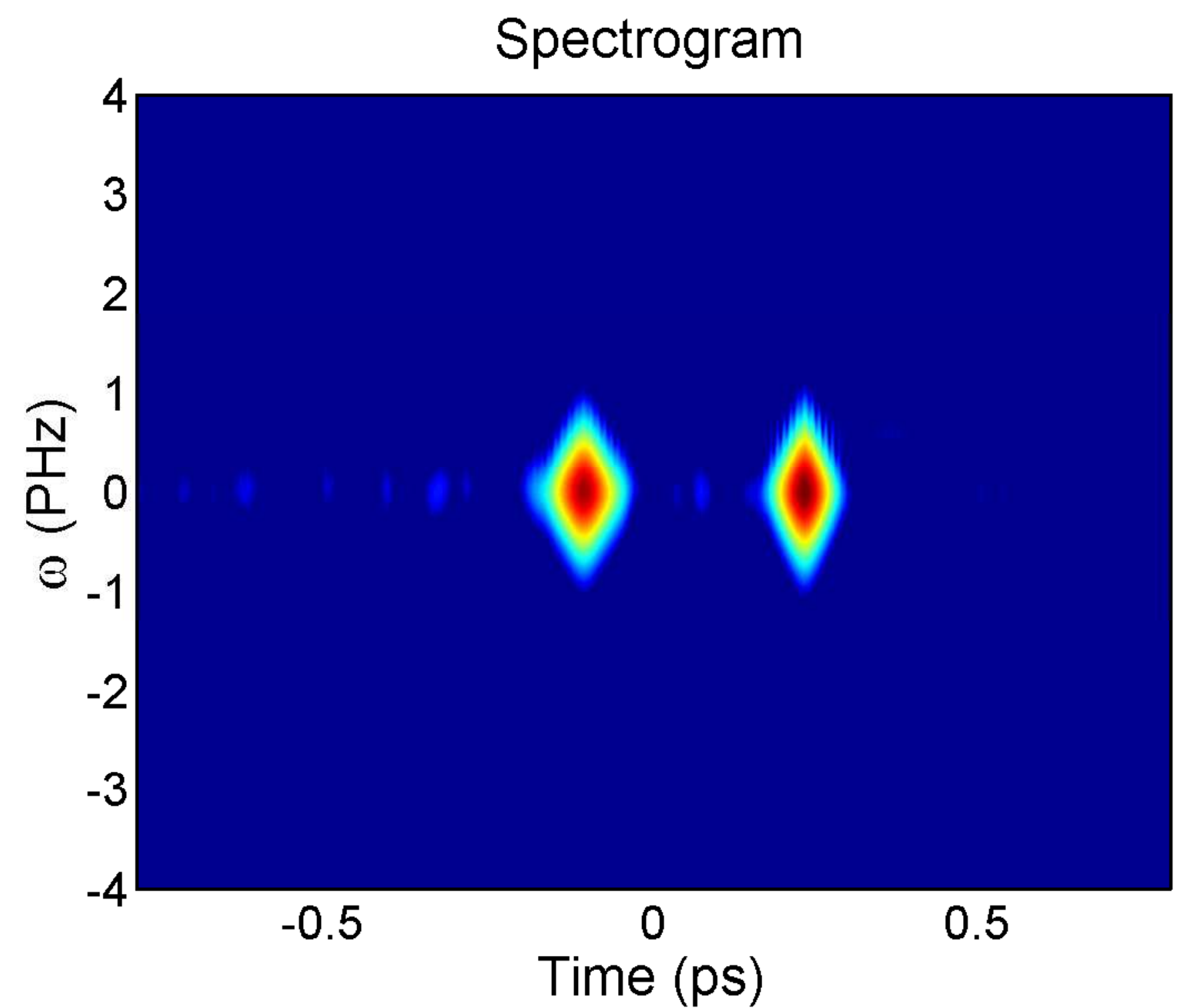}
\includegraphics[width=130pt]{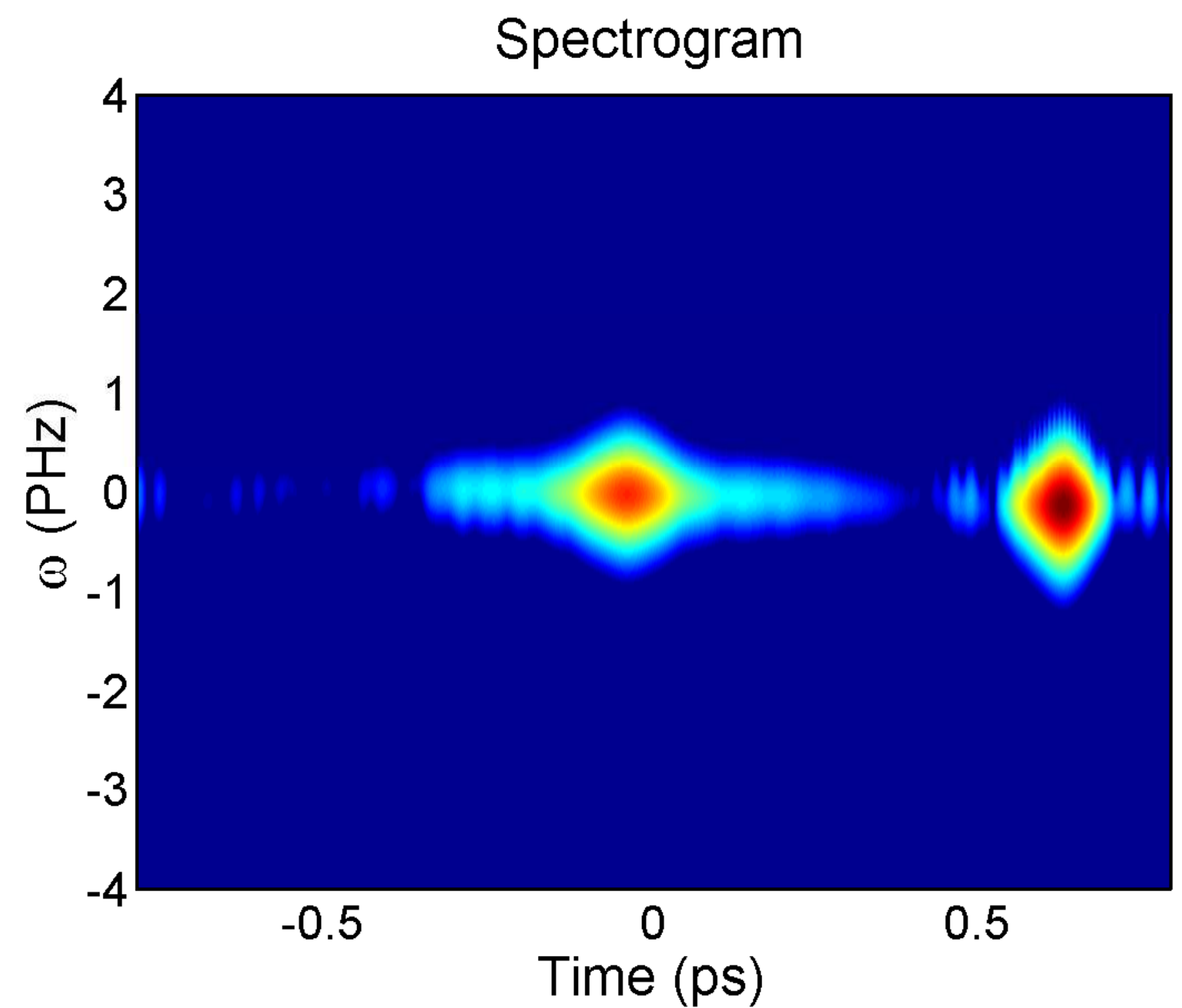}
\includegraphics[width=130pt]{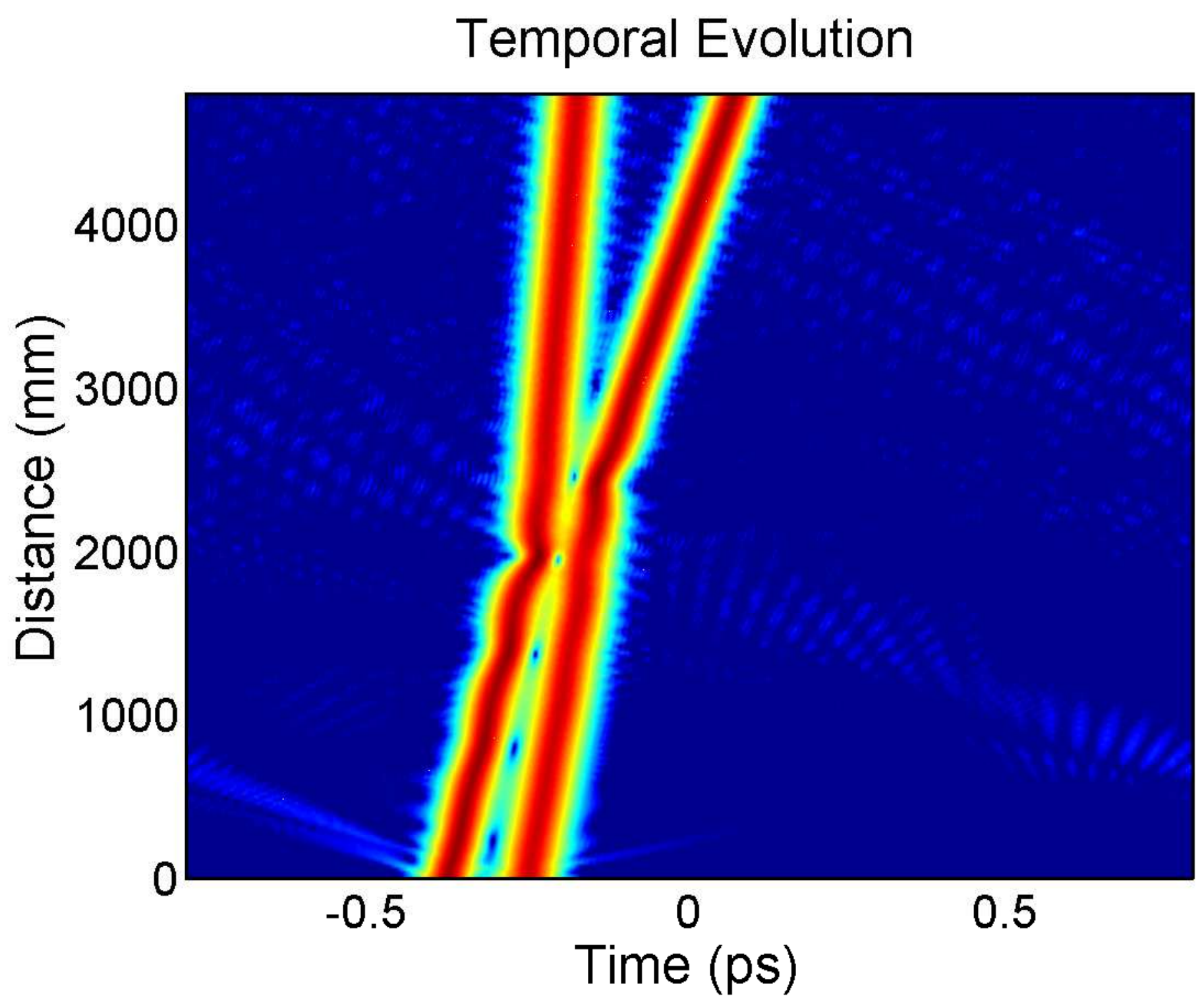}
\includegraphics[width=130pt]{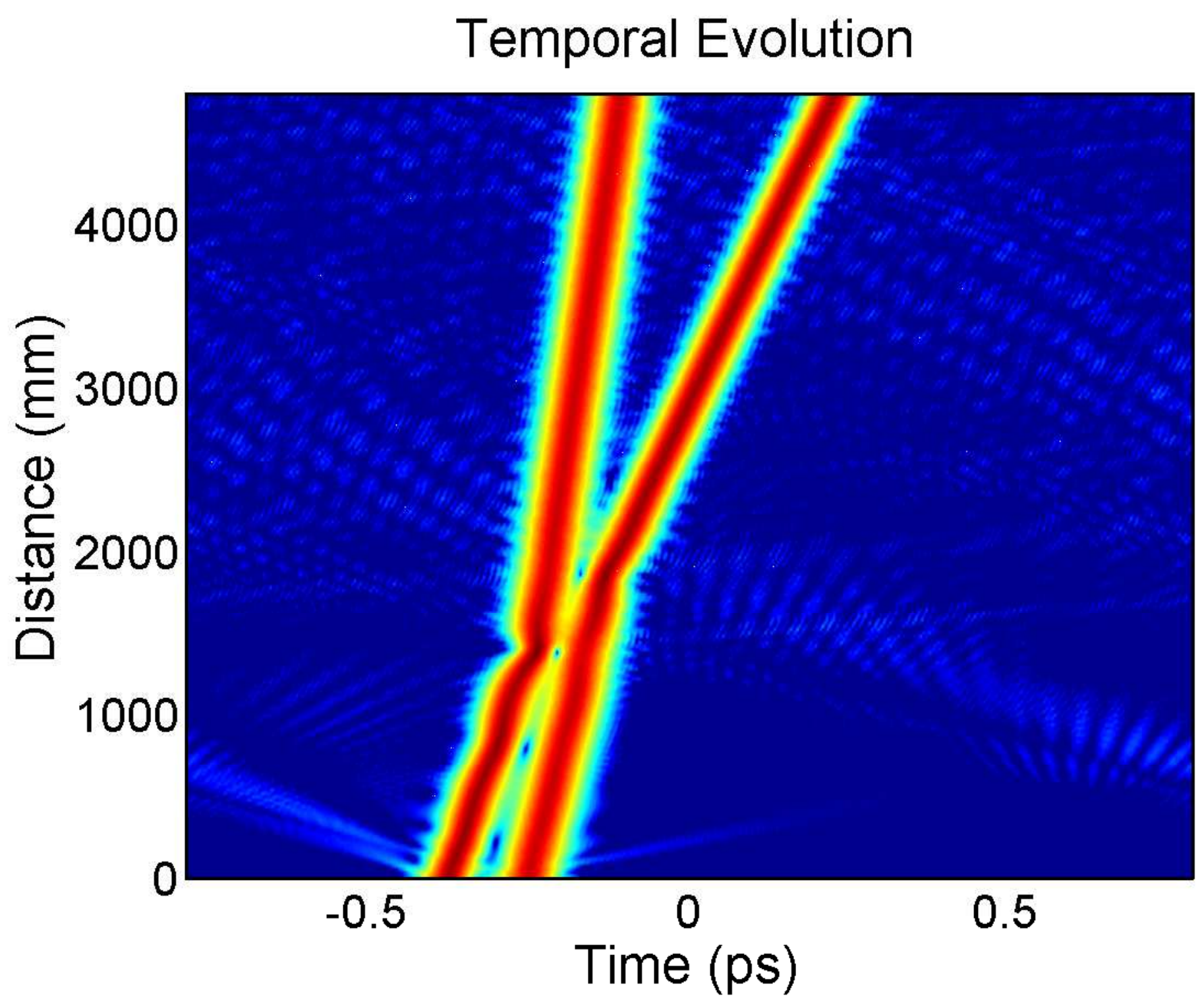}
\includegraphics[width=130pt]{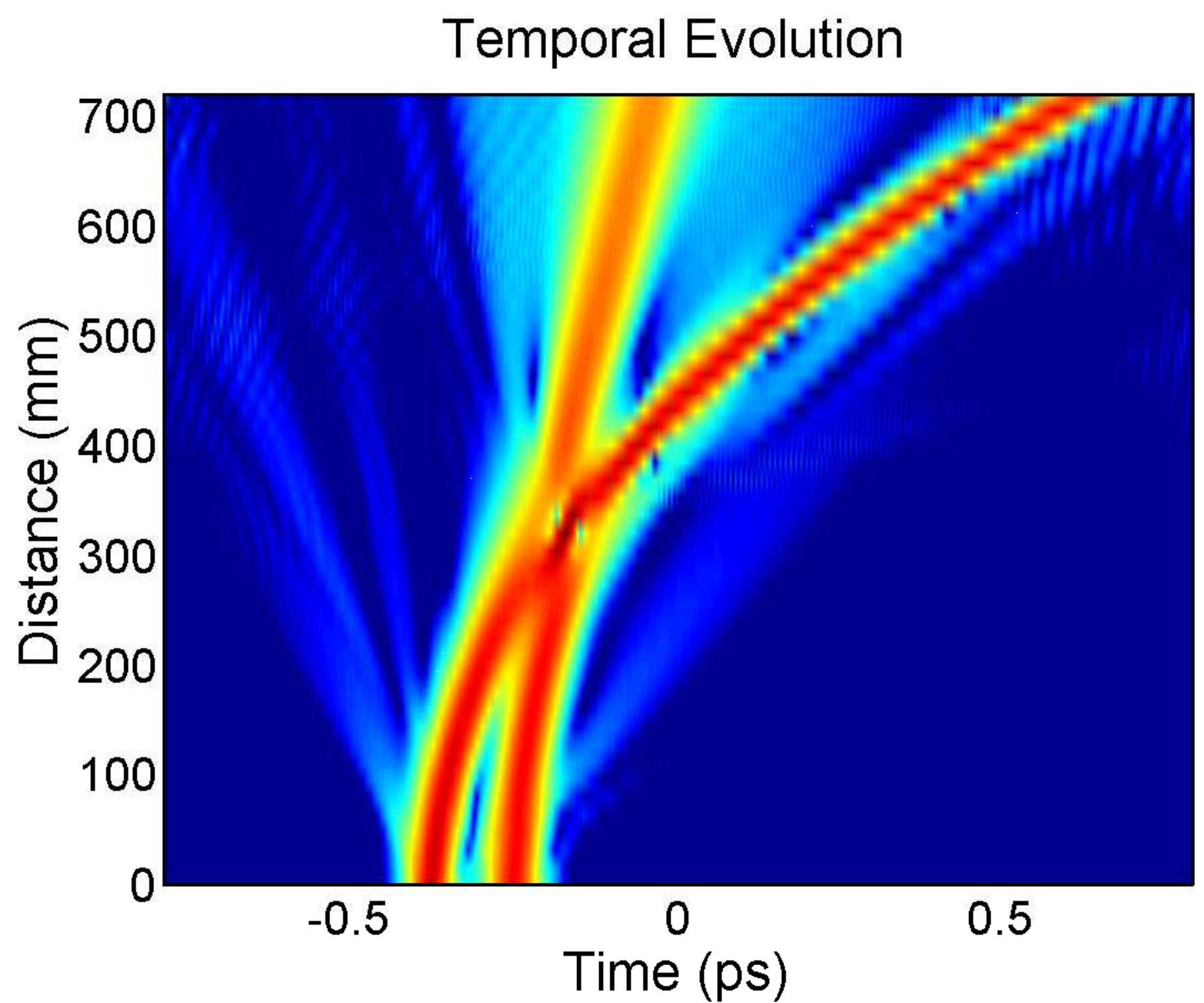}
\includegraphics[width=130pt]{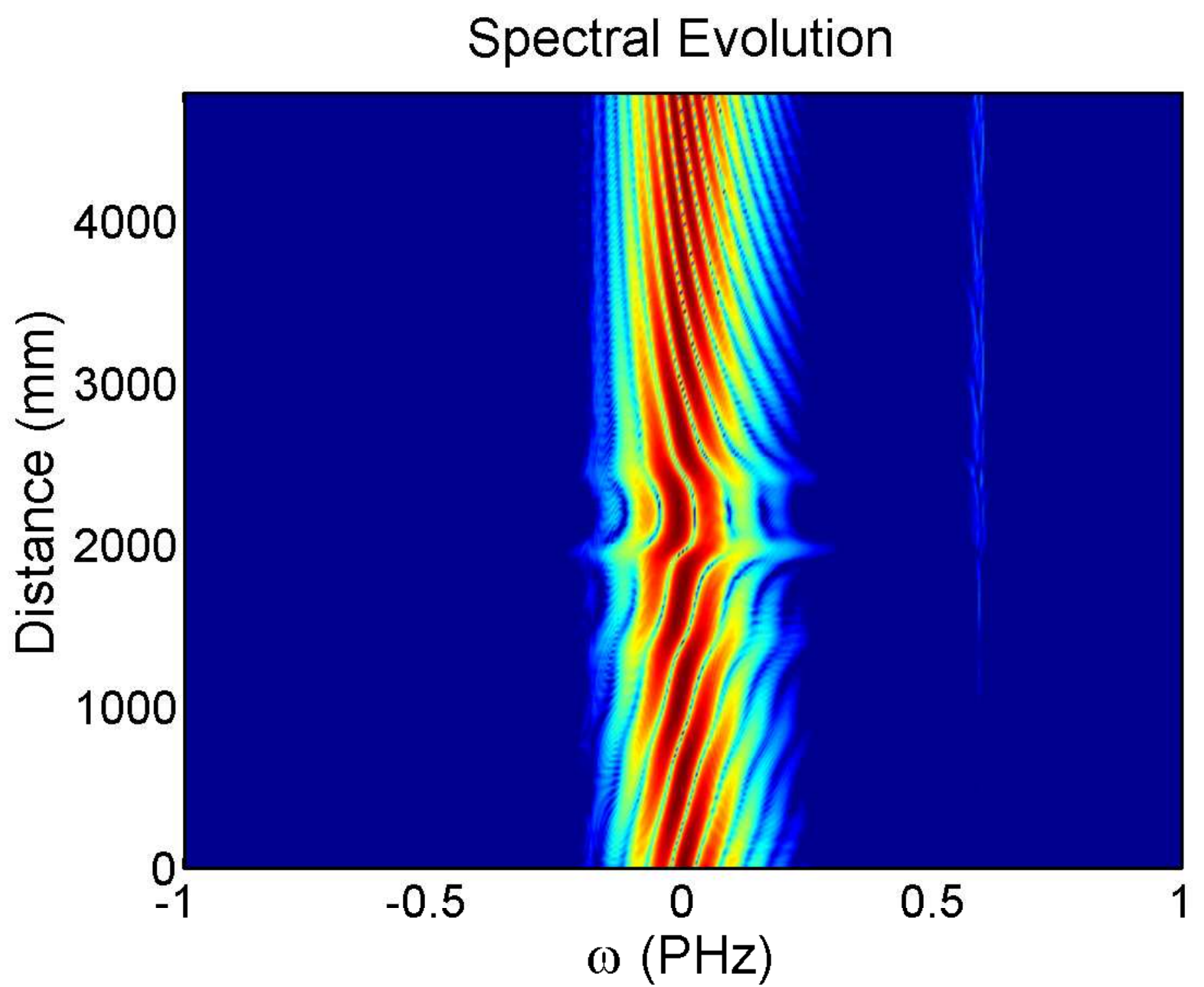}
\includegraphics[width=130pt]{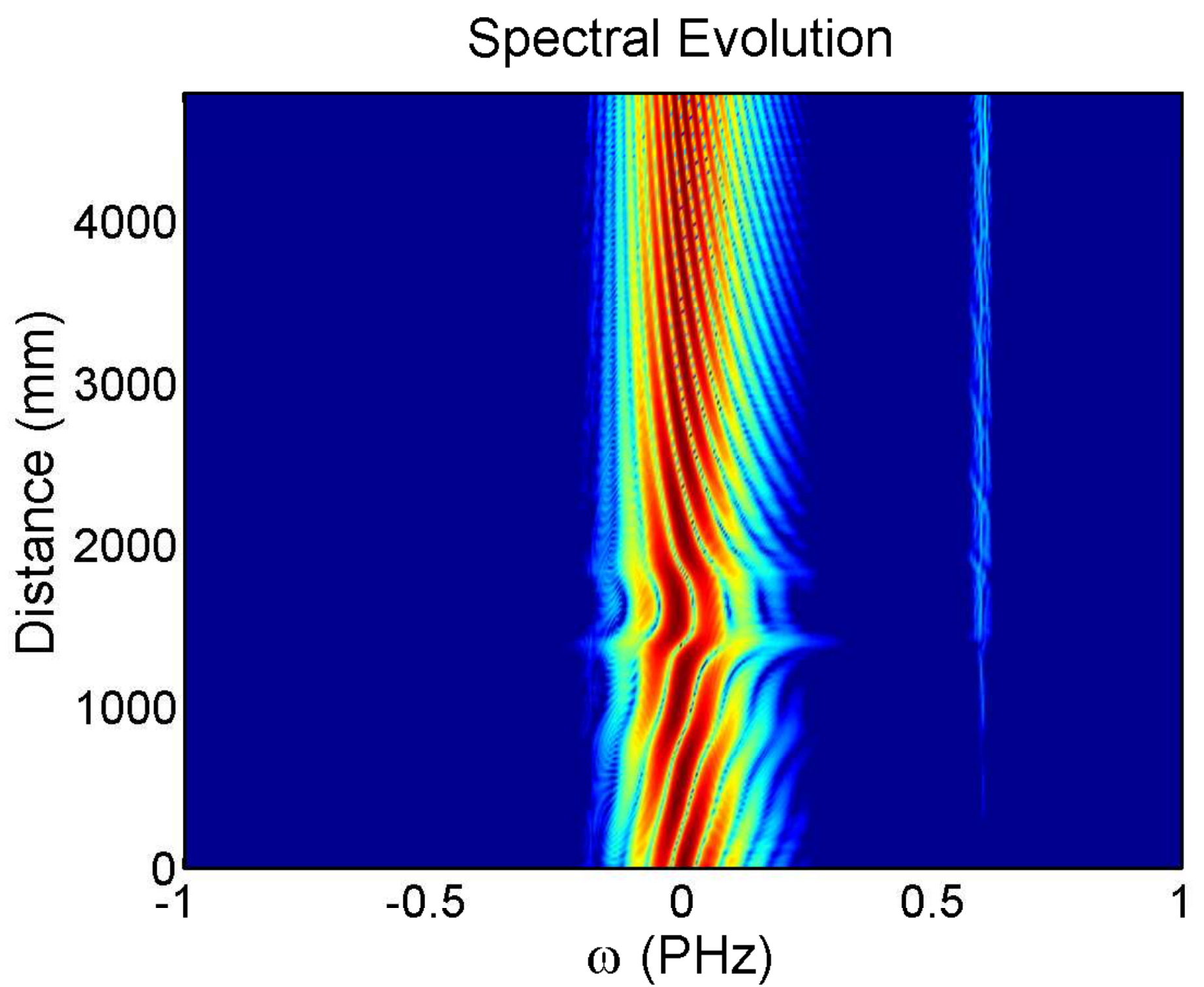}
\includegraphics[width=130pt]{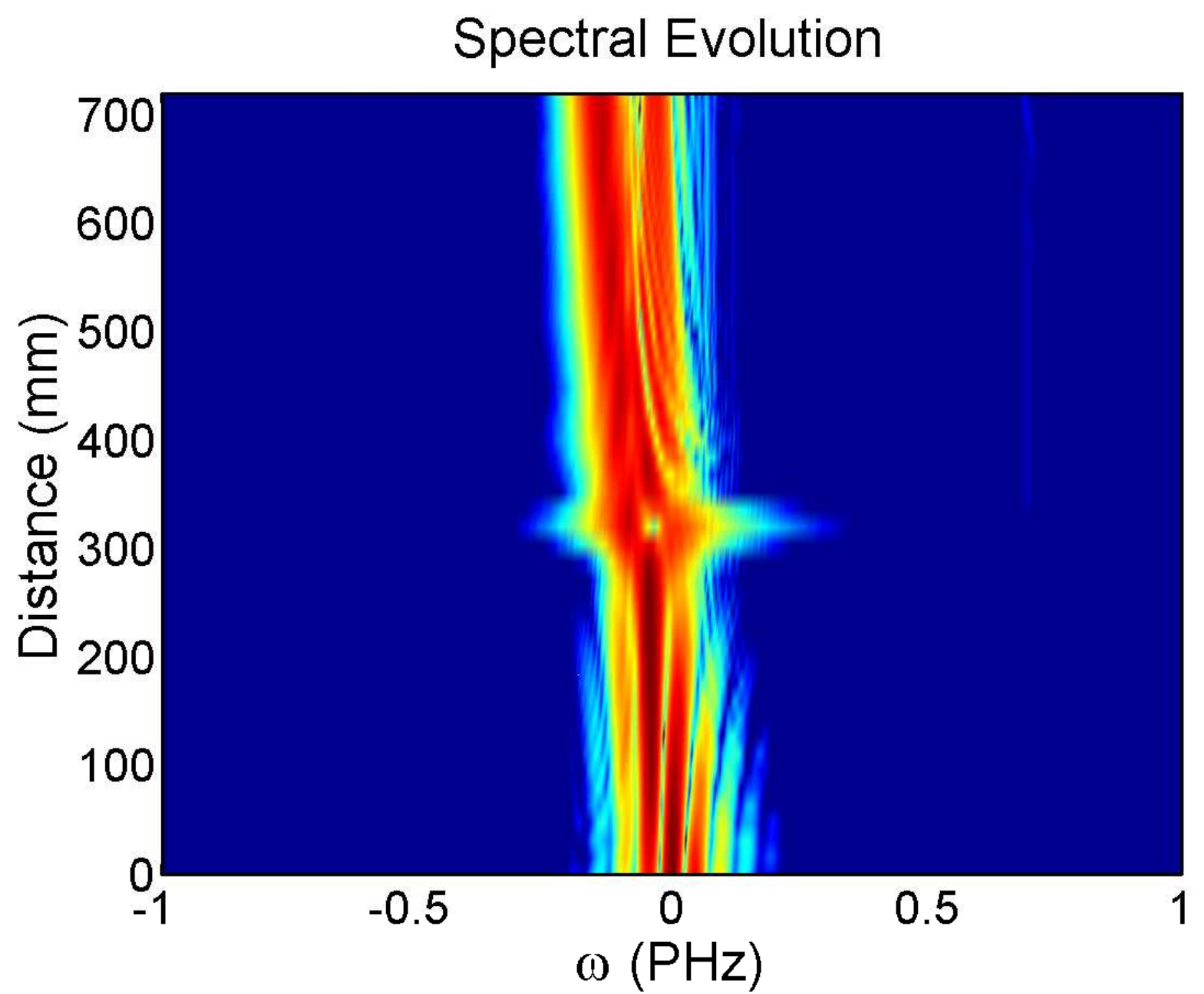}
\caption{\small {\it (Color online) from upper to bottom row: output signal, it's spectrogram, and temporal and phase evolution during the propagation of initial pulse in the form of superposition of two classical NLS solitons in PCF for (from left to right) NLS+HD, NLS+HD+SS and NLS+HD+SS+RS systems respectively.}}
\end{figure}

The second group of experiments was made to analyze solitons interactions and especially of their collisions. Initial field was taken as a superposition of two exact NLS solitons with the same group speeds but with slightly different amplitudes. In case of the classical NLSE such a combination oscillates near it's initial state (see Fig.2). When higher order dispersion or additional nonlinearity is added, the two-soliton state is no longer stable and decomposes for individual peaks moving with slightly different group speeds. Difference in peak speeds may lead to their collision. As for the previous group of experiments, for all 3 systems a), b) and c) solitons after the collision turned out to be very close to the family of exact NLS solitons. Nevertheless, here arises qualitative difference between the considered systems: for NLS+HD and NLS+HD+SS systems collisions were almost elastic even when the additional nonlinearity beyond the classical NLSE was added (see Fig.3). Elasticity broke only after addition of Raman scattering. In this case smaller solitons were loosing their energy while the bigger ones - were acquiring. Very similar phenomena are observed for very different physical systems from nonintegrable NLS-like equations to MMT-model (see for example \cite{Zakharov2, Jordan, Zakharov}). Besides, in the presence of Raman scattering both quasi-solitons experienced Raman self-frequency shift as shown on Fig.3.

\begin{figure}[t] \centering
\includegraphics[width=130pt]{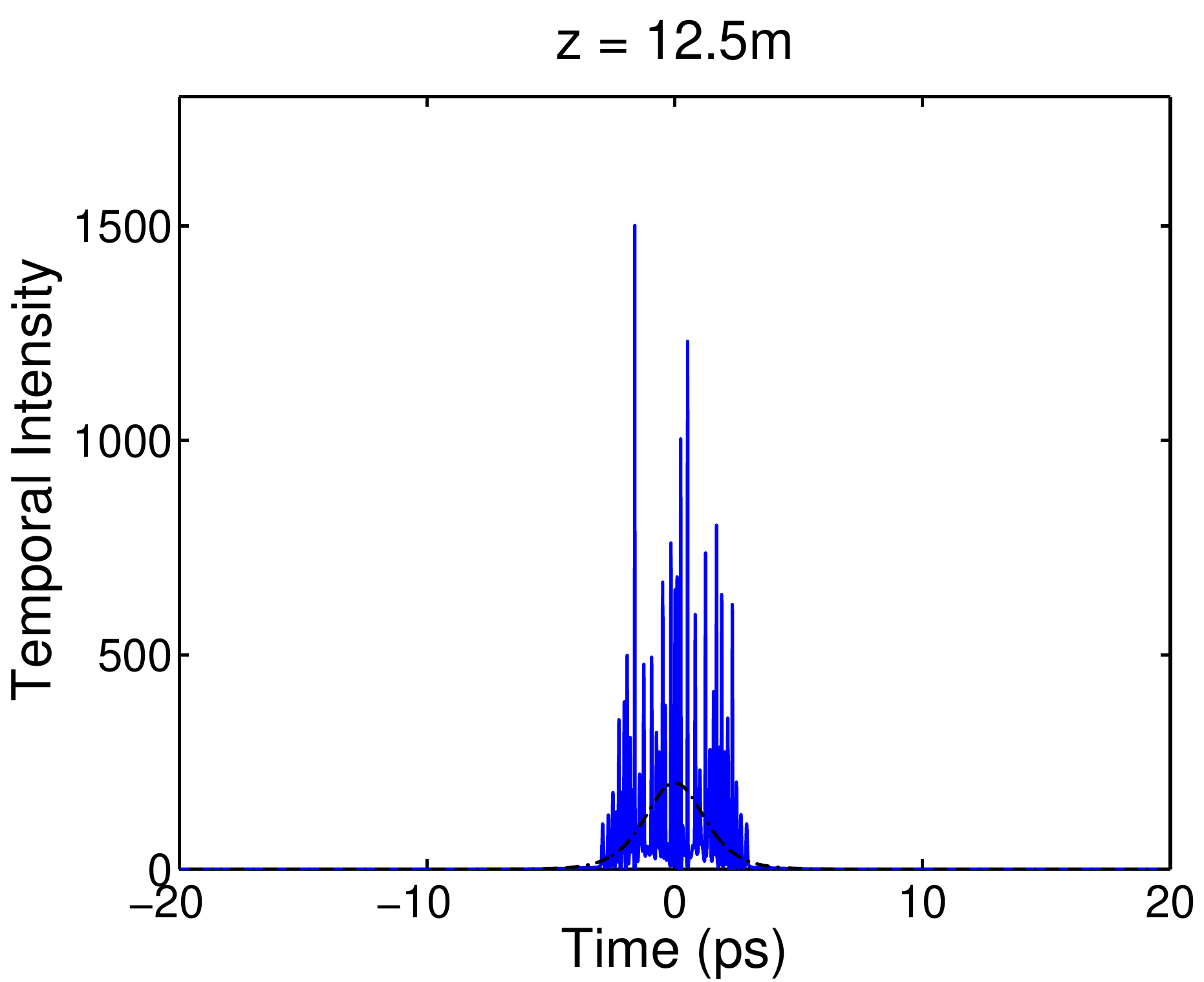}
\includegraphics[width=130pt]{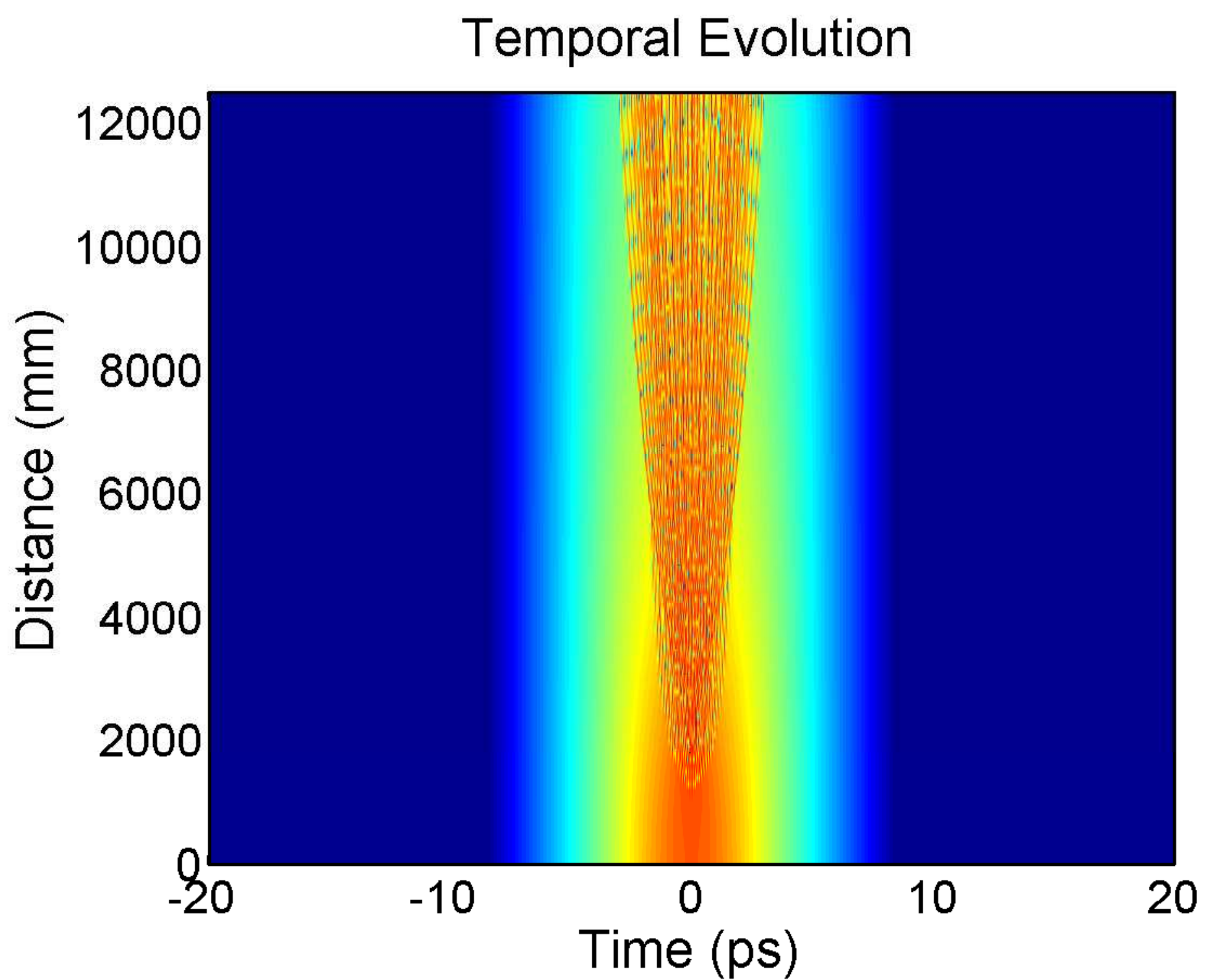}
\includegraphics[width=130pt]{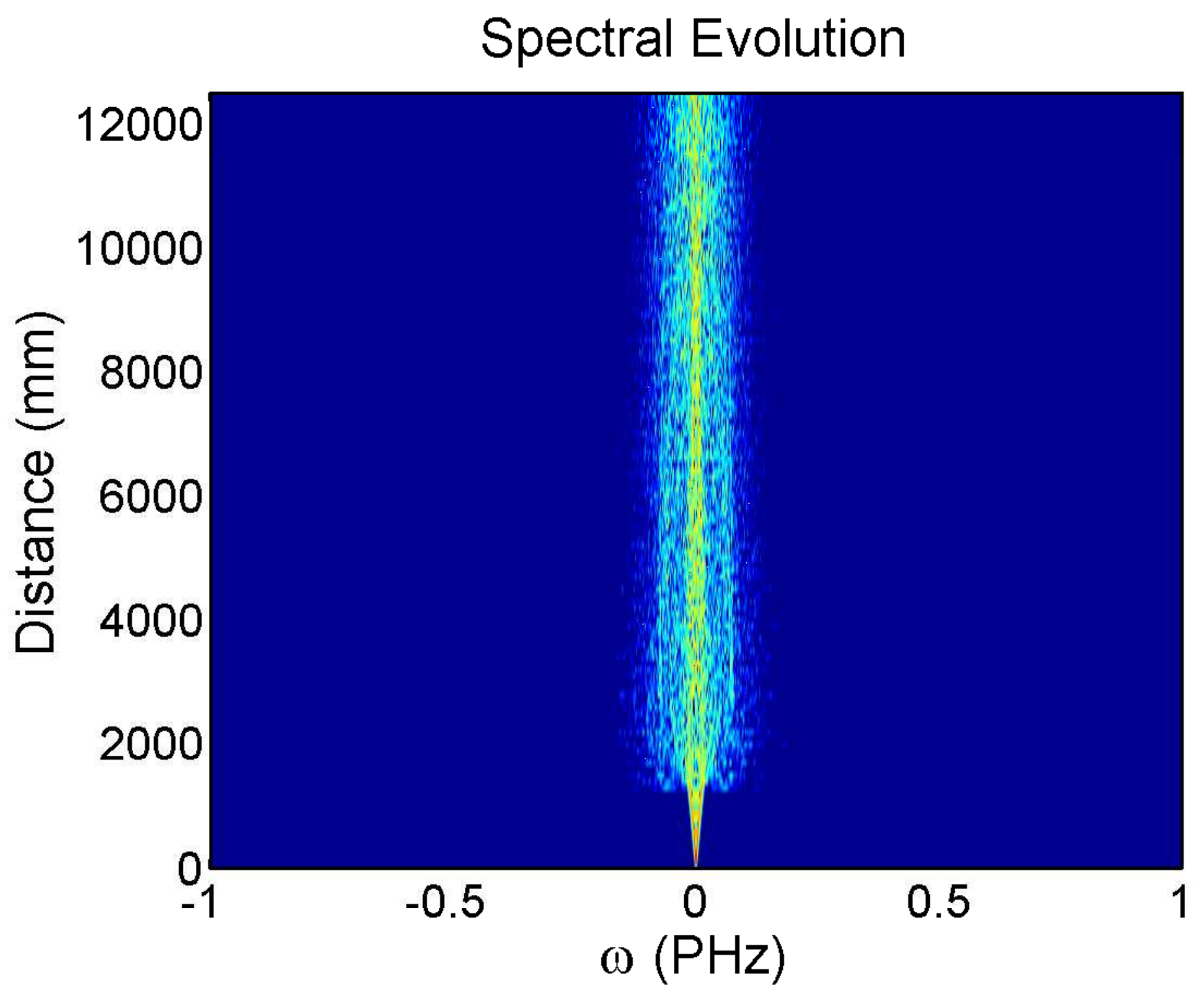}
\caption{\small {\it (Color online) from left to right: output signal and it's temporal and spectral evolution of the MI development from initial sech-shaped pulse with 200W peak power inside the PCF described by the classical NLSE.}}
\end{figure}

Solitons behavior discussed above allows one to make the conclusion that despite the fact that Eq. (\ref{1_OpticsEnvelopeEvolution}) does not possess exact soliton solutions, it allows quasi-soliton states which are very close to classical NLS solitons and have common features with them: almost steady movement, stability with respect to class of quasi-soliton solutions against collisions. From the over hand, elasticity of collisions depends on additional nonlinearity added to the system.

\begin{figure}[t] \centering
\includegraphics[width=130pt]{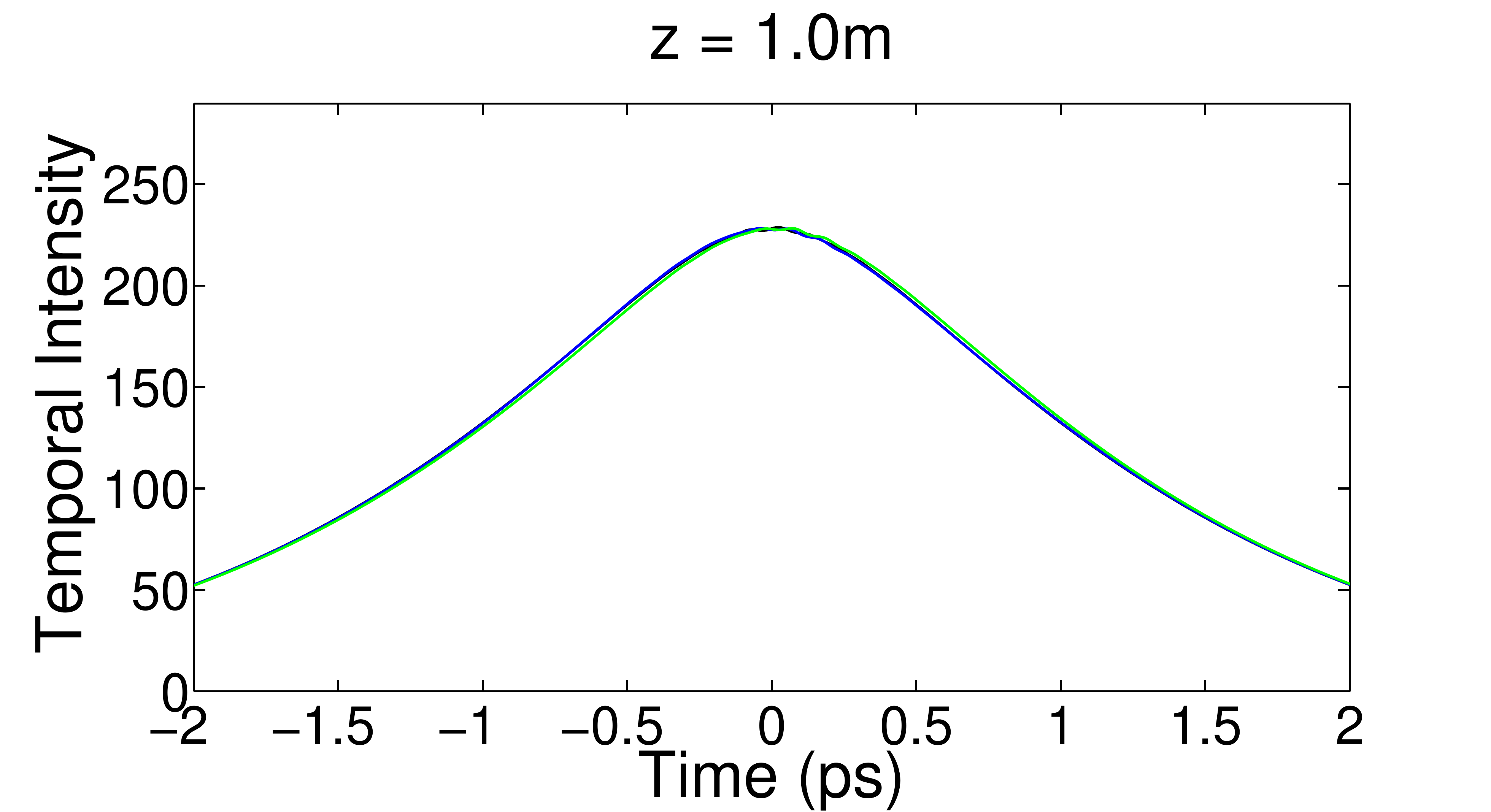}
\includegraphics[width=130pt]{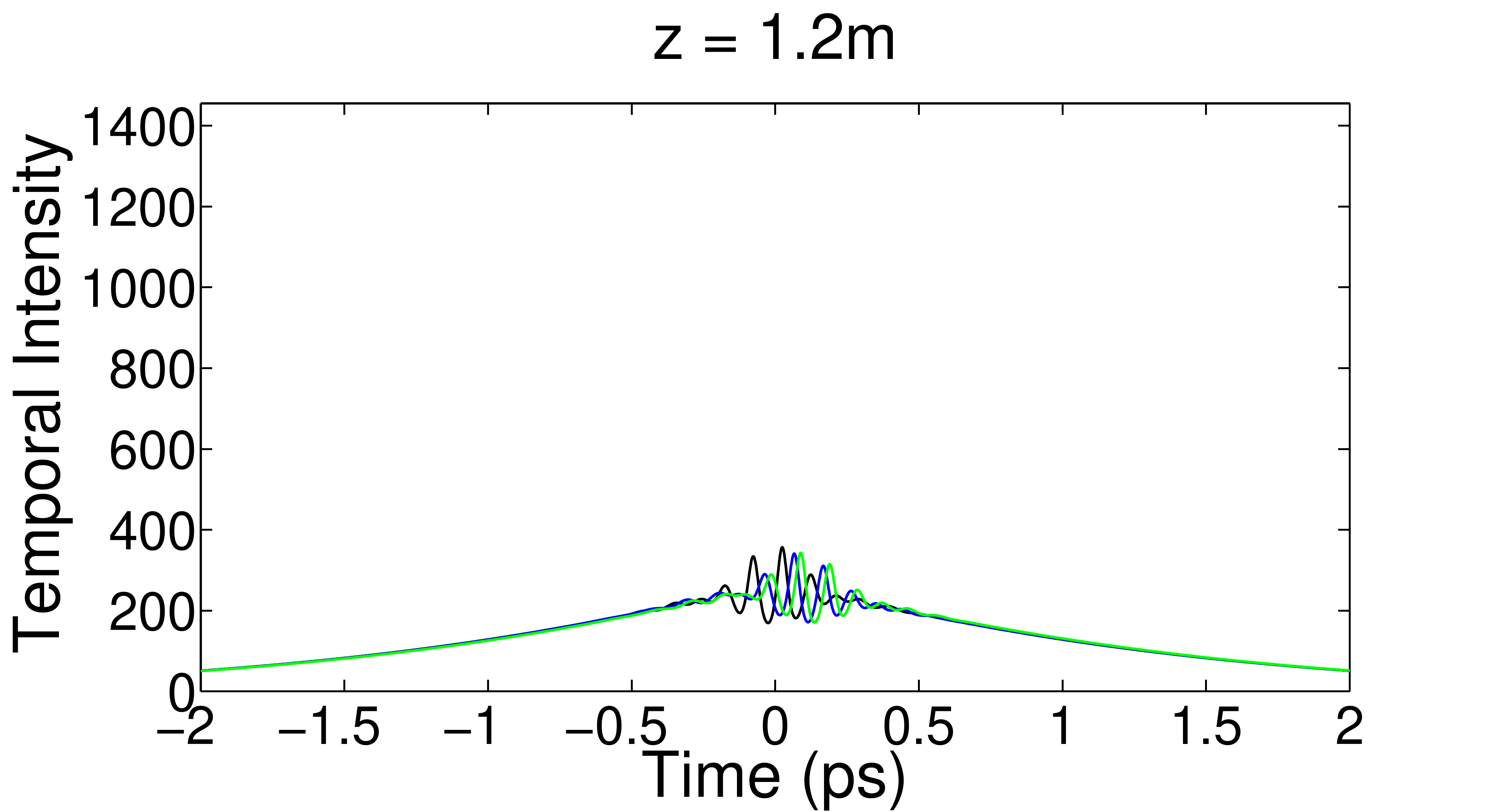}
\includegraphics[width=130pt]{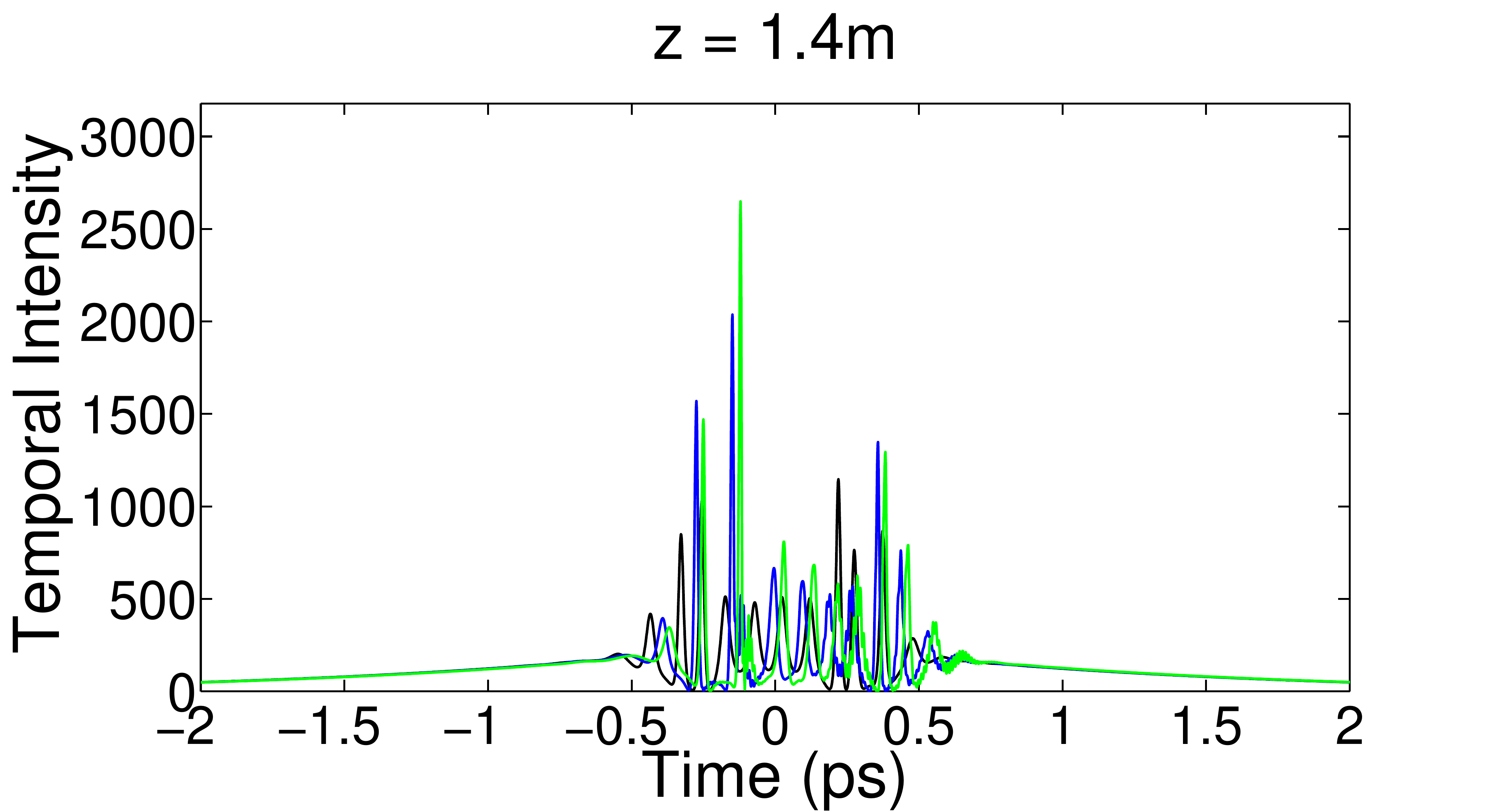}
\includegraphics[width=130pt]{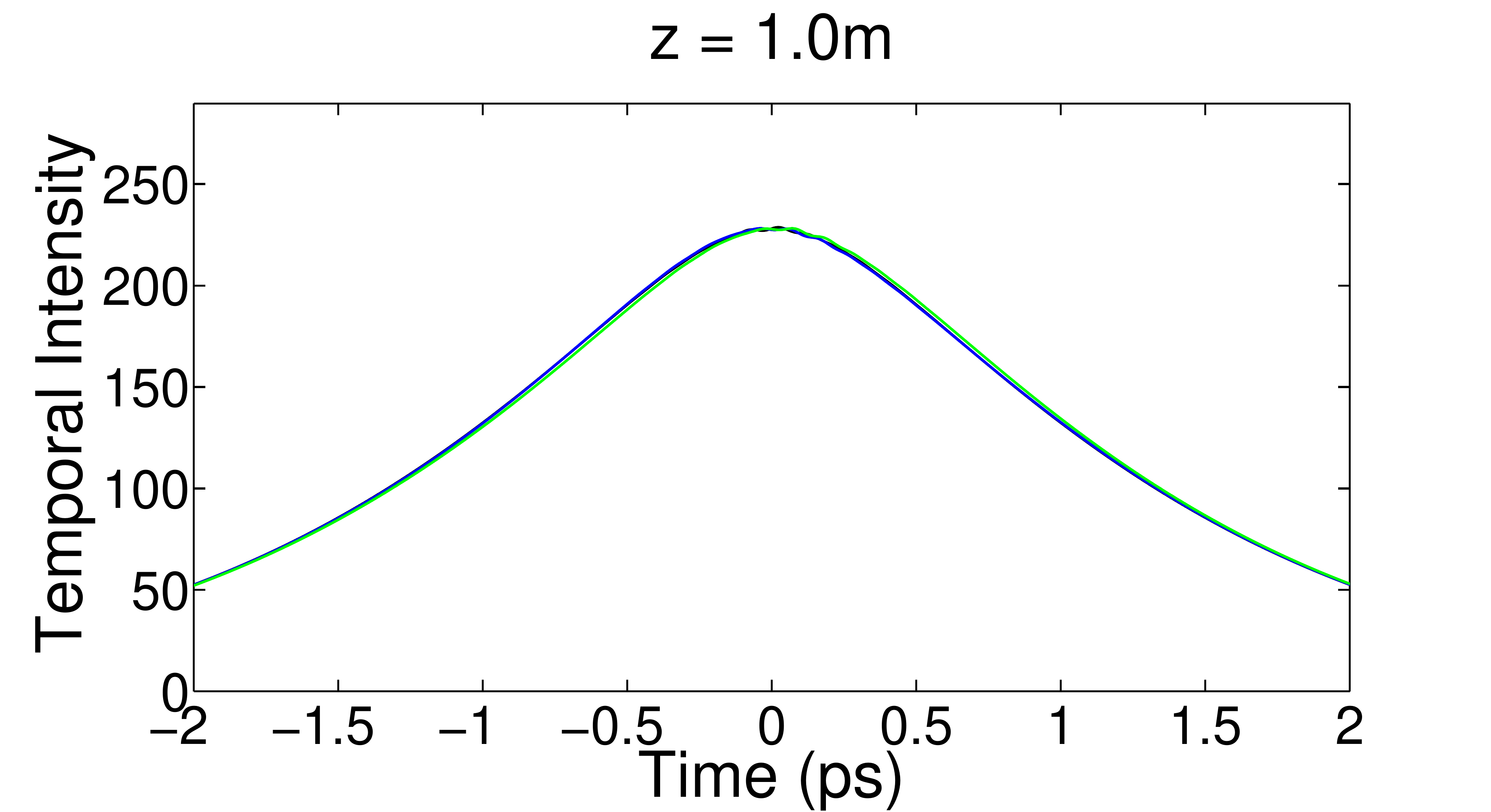}
\includegraphics[width=130pt]{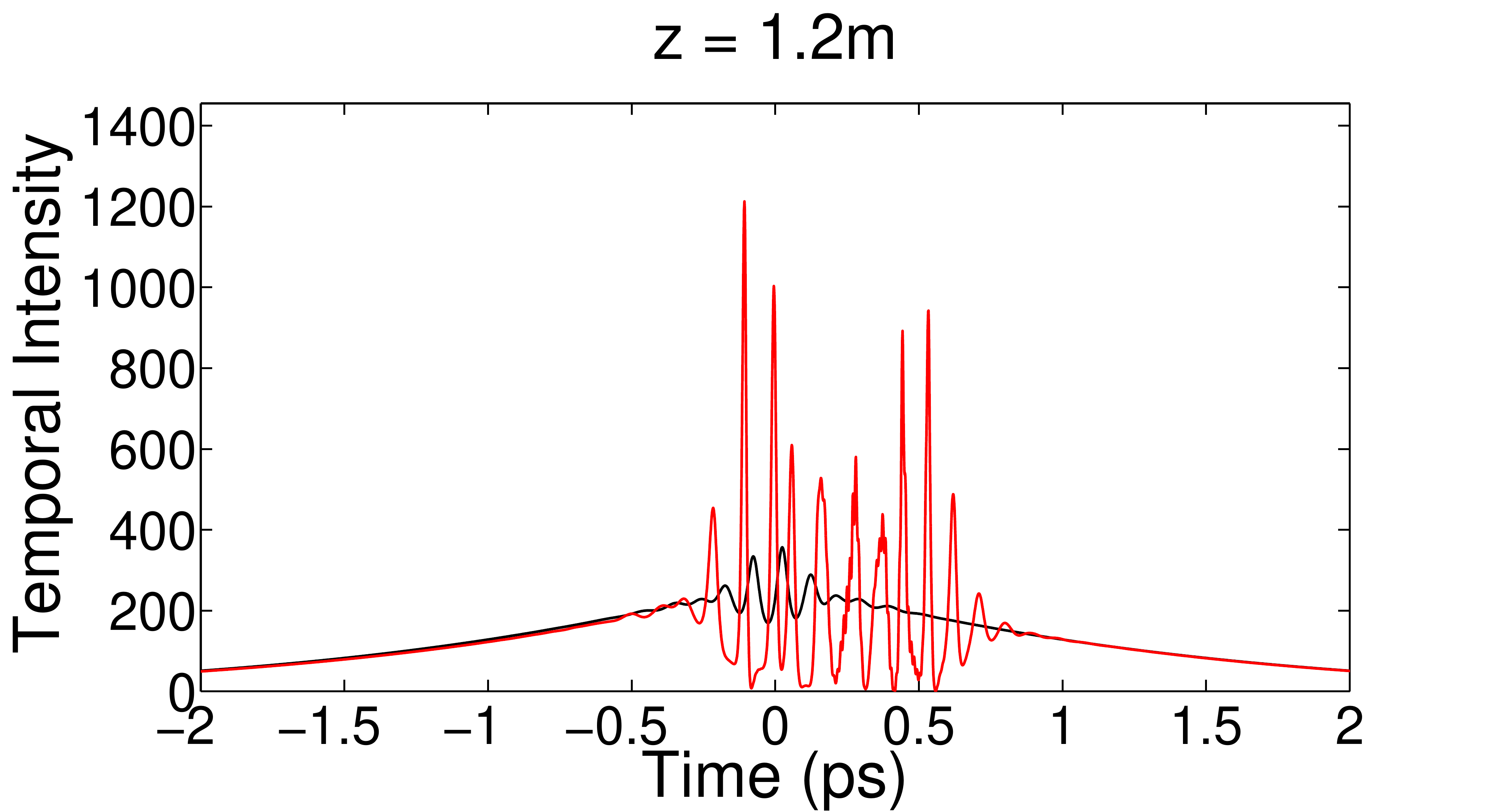}
\includegraphics[width=130pt]{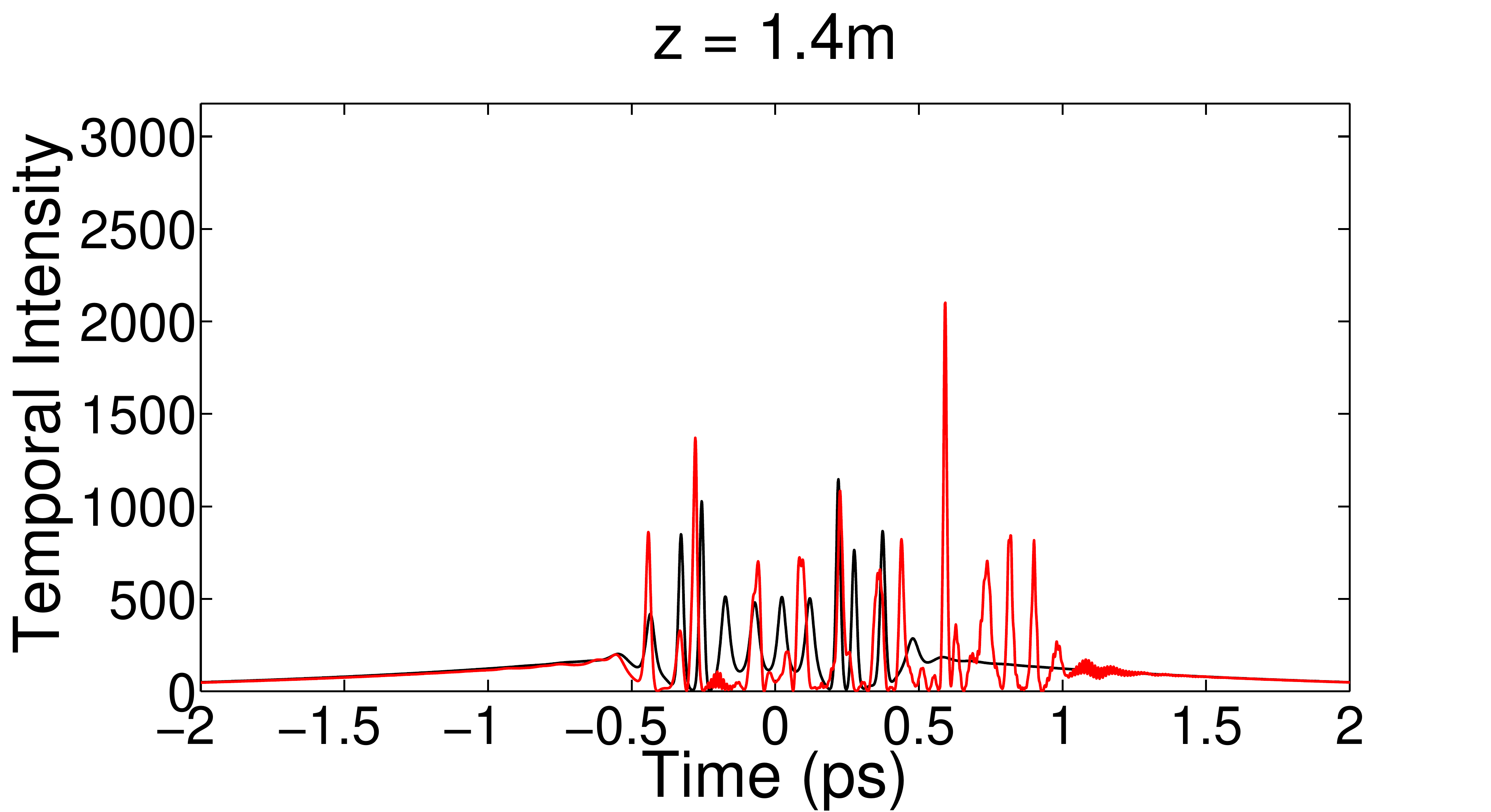}
\caption{\small {\it (Color online) first stage of quasi-solitons formation due to MI development from initial sech-shaped pulse with 200W peak power. The black line corresponds to classical NLS system, the blue one - to NLS+HD, the green one - to NLS+HD+SS, and the red one - to NLS+HD+SS+RS.}}
\end{figure}

The third group of experiments was devoted to MI development. Initial wave field was taken in the form $A(t)=\sqrt{P_{0}}sech(t/T_{0})$, where $P_{0}=200W$ and $T_{0}=1.6ps$ were peak power and pulse duration respectively. In case of the classical NLSE such peak decomposes for several peaks condensed near the center of the initial pulse (see Fig.4). During the decomposition of the pulse it's spectra broadens from few angstroms to several nanometers. When the symmetry $t\to -t$ is broken due to higher order dispersion or additional nonlinearity, the condensation breaks: peaks start to move with slightly different group speeds. In the literature (see \cite{DGC} for example) such process is referred as soliton fission. From the other hand, integrated effects of higher order dispersion and additional nonlinearity dramatically increase the pulse spectra broadening from several nanometers to hundreds of nanometers, which is often called as supercontinuum generation. As was shown before \cite{DGC}, the solitonic components are located in the red-shifted area of the pulse spectra, while the dispersive waves components - in the blue-shifted area.

\begin{figure}[t] \centering
\includegraphics[width=130pt]{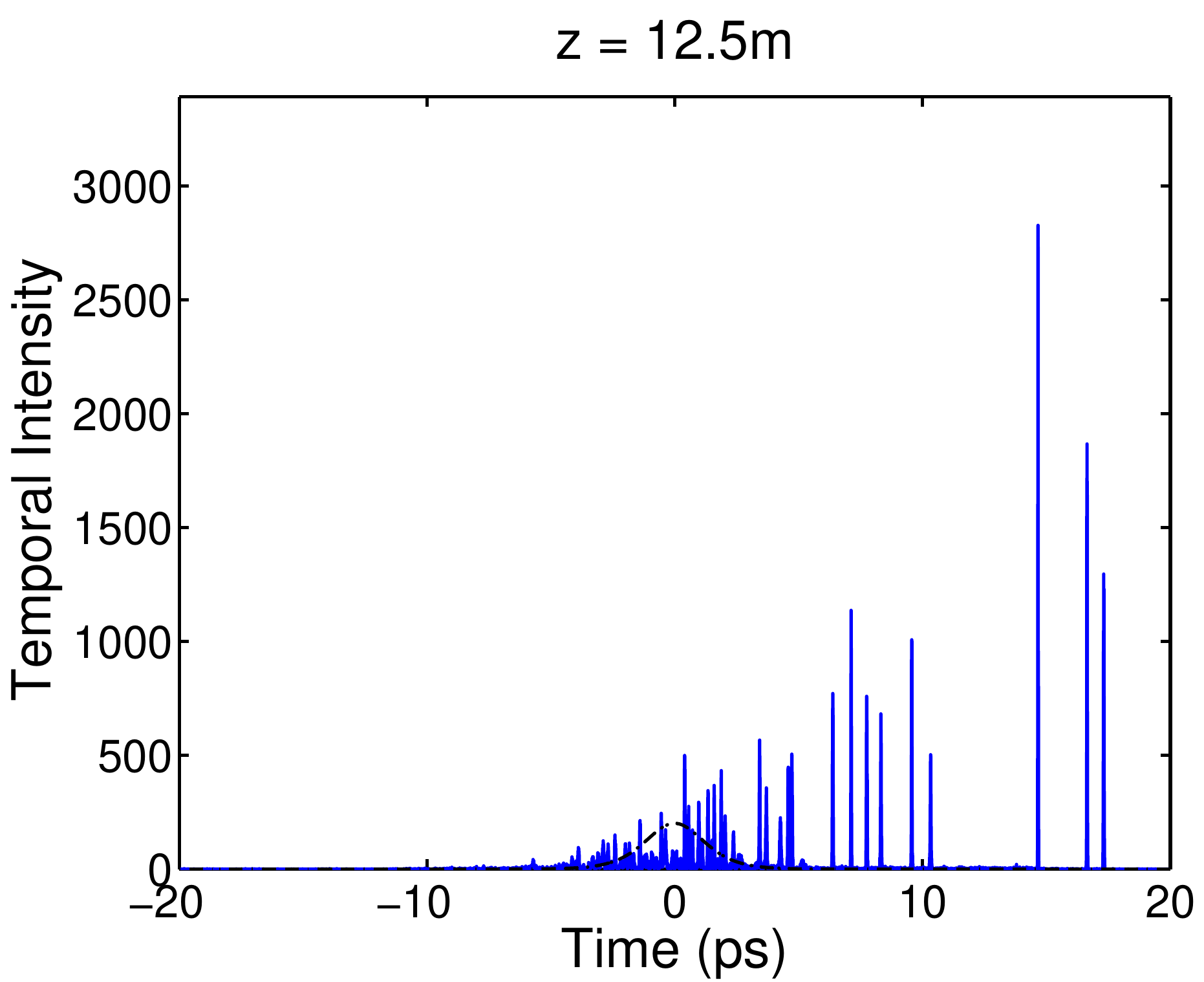}
\includegraphics[width=130pt]{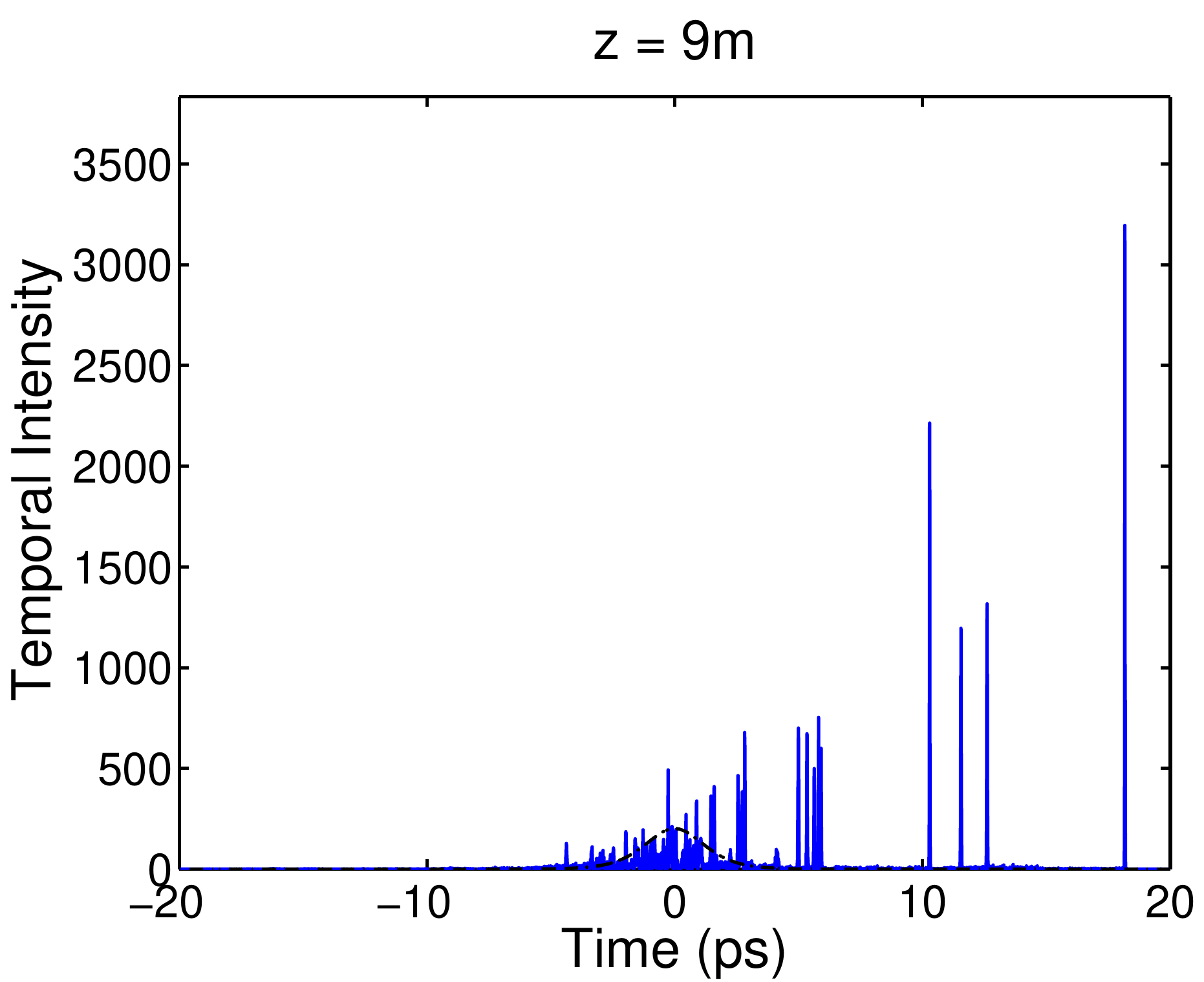}
\includegraphics[width=130pt]{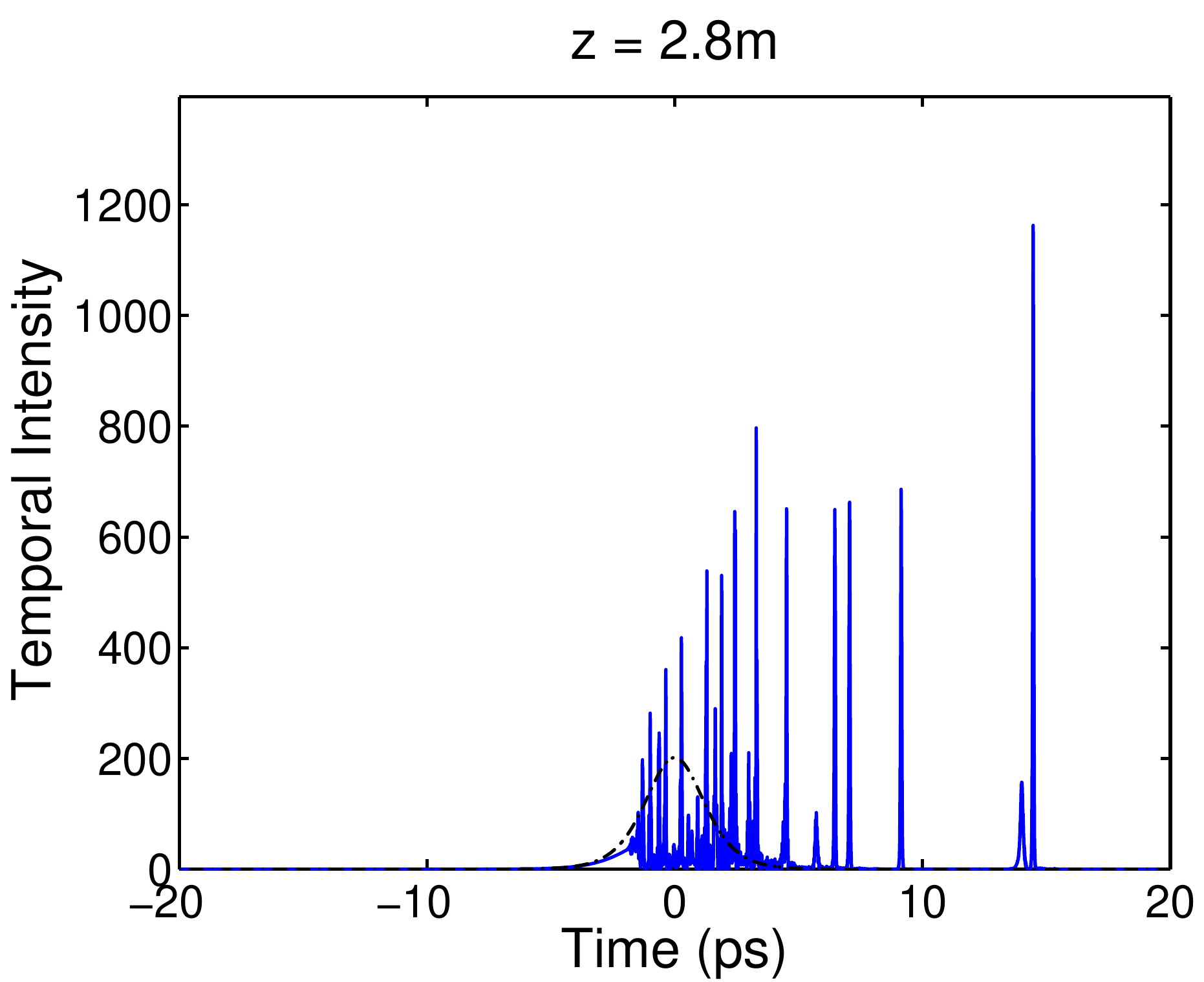}
\includegraphics[width=130pt]{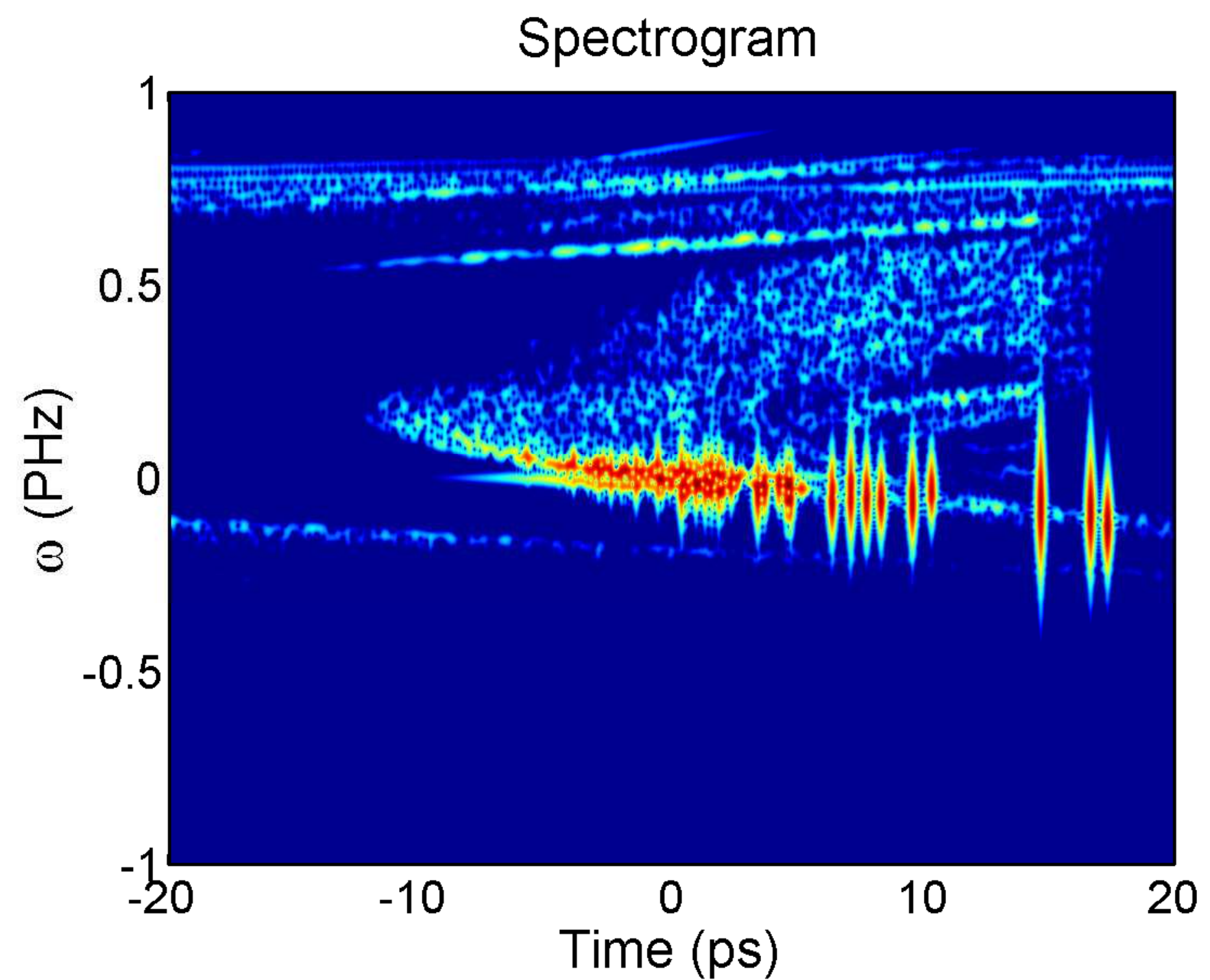}
\includegraphics[width=130pt]{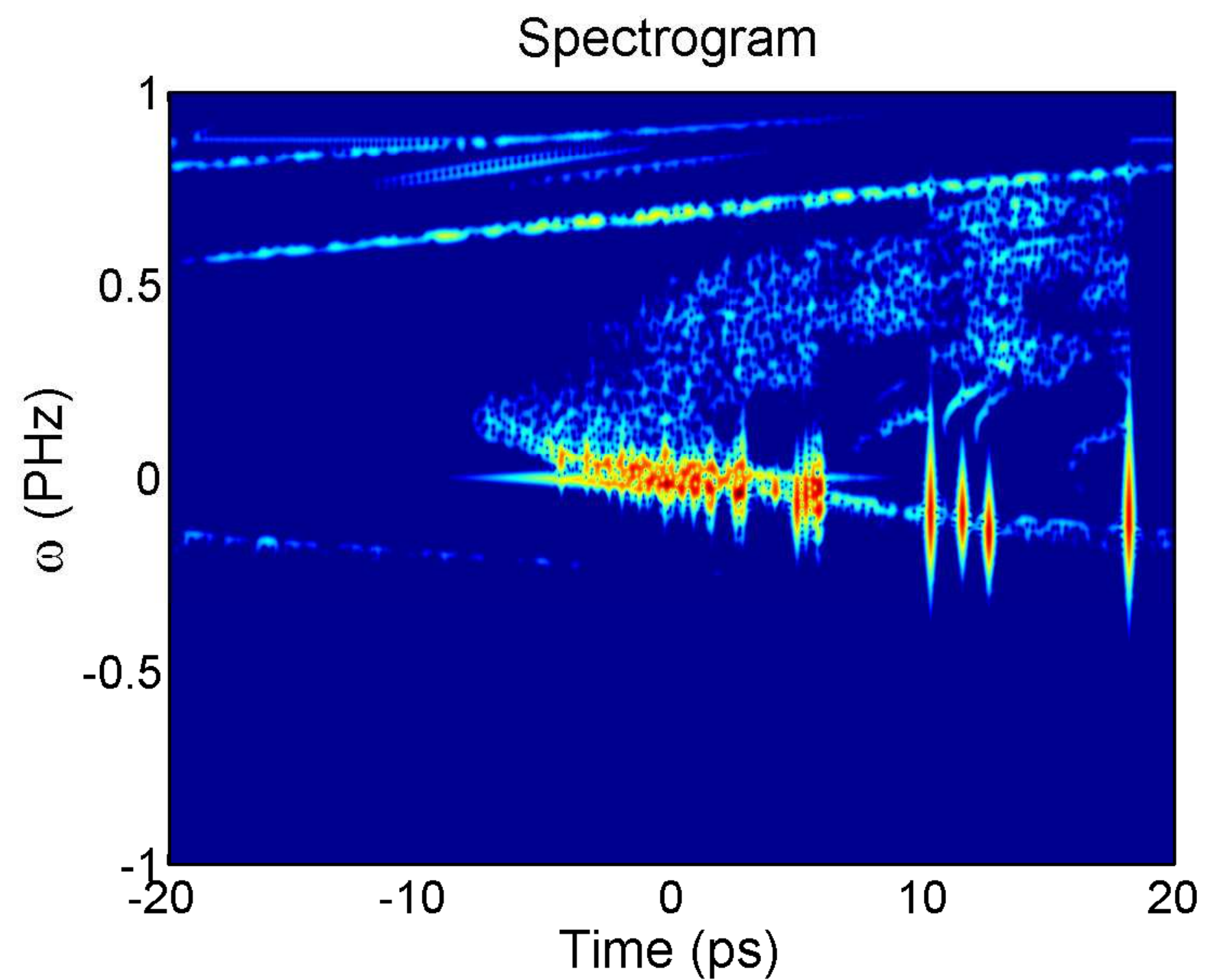}
\includegraphics[width=130pt]{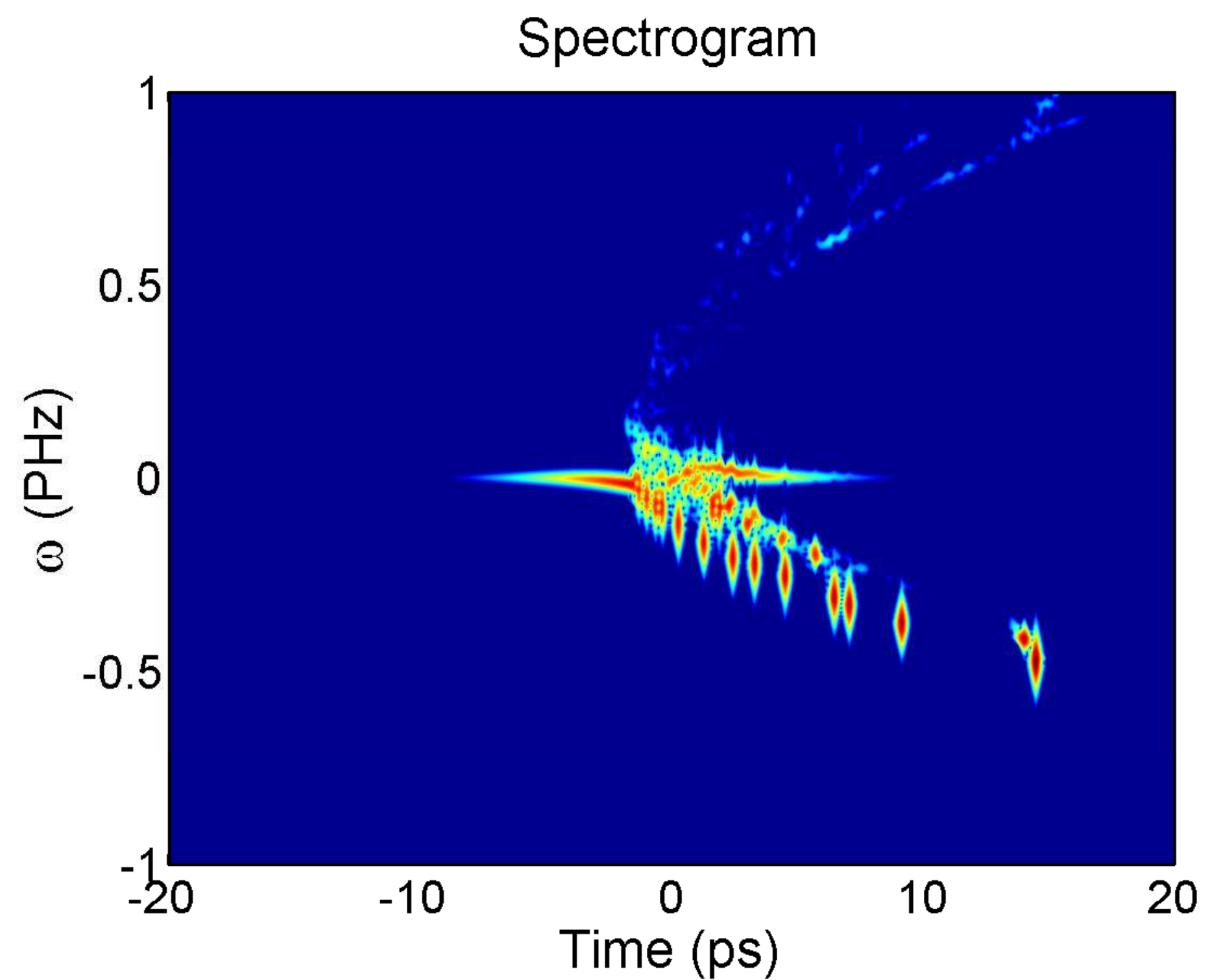}
\includegraphics[width=130pt]{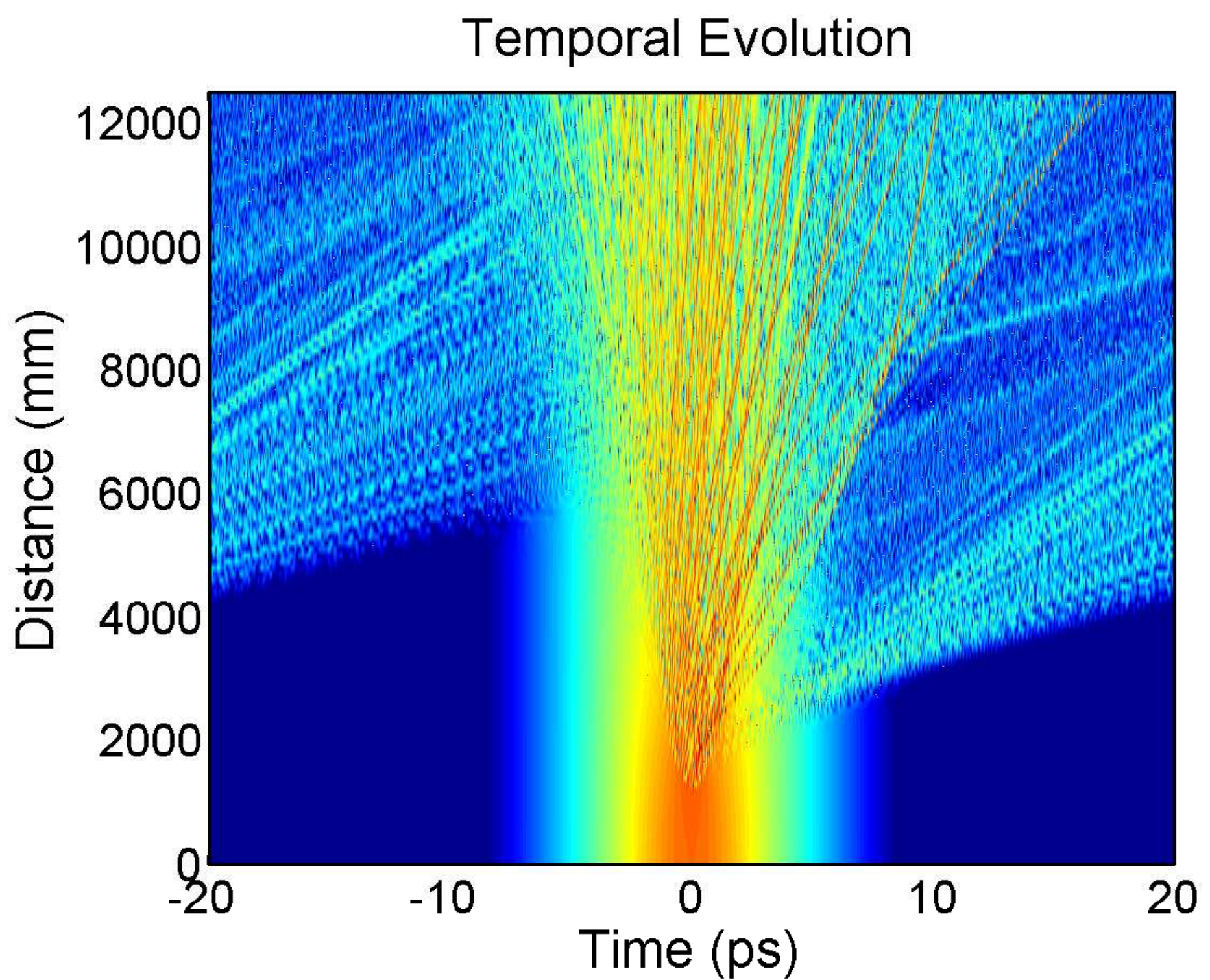}
\includegraphics[width=130pt]{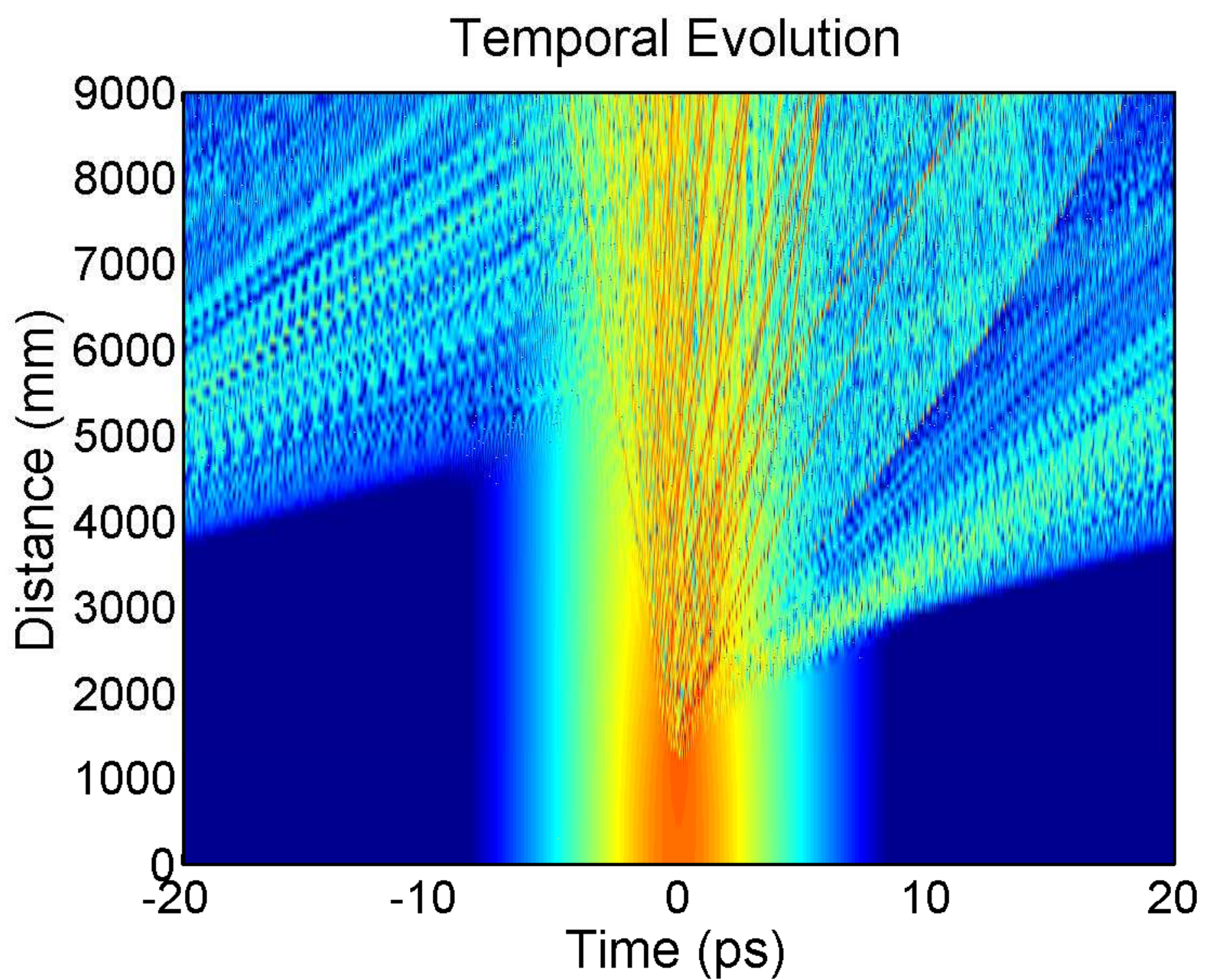}
\includegraphics[width=130pt]{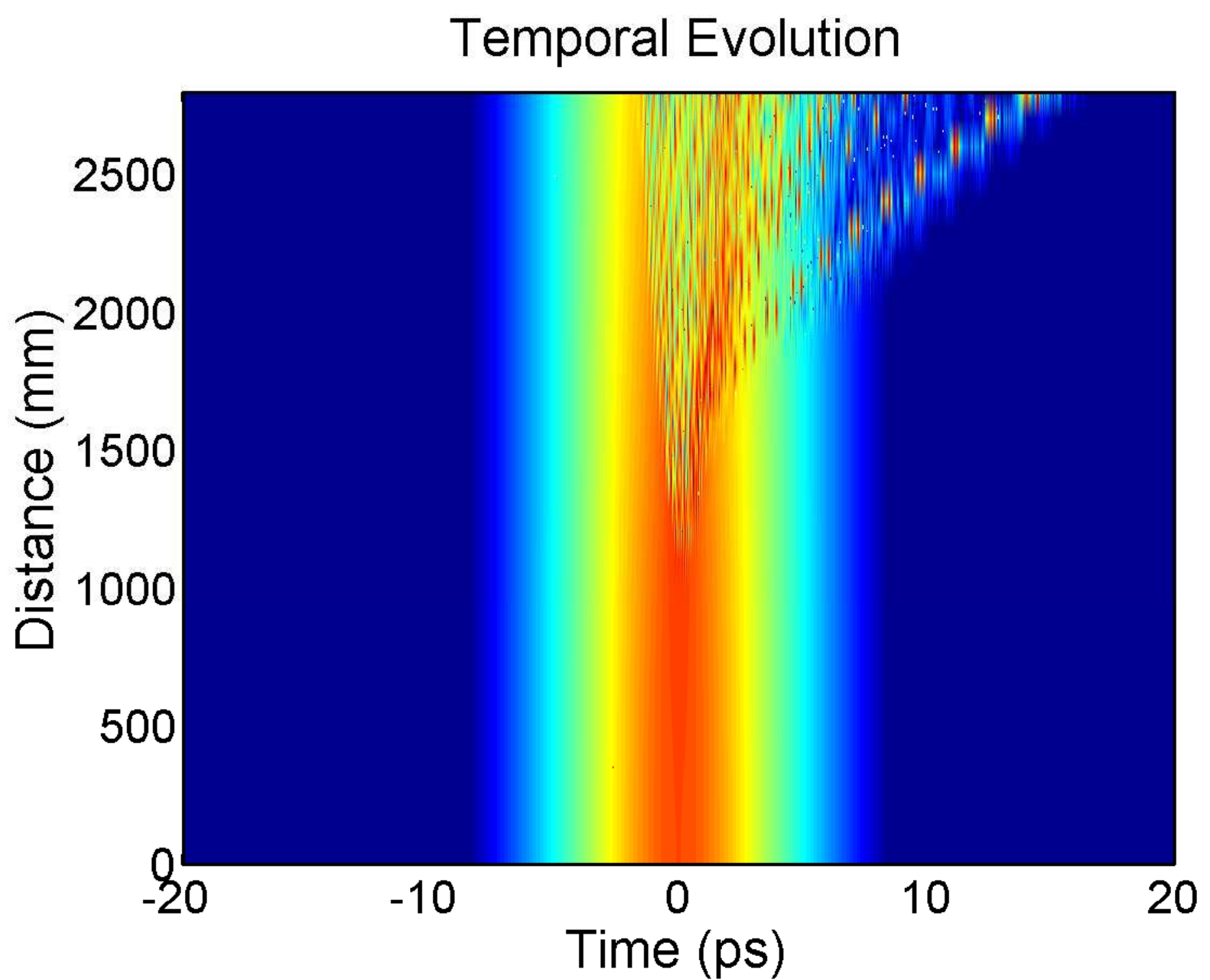}
\includegraphics[width=130pt]{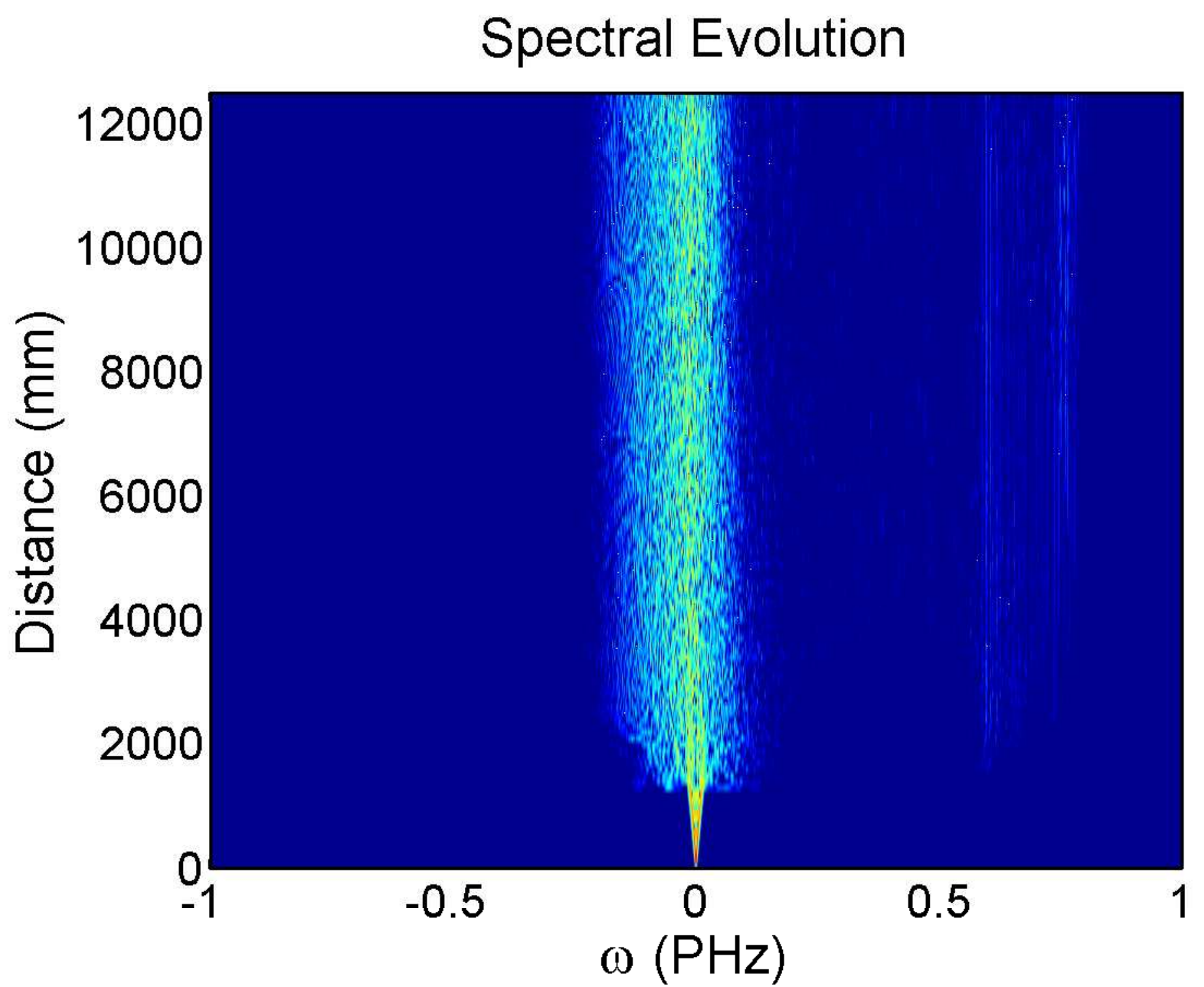}
\includegraphics[width=130pt]{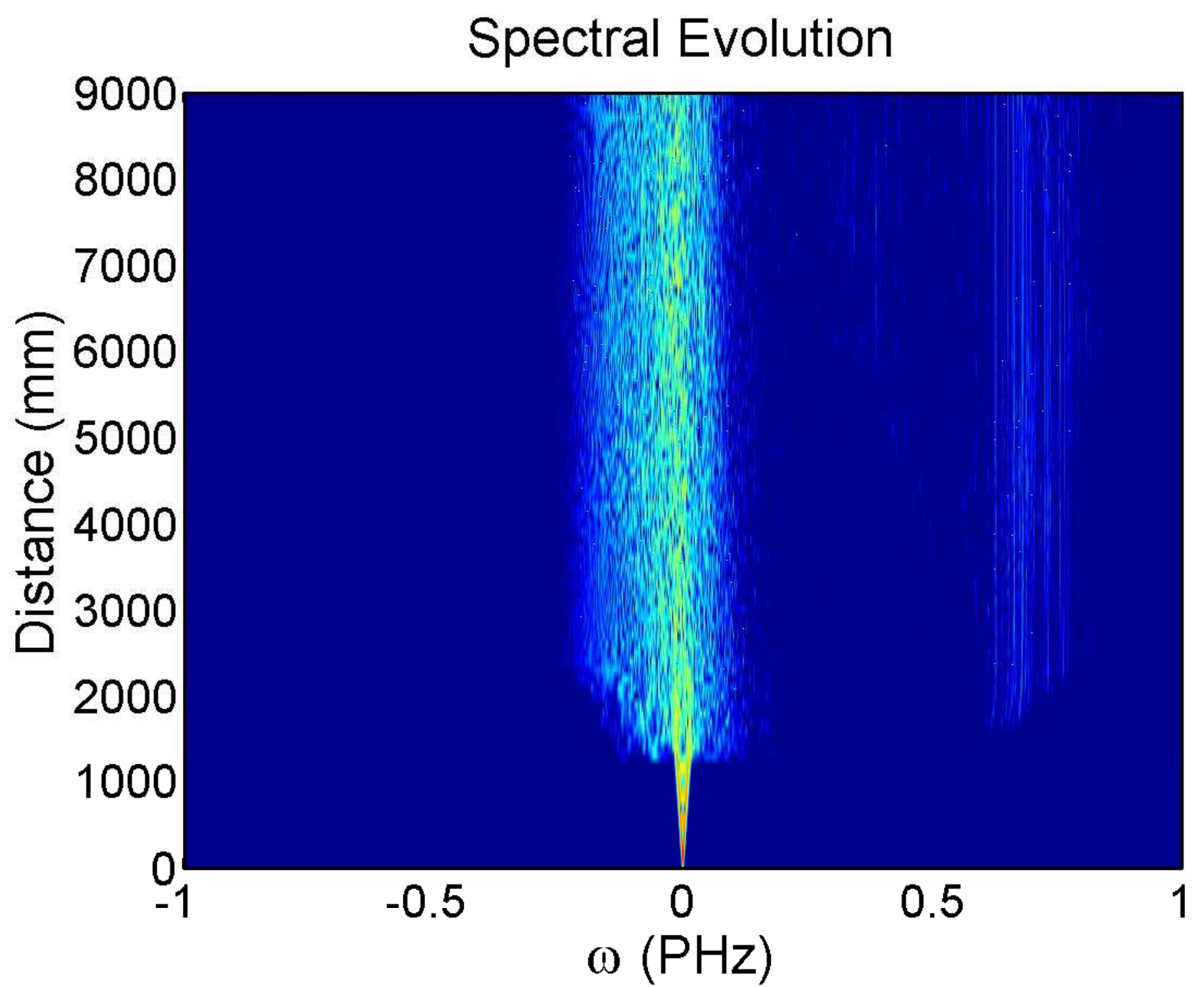}
\includegraphics[width=130pt]{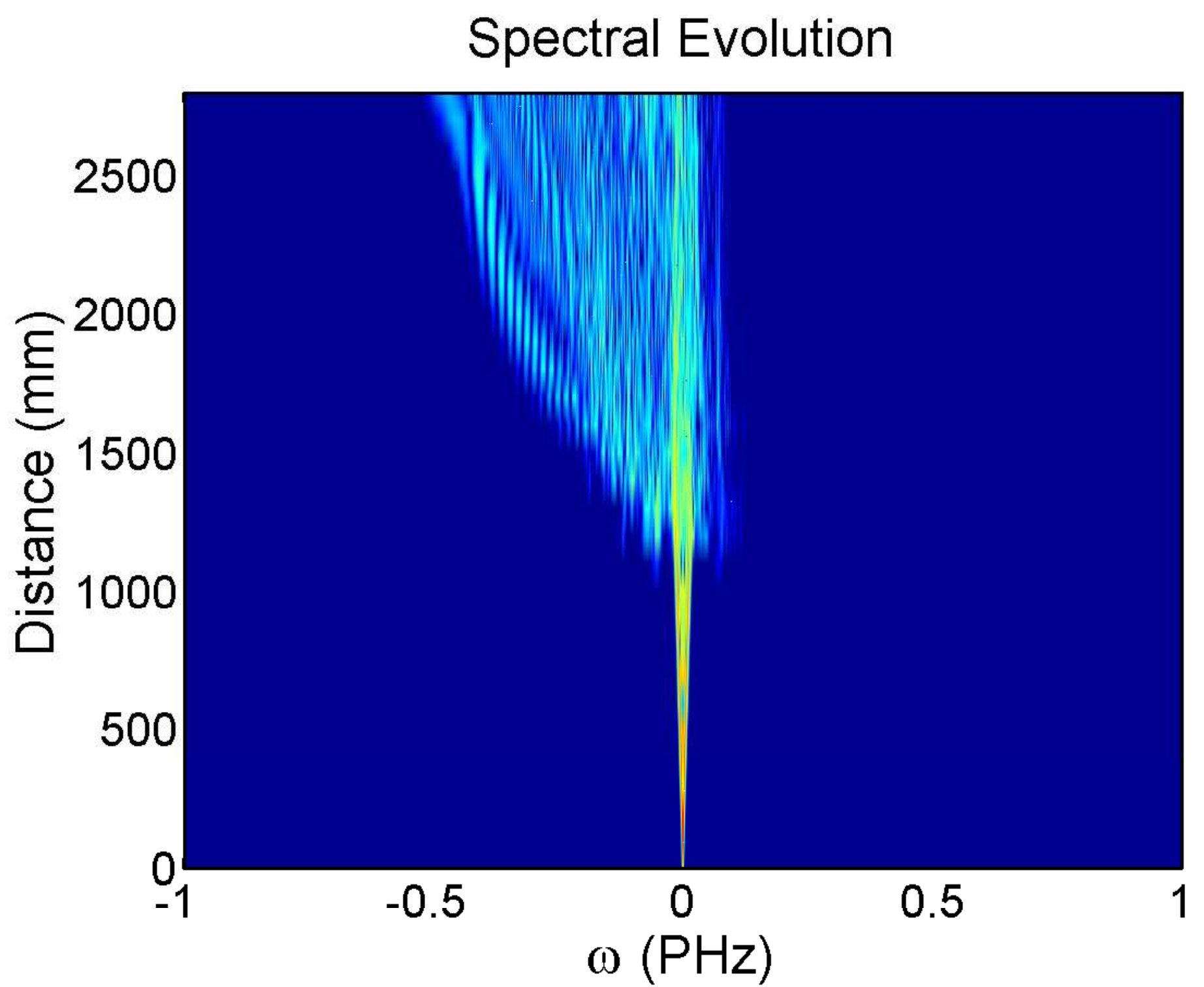}
\caption{\small {\it (Color online) from upper to bottom row: output signal, it's spectrogram, and temporal and phase evolution of the MI development from initial sech-shaped pulse with 200W peak power inside the PCF described by (from left to right) NLS+HD, NLS+HD+SS and NLS+HD+SS+RS systems respectively.}}
\end{figure}

Let us consider the process of quasi-solitons formation and their further fission from it's beginning. As shown on Fig.5, the start of such decomposition for all nonlinear systems considered in this work was well described by the classical NLS equation. Higher order dispersion as well as the self-steepening were found to have no influence on the speed of quasi-solitons formation. Thus, the role of these terms in soliton fission process was reduced to the symmetry breaking, i.e. bringing different group velocity additions to different quasi-solitons, and also to small dispersive waves generation. From the other hand, Raman scattering drastically amplified appearance of quasi-solitons from the very beginning of their formation, though the start of quasi-solitons formation itself was well described by the classical NLS equation as before.

As the quasi-solitons are emerged, we are coming to the next stage of MI development - quasi-solitons interaction (see Fig.6). During this stage quasi-solitons chaotically interact with each over and with dispersive waves. We think that soliton-to-soliton interactions are governed by the same general rules as were found for the soliton-to-soliton collisions: quasi-solitons interactions are almost elastic for NLS+HD and NLS+HD+SS systems, and inelastic for fully generalized NLS equation with Raman scattering taken into account. In the latter case smaller quasi-solitons loose their energy while the bigger ones - acquire. Therefore, Raman scattering not only affects the speed of quasi-soliton formation, but also brings to the system additional and effective mechanism of the energy exchange between them. The other important conclusion that could be made from this consideration is that the predictions of the output spectra based on application of the kinetic theory to conservative generalized NLS equation with higher order dispersion or self-steepening taken into account (see \cite{Picozzi, Picozzi2} for more information) should be only a rough approximations to the real optical fibers.

It is necessary to pay attention here to the generation of dispersive waves during MI development which is clearly seen on Fig.6. In fact, such generation occurs even for a single quasi-soliton moving along the PCF in the presence of higher order dispersion (see \cite{DGC} for details). The absence of such effect on Fig.1 take place because the generated dispersive waves turned out to be too small to give contribution to spectrogram or frequency evolution pictures. This situation changes for soliton-to-soliton collisions (see Fig.3) where the dispersive waves are clearly seen on the far right parts of frequency evolution plots but are not visible on the corresponding spectrograms. Interesting, that for the both sets of simulations with soliton-to-soliton collisions and MI development the additional nonlinearity played the role of dumping for the dispersive waves generation, from only a slight influence in case of self-steepening to almost complete dumping in case of Raman scattering.

In the last stage of MI development noninteracting quasi-solitons and dispersive waves are moving with different velocities along the fiber. While in case of higher dispersion and self-steepening terms taken into account the quasi-solitons move almost without change in their shape and speed, the Raman scattering also affects this stage through the Raman self-frequency shift as shown on Fig.6. It is noteworthy to point out that in complete correspondence with \cite{DGC}, earlier emerged quasi-solitons have greater self-frequency shift from the initial wave carrier frequency (see spectrogram plots on Fig.6).

\begin{figure}[t] \centering
\includegraphics[width=130pt]{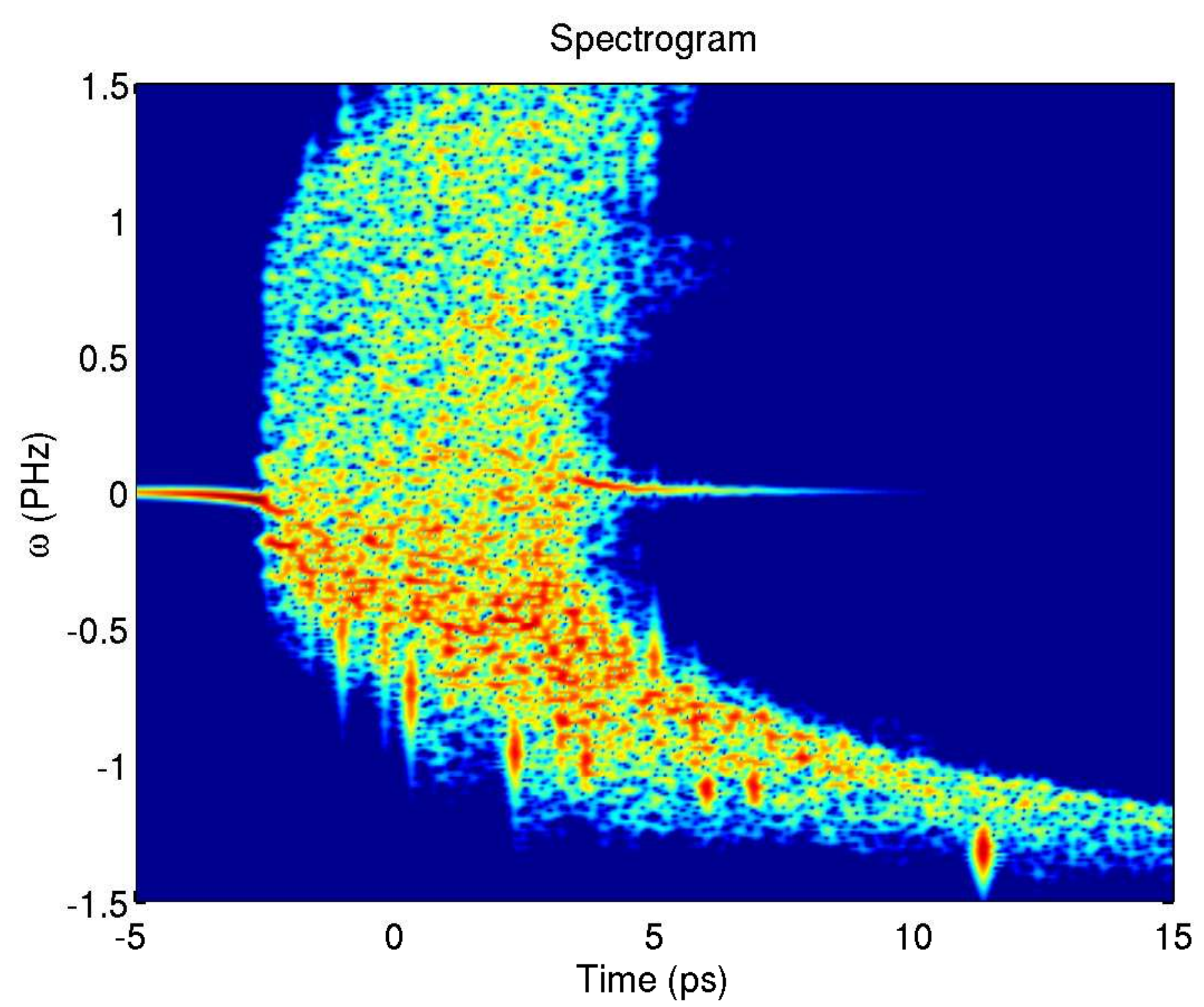}
\includegraphics[width=130pt]{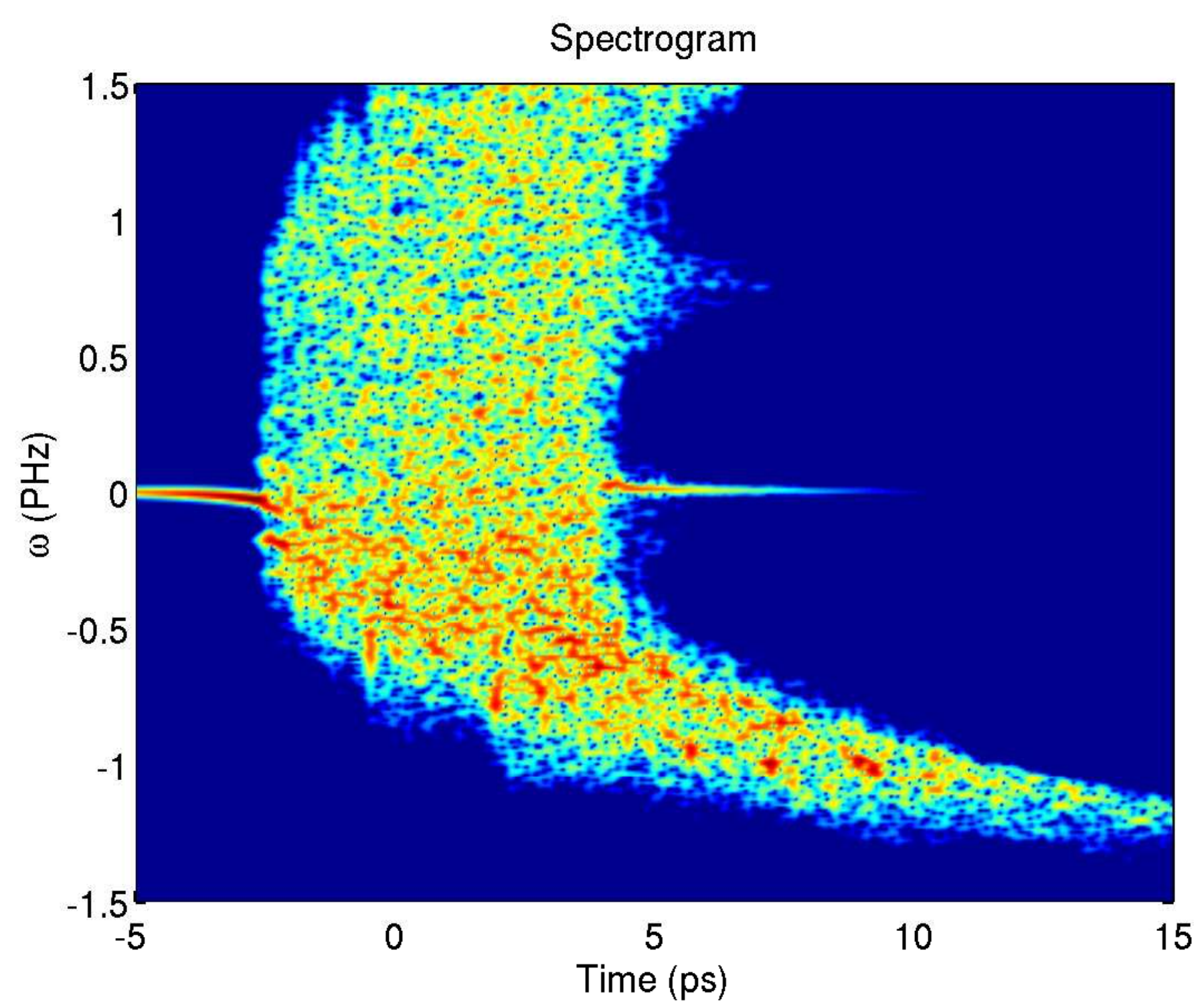}
\includegraphics[width=130pt]{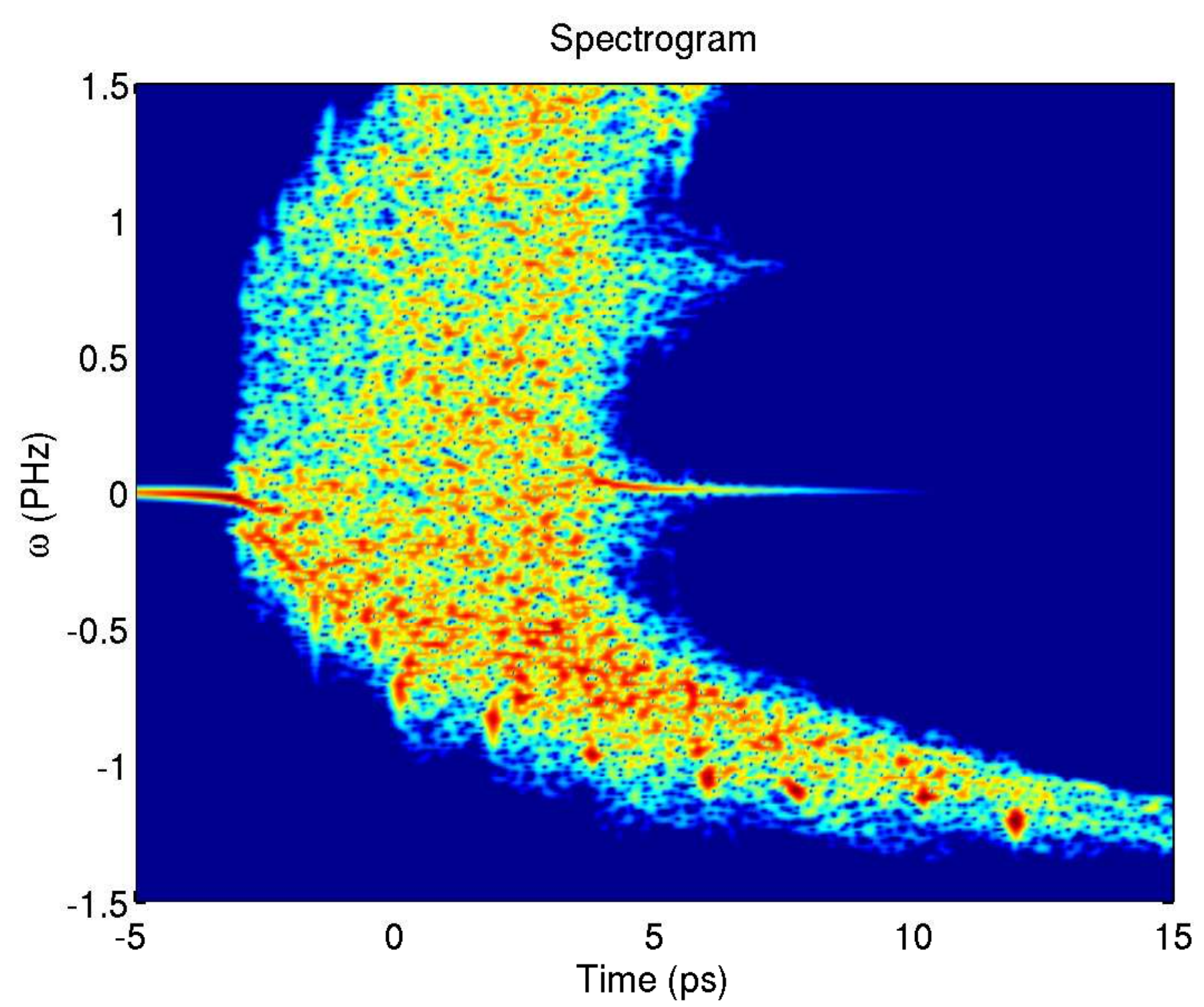}
\caption{\small {\it (Color online) spectrograms of output signals of the MI development from initial sech-shaped pulse with 13kW peak power inside the PCF described by (from left to right) NLS+HD, NLS+HD+SS and NLS+HD+SS+RS systems respectively.}}
\end{figure}

In our numerical simulations the applicability of this scenario of MI development was verified for all considered nonlinear systems NLS+HD, NLS+HD+SS and NLS+HD+SS+RS for initial pulses with peak power from 50W to 13kW. No qualitative changes were noticed. Enlarged at center parts output spectrograms for NLS+HD, NLS+HD+SS and NLS+HD+SS+RS systems for initial pulse with peak power of 13kW with clear sings of quasi-solitons are shown on Fig.7. Thereby, the hypothesis that the MI development qualitatively changes from the situation when quasi-solitons affect MI deeply at the initial pulse peak power less than 100W to the situation when no sings of quasi-solitons are present at the initial pulse peak power more than 1kW (compare with \cite{Picozzi}) is shown to be incorrect.

\begin{figure}[t] \centering
\includegraphics[width=130pt]{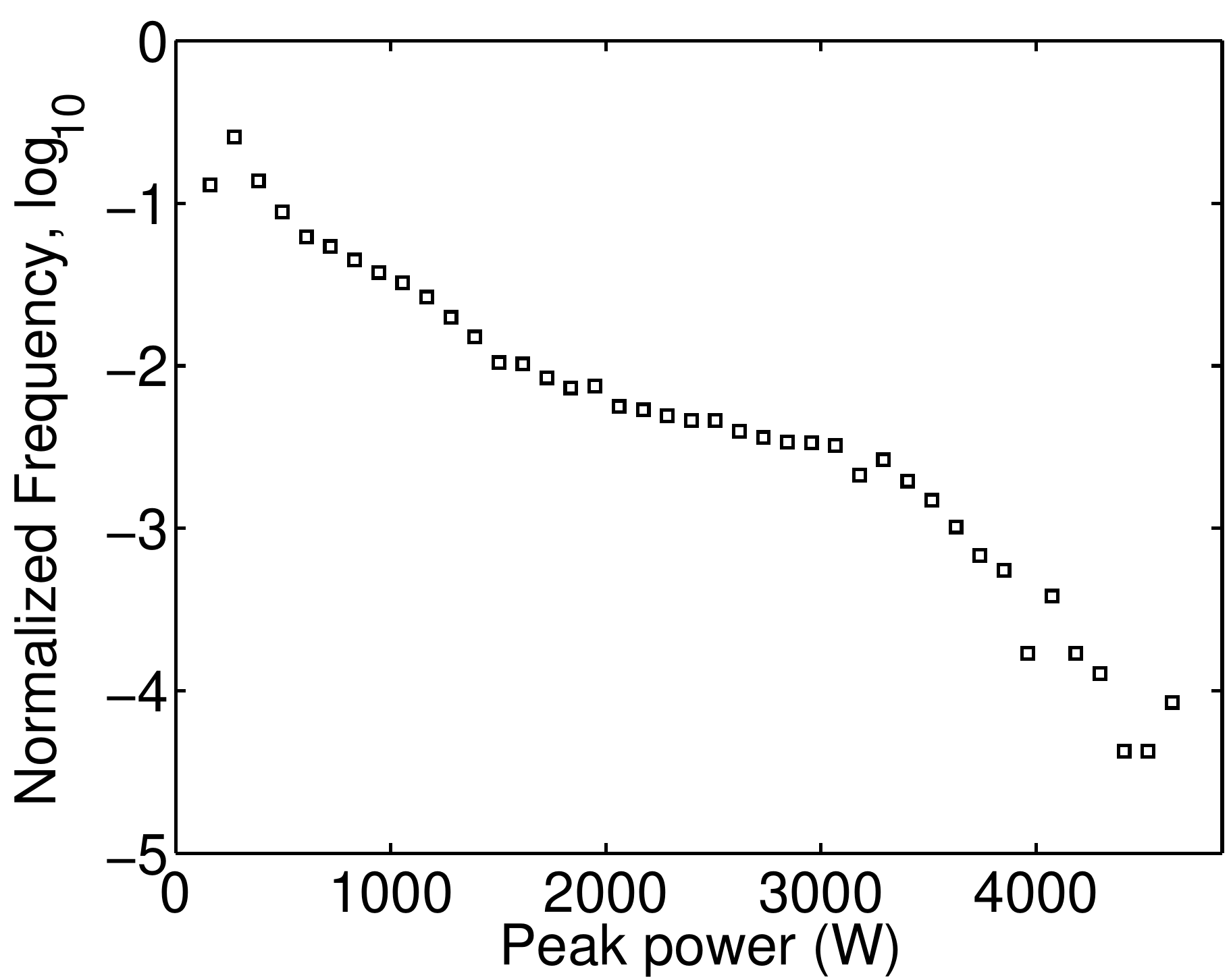}
\includegraphics[width=130pt]{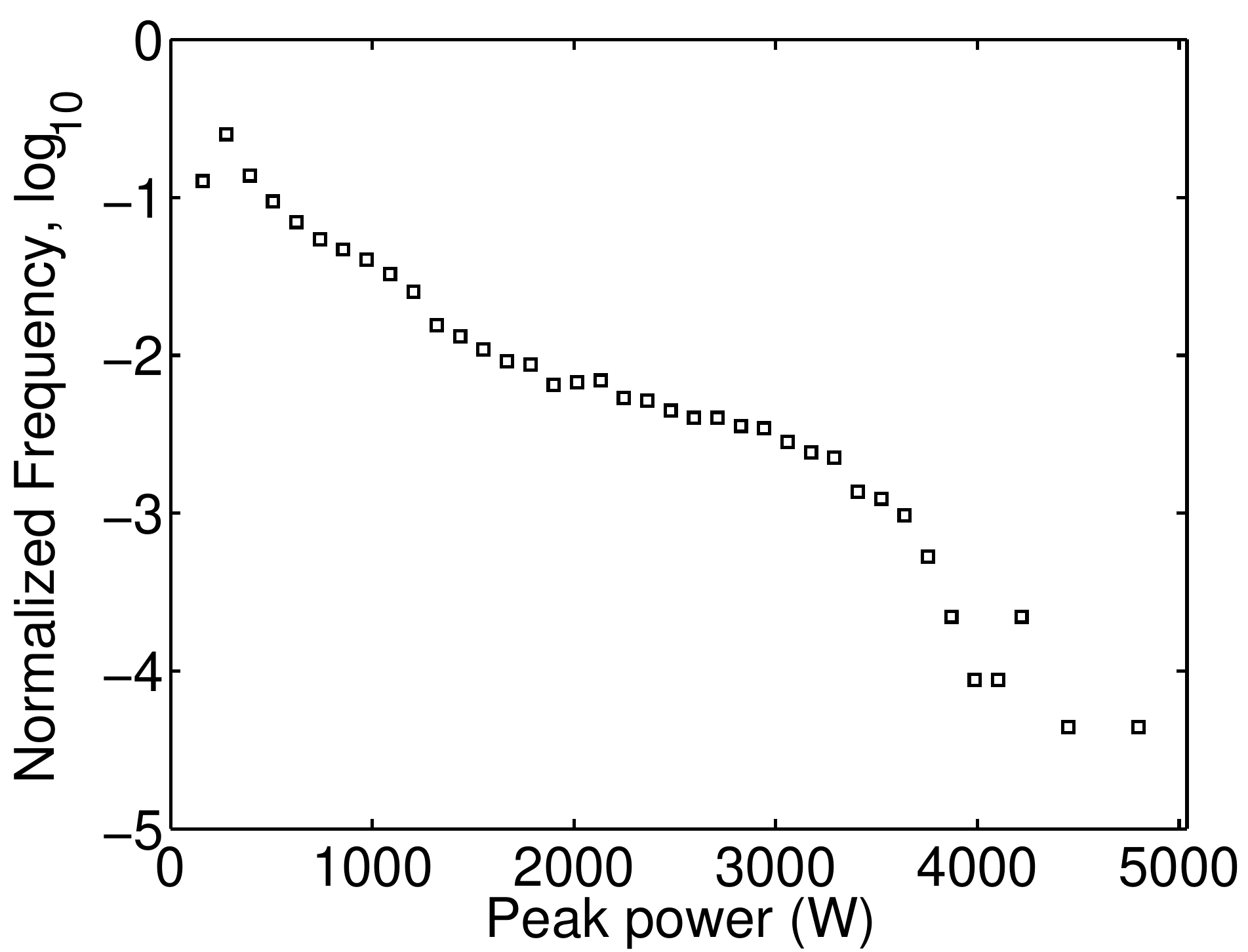}
\includegraphics[width=130pt]{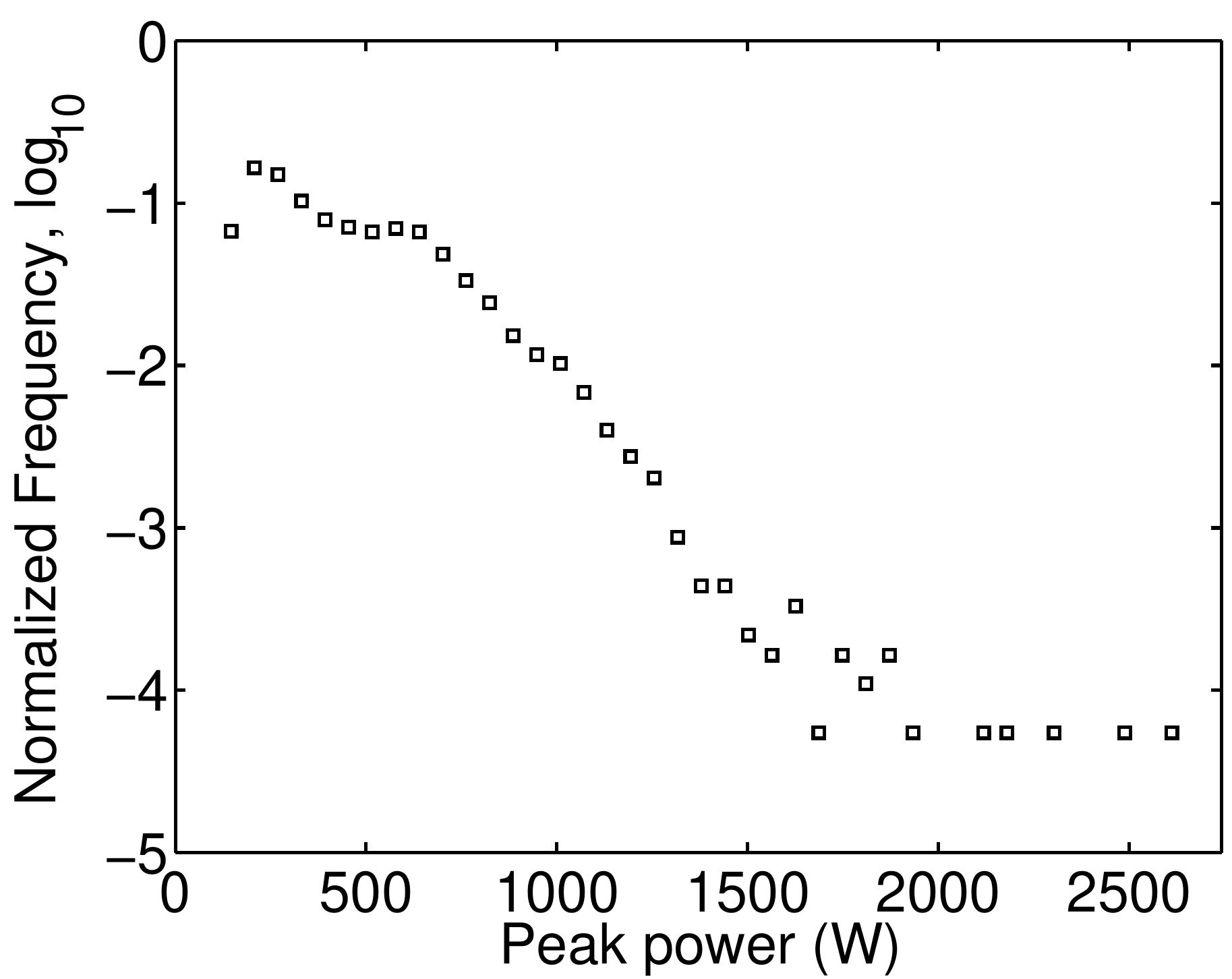}
\caption{\small {\it Normailized appearance frequency in logarithmic scale depending on quasi-solitons peak power for the MI development from initial sech-shaped pulse with 200W peak power inside the PCF described by (from left to right) NLS+HD, NLS+HD+SS and NLS+HD+SS+RS systems respectively.}}
\end{figure}

To check whether Raman scattering affects the frequency of rogue waves appearance, for each of the 3 systems NLS+HD, NLS+HD+SS and NLS+HD+SS+RS there were made ensembles of 1000 simulations with different noise seeds. The initial pulse was chosen to be the same as it was for the third group of experiments devoted to MI development with peak power $P_{0}=200W$. Input pulse noise was included in the frequency domain through one photon per mode spectral density $h\nu/\Delta\nu$ on each spectral discretization bin $\Delta\nu$. The thermal Raman noise was not included because, as was shown before, it doesn't have significant influence on the statistical properties of rogue waves (see \cite{Dudley1, Chang}). Individual peaks of each output realization were isolated using the similar technique as was developed in \cite{Solli}. As shown on Fig.8, the frequency distribution of the pulse peak power exponentially decays with the wave amplitude for NLS+HD and NLS+HD+SS systems, while in presence of Raman scattering the existence of long non-exponential tail in the region of extra-high peak powers was found. In other words, the addition of Raman scattering yielded in great increase of rogue waves appearance frequency. We suppose that the mechanism of the rogue waves appearance in the presence of Raman scattering lies in the nonlinear energy transfer from smaller quasi-solitons to the bigger ones, as was shown in the soliton-to-soliton collisions experiments.

The author thanks J.Dudley for usage of part of the code concerning Raman scattering and also F.Dias for valuable discussions concerning the numerical simulations. The work of the author was supported by the MANUREVA project, the Program of Presidium of RAS "Fundamental problems of nonlinear dynamics", program of support for leading scientific schools of Russian Federation, and also RFBR Grants 07-01-92165-NTsNI\_a and 09-01-00631-a.

\end{document}